\newcommand*{\ATLASLATEXPATH}{}
\pdfoutput=1
\documentclass[cernpreprint, texlive=2016, USenglish]{\ATLASLATEXPATH atlasdoc}
\usepackage{\ATLASLATEXPATH atlaspackage}
\usepackage{\ATLASLATEXPATH atlasbiblatex}

\usepackage{\ATLASLATEXPATH atlasphysics}

\newcommand{\PbPb}{\mbox{Pb+Pb}}
\newcommand{\pbpb}{\mbox{Pb+Pb}}

\newcommand{\pTmin}{\mbox{$p_{\mathrm{T,min}}$}}
\newcommand{\pTmax}{\mbox{$p_{\mathrm{T,max}}$}}

\newcommand{\Ncone}{\mbox{$N_{\mathrm{cone}}$}}

\newcommand{\Njet}{\mbox{$N_{\mathrm{jet}}$}}
\newcommand{\Njetmeas}{\mbox{$N_{\mathrm{jet}}^{\mathrm{meas}}$}}       % added by DPS
\newcommand{\Nch}{\mbox{$N_{\mathrm{ch}}$}}

\newcommand{\nchUE}{\mbox{$n_{\mathrm{ch}}^{\mathrm{UE}}$}}
\newcommand{\nchmeas}{\mbox{$n_{\mathrm{ch}}^{\mathrm{meas}}$}}
\newcommand{\nchsub}{\mbox{$n_{\mathrm{ch}}^{\mathrm{sub}}$}}
\newcommand{\nchunf}{\mbox{$n_{\mathrm{ch}}^{\mathrm{unfolded}}$}}
\newcommand{\nch}{\mbox{$n_{\mathrm{ch}}$}}

\newcommand{\sqrtsnn}{\mbox{$\sqrt{s_{_\text{NN}}}$}}
\newcommand{\sqrts}{\mbox{$\sqrt{s}$}}

\newcommand{\pythiaeight}{\textsc{Pythia}8}

\newcommand{\powheg}{\textsc{Powheg}}
\newcommand{\hijing}{\textsc{HIJING}}

\newcommand{\RFour}{\mbox{$R = 0.4$}}

\newcommand{\RTwo}{\mbox{$R= 0.2$}}

\newcommand{\ETfcal}{\mbox{$\Sigma E_{\mathrm{T}}^{\mathrm{FCal}}$}}

\newcommand{\Dz}{\mbox{$D(z)$}}

\newcommand{\Dzsub}{\mbox{$D^{\mathrm{sub}}(z)$}}
\newcommand{\Dpt}{\mbox{$D(\pT)$}}

\newcommand{\Dptsub}{\mbox{$D^{\mathrm{sub}}(\pttrk)$}}

\newcommand{\pTch}{\mbox{$p_{{\mathrm{T}}}^{\mathrm{ch}}$}}
\newcommand{\pTtrk}{\mbox{$p_{{\mathrm{T}}}^{\mathrm{ch}}$}}
\newcommand{\pttrk}{\mbox{$p_{\mathrm{T}}^{\mathrm{ch}}$}}

\newcommand{\pTtrue}{\mbox{$p_{\mathrm{T}}^{\mathrm{truth}}$}}

\newcommand{\pPb}{\mbox{$p$+Pb}}
\newcommand{\ptjet}{\mbox{$p_{\mathrm{T}}^{\mathrm{jet}}$}}
\newcommand{\etatrk}{\mbox{$\eta^{\mathrm{ch}}$}}
\newcommand{\phitrk}{\mbox{$\phi^{\mathrm{ch}}$}}

\newcommand{\yjet}{\mbox{$y^{\mathrm{jet}}$}}
\newcommand{\Lres}{\mbox{$L_{\mathrm{res}}$}}
\newcommand{\Rres}{\mbox{$R_{\mathrm{res}}$}}

\renewcommand{\_}{{\tt \char`\_}}  % works properly only in \tt mode (!)
\newcommand{\z}{\mbox{$z$}}

\newcommand{\cnchUE}{\mbox{$\tilde{n}_{\mathrm{ch}}^{\mathrm{UE+fake}}$}}

\graphicspath{{}{}}

\AtlasTitle{Measurement of jet fragmentation in Pb+Pb and \pp\ collisions at $\sqrtsnn = 5.02$~\TeV\ with the ATLAS detector}

\AtlasAbstract{%
This paper presents a measurement of jet fragmentation functions in
   0.49~nb$^{-1}$ of \PbPb\ collisions and 25~pb$^{-1}$ of \pp\ collisions at $\sqrtsnn = 5.02$~\TeV\ collected in 2015 with the ATLAS detector at the LHC.  These
   measurements provide insight into the jet quenching process
   in the quark--gluon plasma created in the aftermath of ultra-relativistic collisions
   between two nuclei.  The modifications to the jet fragmentation functions are quantified by
   dividing the measurements in \PbPb\ collisions by baseline 
measurements in \pp\ collisions. This ratio is studied as a function of the transverse momentum of the jet,
   the jet rapidity, 
   and the centrality of the collision.  In both collision systems, the jet fragmentation functions are measured for jets with transverse momentum between 126~\GeV\ and 398~\GeV\ and with an absolute value of jet rapidity less than 2.1. 
   An enhancement of particles carrying a small fraction of the jet momentum is observed, which increases
   with centrality and with increasing jet transverse momentum.  Yields of particles
   carrying a very large fraction of the jet momentum are also observed to be enhanced.  
Between these two enhancements of the fragmentation functions a  suppression
of particles carrying an intermediate fraction of the jet momentum is observed in \pbpb\ collisions.
   A small dependence of the modifications on jet rapidity is observed.

}

\AtlasRefCode{HION-2017-04}

\PreprintIdNumber{CERN-EP-2018-096}

\AtlasJournalRef{Phys. Rev. C 98 (2018) 024908}
\AtlasDOI{10.1103/PhysRevC.98.024908}
\hypersetup{pdftitle={ATLAS document},pdfauthor={The ATLAS Collaboration}}

\begin{document}

\maketitle

\tableofcontents

%-------------------------------------------------------------------------------
\section{Introduction}
\label{sec:intro}
Ultra-relativistic nuclear collisions at the Large Hadron Collider (LHC) produce hot dense matter
called the quark--gluon plasma (QGP); 
recent reviews can be found in Refs.~\cite{Roland:2014jsa,Busza:2018rrf}.
Hard-scattering processes occurring in these collisions produce jets which 
traverse and interact with the QGP.
The study of modifications of jet rates and properties in
heavy-ion collisions compared to \pp\ collisions provides information about 
the properties of the QGP.

The rates of jet production
are observed to be reduced by approximately a factor of two in lead--lead (\PbPb) collisions at LHC energies
compared to 
expectations from 
the jet production cross-sections measured in \pp\ interactions scaled by the 
nuclear overlap function of \PbPb\ 
collisions~\cite{Abelev:2013kqa,Aad:2014bxa,Khachatryan:2016jfl}.  
Similarly, back-to-back dijet~\cite{Aad:2010bu,Chatrchyan:2011sx,Aaboud:2017eww} 
and photon--jet pairs~\cite{Chatrchyan:2012gt} are observed to have
unbalanced transverse momentum in \pbpb\ collisions compared to \pp\ collisions. 
Related phenomena were first observed at the Relativistic Heavy Ion Collider
where the measurements were made with hadrons rather than reconstructed 
jets~\cite{Adcox:2001jp,Adler:2002tq,Abelev:2009gu}.
These observations imply that some of the energy of the parton showering process is
transferred outside of the jet through its interaction with the QGP.  This has been termed ``jet quenching''.

The distribution of particles within the jet are affected by this mechanism of energy loss. Several related observables sensitive to the properties of the medium can be constructed. Measurements of the jet shape~\cite{Chatrchyan:2013kwa} and  the fragmentation 
functions were made in 2.76~\TeV\ \pbpb\ 
collisions~\cite{Aad:2014wha,Chatrchyan:2014ava,Aaboud:2017bzv}.  
In Ref.~\cite{Aaboud:2017bzv}, jet fragmentation functions are 
measured as a function of both the charged-particle transverse momentum, \pT, and 
the charged-particle longitudinal momentum fraction relative to the jet,
\begin{equation}
\z \equiv \pt \cos \Delta R \: / \:\ptjet.
   \label{eq:zdef}
\end{equation}
The fragmentation functions are defined as:
  \begin{equation*}
\Dz \equiv \dfrac{1}{\Njet} \dfrac{d \nch}{d z},
\label{eq:ffuncdef}
  \end{equation*}
and
  \begin{equation*}
\Dpt \equiv \dfrac{1}{\Njet} \dfrac{d \nch}{d \pt}, \label{eq:ffuncdefpt}
  \end{equation*}
where \ptjet\ is the transverse momentum of the jet, 
\nch\ is the number of charged particles in the jet, \Njet\ is the number of jets under consideration,
 and  $\Delta R = \sqrt{(\Delta\eta)^{2}+(\Delta\phi)^{2}}$ with $\Delta \eta$ and $\Delta \phi$ 
defined as the differences between the jet axis and the charged-particle direction in
pseudorapidity and azimuth,\footnote{
  ATLAS uses a  
right-handed coordinate system with its origin at the nominal
interaction point (IP) in the center of the detector and the
$z$-axis along the beam pipe. The $x$-axis points from the IP to the
center of the LHC ring, and the $y$-axis points upward. Cylindrical
coordinates $(r,\phi)$ are used in the transverse plane, $\phi$
being the azimuthal angle around the beam pipe. The pseudorapidity
is defined in terms of the polar angle $\theta$ as
$\eta=-\ln\tan(\theta/2)$. The rapidity is defined as $y = 0.5 \ln [(E+p_z)/(E-p_z)]$ where
$E$ and $p_z$ are the energy and the component of the momentum along the beam direction.} respectively. 
In order to quantify differences between \PbPb\ and \pp\ collisions at the same
collision energy, the ratios of the fragmentation functions are measured:
\begin{equation*}
R_{D(\z)} \equiv \frac{\Dz_{\mathrm{PbPb}}}{\Dz_{pp}},
\label{eq:rdz} 
\end{equation*}
and
\begin{equation*}
R_{D(\pT)} \equiv \ \frac{\Dpt_{\mathrm{PbPb}}}{\Dpt_{pp}}.
\label{eq:rdpt}
\end{equation*}
Relative to jets in \pp\ collisions, it was found in Ref.~\cite{Aaboud:2017bzv} that jets in \pbpb\ collisions 
have an excess of particles with transverse momentum below 4~\GeV\ and an excess of particles carrying 
a large fraction of the jet transverse momentum.  
At intermediate charged-particle \pT, there is a suppression of the charged-particle yield.  
At the same time, an excess of low-\pT\ particles is observed for particles in a wide region around the jet cone~\cite{Khachatryan:2016tfj,Sirunyan:2018jqr}. 
These observations 
may indicate that the energy lost by jets through the jet quenching process is being transferred to soft particles within and around the jet~\cite{Qin:2015srf,Blaizot:2014ula}; measurements of these soft particles have the potential to constrain the models describing such processes.
A possible explanation for the enhancement of particles carrying a large fraction of the jet momentum is
that it is related to the gluon-initiated jets losing more energy than quark-initiated jets. This leads to a higher quark-jet fraction in \pbpb\ collisions than in \pp\ collisions.
The change in flavor composition, combined with the different shapes of the quark and gluon fragmentation functions~\cite{Spousta:2015fca} then lead to the observed excess.

Proton--nucleus 
collisions, which do not generate a large amount of QGP,
are used to differentiate between initial- and final-state effects due to the QGP formed in \PbPb\ collisions. Fragmentation functions in \pPb\ collisions show no evidence of
modification when compared with those in \pp\ collisions~\cite{Aaboud:2017tke}. Thus, any modifications observed in \PbPb\ collisions can be attributed to the presence of the QGP rather than to effects arising from the presence of the large nucleus.

The rapidity dependence of jet observables in \pbpb\ collisions is of great interest,
in part because at fixed \ptjet\ the fraction of quark jets increases with increasing $|\yjet|$
(see, for example, Refs.~\cite{Spousta:2015fca,Aad:2016oit}).  This makes the rapidity
dependence of jet observables potentially sensitive to the different interactions of quarks and gluons with
the QGP.  Previous measurements of the rapidity dependence of jet fragmentation functions at 
$\sqrtsnn = 2.76$~\TeV\ in \pbpb\ collisions found a rapidity dependence of the 
fragmentation function modification with limited significance~\cite{Aaboud:2017bzv}.

In this paper, the fragmentation functions and the $R_{D(\z)}$ and $R_{D(\pT)}$ ratios are measured in \pbpb\ and \pp\
collisions at 5.02~\TeV\ using 0.49~nb$^{-1}$ of \pbpb\ collisions and 25~pb$^{-1}$ of \pp\ collisions collected in 2015.
Jets are measured over a rapidity range of $|\yjet| <$~2.1 using the \antikt\ reconstruction algorithm~\cite{Cacciari:2008qp}
with radius parameter \RFour\ .  The measurement is presented in intervals of \ptjet, \yjet, and collision centrality.
These data extend the previous studies at $\sqrtsnn = 2.76$~\TeV\ in two ways.
First, an increase in the peak energy density of the medium is expected.  
Second, 
the \pbpb\ integrated luminosity in the current dataset is 
3.5 times the integrated luminosity available at 2.76~\TeV, and the increase in the collision energy also increases the jet cross-sections.
These two factors allow a measurement of the dependence of jet fragmentation functions on
the transverse momentum of the jet over a wider range than was previously possible.

%-------------------------------------------------------------------------------
\section{Experimental setup}
\label{sec:setup}
The measurements presented in this paper were performed using the ATLAS inner detector, calorimeter, trigger and data acquisition systems~\cite{Aad:2008zzm}.
The calorimeter system consists of a sampling liquid argon (LAr) 
electromagnetic (EM) calorimeter covering $|\eta|<3.2$, a
steel/scintillator sampling hadronic calorimeter covering $|\eta| <1.7$, 
LAr hadronic calorimeters covering $1.5 < |\eta| < 3.2$, and
two LAr forward calorimeters (FCal) covering $3.1 < |\eta| <
4.9$~\cite{Aad:2008zzm}. 
The EM calorimeters are segmented longitudinally in shower depth into three
layers with an additional presampler layer. They have segmentation in $\phi$ and $\eta$ that varies with layer and pseudorapidity. 
The hadronic calorimeters have three sampling layers longitudinal
in shower depth.

The inner detector measures charged particles 
within the pseudorapidity interval 
$|\eta|<2.5$ using a combination of silicon pixel detectors, silicon
microstrip detectors (SCT), and a straw-tube transition radiation
tracker (TRT),
all immersed in a 2~T axial magnetic field~\cite{Aad:2008zzm}. Each of the three detectors is
composed of a barrel and two symmetric endcap sections.
The pixel detector 
is composed of four layers: the ``insertable B-layer''~\cite{ibl1,ibl2}
and three layers  with a pixel size of $50~\mu {\mathrm m} \, \times \, 400~\mu {\mathrm m}$. 
The SCT barrel section contains four layers of modules with 80~$\mu {\mathrm m}$ 
pitch sensors on both sides and each endcap consists of nine
layers of double-sided modules with radial strips having a mean pitch
of $80~\mathrm{\mu m}$. The two sides of each SCT layer in both the barrel
and the endcaps have a relative stereo angle of 40~mrad. The
TRT contains up to 73 (160) layers of
staggered straws interleaved with fibres in the barrel (endcap).

The zero-degree calorimeters (ZDCs) are located symmetrically at $z = \pm140$~m and cover $|\eta| > 8.3$.
They are constructed from tungsten absorber plates and \v Cerenkov light is transmitted via quartz fibers.
In \PbPb\ collisions the ZDCs primarily measure ``spectator'' neutrons, i.e.\ neutrons that do not interact
hadronically when the incident nuclei collide. A ZDC coincidence trigger is implemented by requiring
the pulse height from each ZDC to be above a threshold set to accept the single-neutron peak.

A two-level trigger system is used to select the \PbPb\ and \pp\ collisions. 
The first trigger level (L1) is hardware-based and implemented with custom electronics. 
The second level is the software-based high-level trigger (HLT) and is used to
further reduce the accepted event rate. Minimum-bias \PbPb\ events are recorded using a trigger defined by the logical OR of a L1 total energy trigger and the ZDC coincidence trigger. The total energy trigger required the total transverse energy measured in the calorimeter system to be greater than 50~\GeV\ in \pbpb\ collisions.  Jet events are selected by the HLT, after requiring the identification of a jet by the L1 jet trigger in \pp\ collisions or the total energy trigger with a threshold of 50~\GeV\ in \PbPb\ collisions. The L1 jet trigger utilized in \pp\ collisions required 
a jet with transverse momentum greater than 20~\GeV.  
The HLT jet trigger used a jet reconstruction algorithm similar to that used in the offline 
analysis (the offline jet reconstruction is discussed in Section~\ref{sec:reconstruction}). 
It selected events containing jets with transverse energy 
of at least 75~\GeV\ in \PbPb\ collisions and at least 
85~\GeV\ in \pp\ collisions. In \pp\ collisions, the 85~\GeV\ threshold jet trigger sampled the full 
delivered luminosity. The 75~\GeV\ threshold jet trigger used in \PbPb\ 
collisions was prescaled\footnote{The prescale indicates which fraction of events that passed the trigger selection was selected for recording by the data  acquisition system.} in a small part of the \PbPb\ data-taking period; 
however, the trigger
sampled more than 99\% of the total integrated luminosity. The measurement is performed in the 
jet transverse momentum region where the triggers are fully efficient.
%-------------------------------------------------------------------------------
\section{Data sets and event selection}
\label{sec:data}
The \PbPb\ and \pp\ data used in this analysis were recorded in 2015.
 The data samples consist of 
25~pb$^{-1}$ of $\sqrts = 5.02$~\TeV\ \pp\ data and 0.49~nb$^{-1}$ of $\sqrtsnn = 5.02$~\TeV\
\pbpb\ data. In \pbpb\ and \pp\ collisions, events are required to have a reconstructed vertex
within 150~mm of the nominal interaction point along the beam axis.
Only events taken during stable beam conditions and satisfying detector and data-quality requirements,
which include the calorimeters and inner tracking detectors being in nominal operation, are considered.

In \PbPb\ collisions, the event centrality reflects the overlap area of the two colliding nuclei and is characterized by \ETfcal, the total transverse energy deposited in the
FCal~\cite{Aaboud:2017tql}. The centrality intervals used in this analysis are defined according to successive percentiles of the \ETfcal\ distribution obtained from minimum-bias triggered \pbpb\ events ordered from the most central (highest \ETfcal) to the most peripheral collisions (lowest \ETfcal): 0--10\%, 10--20\%, 20--30\%, 30--40\%, 40--60\%, 60--80\%.

In addition to the jet-triggered sample, a separate \PbPb\ data sample was recorded with the minimum-bias trigger and two total transverse-energy triggers requiring 1.5~\TeV\ and 6.5~\TeV\ to enhance the rate of more central \PbPb\ events. This data sample is used to produce a \PbPb\ Monte Carlo (MC) events with conditions that match those registered while the data were recorded.

The performance of the detector and of the analysis procedure in \PbPb\ collisions is evaluated using 
1.8~$\times10^{7}$ 5.02~\TeV\ MC events. These were produced from minimum-bias \PbPb\ data events overlaid with 
hard-scattering dijet \pp\ events generated with \powheg{}+\pythiaeight~\cite{Nason:2004rx,Sjostrand:2014zea} using a  
set of tuned parameters called the A14 tune~\cite{ATLAS2014021} and the NNPDF23LO parton distribution function (PDF) set~\cite{Ball:2012cx}. 
The detector response was simulated using \textsc{Geant}4~\cite{Agostinelli:2002hh,Aad:2010ah} 
and the simulated hits were combined 
with those from the data event.
A weight is assigned to each MC event such that the event sample obtained from the minimum-bias trigger has the same centrality distribution as the sample collected by
the jet trigger. A separate sample of 1.8~$\times10^{7}$ simulated 5.02~\TeV\ \pythiaeight\ \pp\ hard-scattering 
events, generated
 with the same tune and PDFs as for the \pbpb\ MC sample, is used to evaluate the performance for measuring fragmentation functions in the \pp\ data. The contribution from additional collisions in the same bunch crossing is not included in the MC simulation.  
A sample of \pbpb\ events generated with \hijing\ version 1.38b~\cite{Wang:1991hta} 
is also used to evaluate the performance of
the track reconstruction.

%-------------------------------------------------------------------------------
\section{Jet and track selection}
\label{sec:reconstruction}
The jet reconstruction, underlying event (UE) determination, and subtraction procedures closely follow those used by
\mbox{ATLAS} for jet measurements in \pp\ and \PbPb\ collisions at 
$\sqrtsnn=2.76$~\TeV~\cite{Aad:2014bxa}.
The \antikt\ algorithm is first run in four-momentum recombination mode, on
$\Delta \eta \times \Delta \phi = 0.1\times 0.1$  calorimeter towers with the \antikt\ radius parameter $R=$~0.2 and $R=$~0.4. The energies in the towers are obtained by summing the
	energies of calorimeter cells at the electromagnetic energy scale within the tower boundaries.  Then, an iterative procedure is used to estimate the $\eta$-dependent UE transverse energy density on an event-by-event basis using the energy measurements in all calorimeter towers in the event while excluding the regions populated by jets.
	The resulting UE transverse energy density is modulated taking into account the presence of the azimuthal anisotropy of particle production~\cite{ATLAS:2012at}. The modulation includes contributions of the second-, third-, and fourth-order azimuthal anisotropy harmonics.
	Higher-order harmonics introduce negligible variation of the reconstructed jet energy.
	The UE transverse energy is subtracted from each calorimeter cell within the towers
	included in the reconstructed jet, and the four-momentum of the jet is updated accordingly.
	  Then, a jet $\eta$- and \pT-dependent  correction factor to the \ptjet\ 
	  derived from the simulation samples is applied to correct for the calorimeter energy
	  response~\cite{Aaboud:2017jcu}. An additional correction based on \textit{in situ} studies of jets recoiling against photons, $Z$ bosons, and jets in other regions of the calorimeter is 
	  applied~\cite{ATL-PHYS-PUB-2015-036,ATLAS-CONF-2016-110}. The same jet reconstruction procedure without the
	  azimuthal modulation of the UE is also applied to $pp$ collisions.

	  Jets are required to have a rapidity within $|\yjet|<$~2.1 so that all \RFour\ jet cones are 
	  contained within the inner detector's acceptance.
To prevent neighboring jets from distorting the measurement of the fragmentation functions, 
jets are rejected if there is another jet with higher \ptjet\ anywhere
within a distance $\Delta R < 1.0$. 
A correction is applied to reduce the effects of the broadening of the jet direction
measurement for \RFour\ jets due to the UE. The correction uses jets reconstructed with a smaller distance parameter, \RTwo\, since their angular resolution evaluated in MC studies is found to be less affected by the UE fluctuations than that of larger-$R$ jets. The jet direction is redefined as that of the closest \RTwo\ jet with $\ptjet>35$~\GeV\ and matching the original jet direction within $\Delta R =0.3$ of the \RFour\ jet, when such a matching jet is found. If no matching \RTwo\ jet is found the axis remains unchanged.

Charged-particle tracks are reconstructed from hits in the inner detector using the 
track reconstruction algorithm with settings optimized for the high hit density in heavy-ion 
collisions~\cite{Aaboud:2017all}.
Tracks used in this analysis  are required to have a total of at least 9 (11) hits in the silicon pixel and microstrip detectors for charged particles with pseudorapidity |\etatrk| $\leq$ 1.65 (|\etatrk| > 1.65).  At least one hit is required in one of the two innermost pixel layers. 
If the track trajectory passed through an active module in the innermost layer, then 
a hit in this layer is required. Furthermore, a track must 
have no more than two holes in the Pixel and SCT detectors together, where 
a hole is defined by the absence of a hit predicted by the track 
trajectory. 
All charged-particle tracks used in this analysis are required to have reconstructed transverse momentum $\pttrk > 1$~\GeV. In order to suppress the contribution from
secondary particles, the distance of closest approach of the track to the primary 
vertex in the transverse plane is required to be less than a value which varies from  0.45~mm at $\pttrk=4$~\GeV\ to 0.2~mm at $\pttrk=20$~\GeV, and at that point the track must be less than 1.0~mm from the primary vertex in the longitudinal direction.

\begin{figure}[ht]
\centerline{
\begin{tabular}{cc}
   \includegraphics[width=0.48\textwidth]{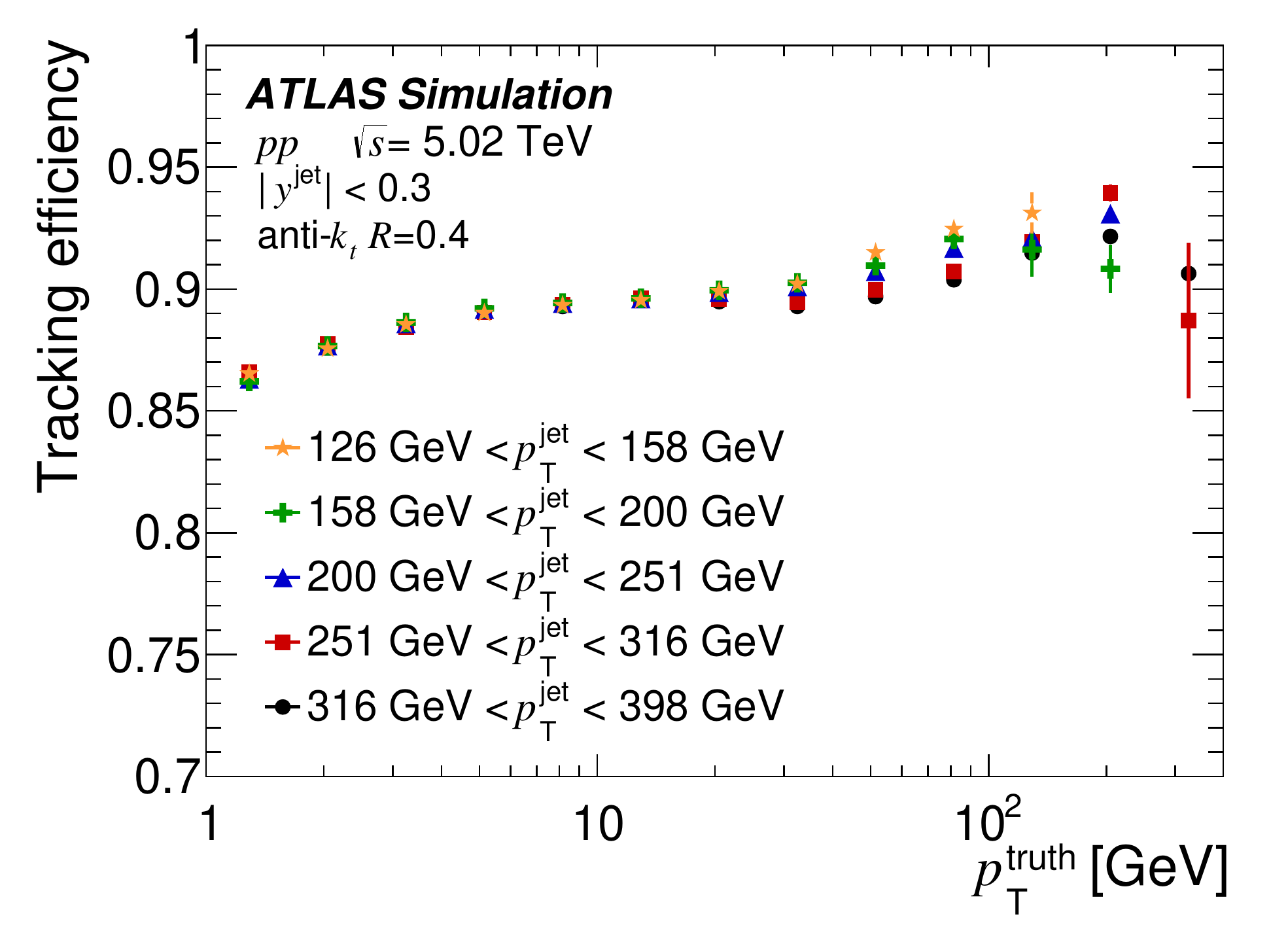} &
   \includegraphics[width=0.48\textwidth]{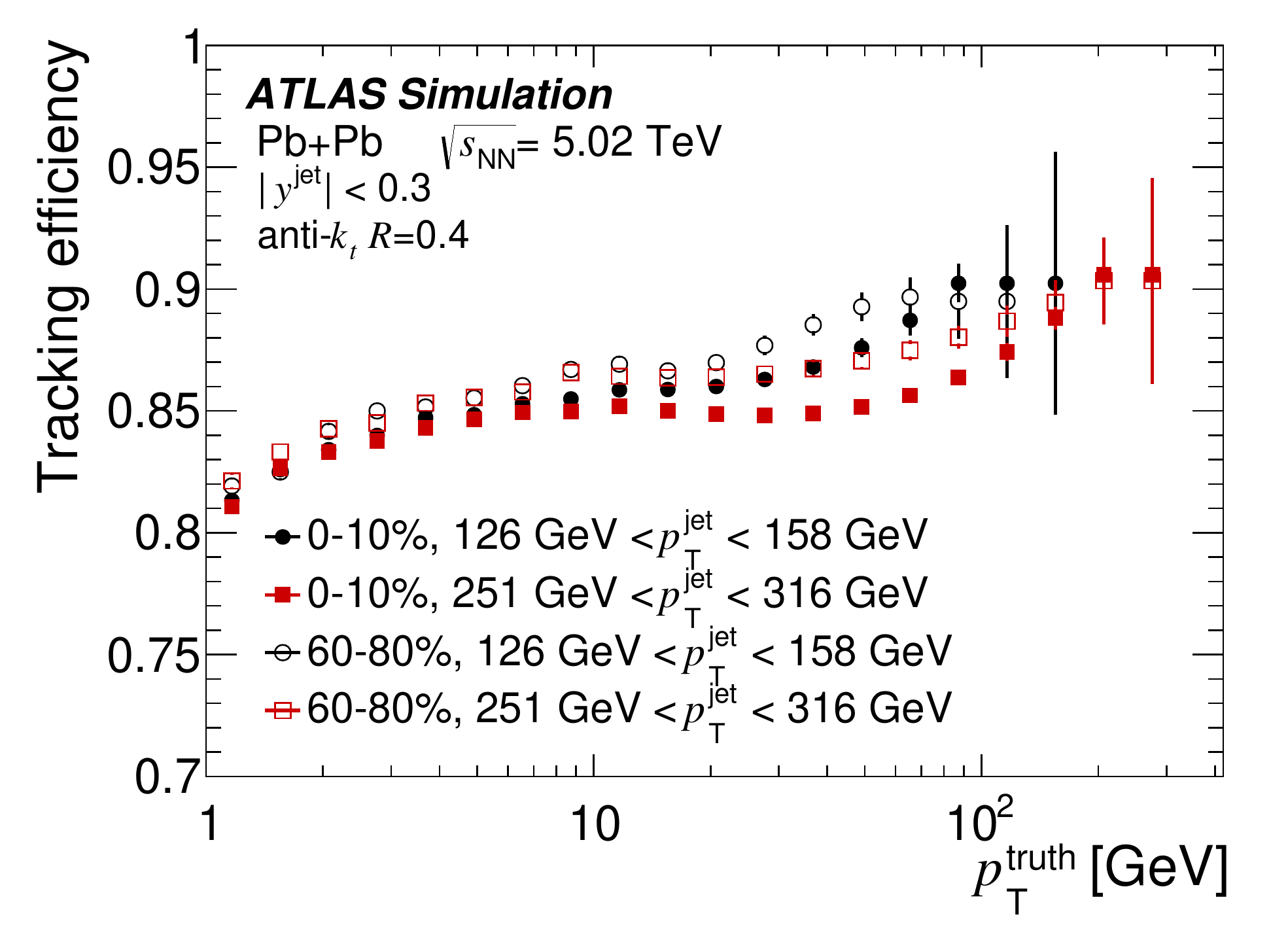} \\
  \\ 
\end{tabular} }
\caption{Tracking efficiency, $\varepsilon$, smoothed using a third-order polynomial in $\ln(\pTtrue)$ as a function of \pTtrue\ in \pp\ collisions in five different jet-\pT\ intervals (left) and in \PbPb\ collisions (right) in two different jet-\pT\ intervals and for 0--10\% and 60--80\% centrality intervals. In both plots the efficiency is evaluated for tracks within jets with $|\yjet|<0.3$.}
\label{fig:trackingeff}
\end{figure}

The efficiency, $\varepsilon(\pTtrue, \ptjet, \yjet)$, for reconstructing charged particles within jets in \PbPb\ and \pp\ collisions is evaluated from the matching of reconstructed tracks to generator-level primary particles\footnote{Primary particles are defined as particles with a mean lifetime $\tau>0.3\times 10^{-10}$~s either directly produced in \pp\ interactions or from subsequent decays of particles with a shorter lifetime. All other particles are considered to be secondary.} using MC samples described above.
The matching is based on contributions of generator-level particles to the hits in the detector layers. A reconstructed track is matched to a generator-level particle if it contains hits produced primarily by this particle~\cite{Aad:2010ah}. The efficiency is evaluated separately in four 
$|\yjet|$ intervals and each interval of reconstructed \ptjet\ used in the measurement. Furthermore, the efficiency is evaluated separately for each centrality interval
in the case of \PbPb\ collisions. The charged-particle reconstruction efficiencies as a function of the generator-level primary particle transverse momentum, \pTtrue, are shown in Figure~\ref{fig:trackingeff} for jets with 
$|\yjet|<0.3$ in \pp\ and \PbPb\ collisions.
In order to remove fluctuations in the efficiency due to the limited MC sample size, the \pTtrue\ dependence of the efficiencies is parameterized and smoothed using a third-order polynomial in $\log(\pTtrue)$ that gives a good description of the efficiency in the full range of \pTtrue. The efficiencies shown in 
Figure~\ref{fig:trackingeff} exhibit only a modest variation with \pTtrue, centrality, and \ptjet. A small almost continuous increase of the efficiency with the increasing \pTtrue\ is observed. 
The efficiency over the 20--100~\GeV\  \pTtrue\ range is smaller for high \ptjet\ compared to low \ptjet\ by about 2\% and 5\% in \pp\ and \PbPb\ collisions, respectively. 
This behavior is attributed to the higher probability to lose tracks in the dense core of high-\pT\ jets than to lose tracks that are more isolated~\cite{ATL-PHYS-PUB-2016-007}. The efficiency is lower
in more central \PbPb\ collisions due to the higher hit density. The efficiency exhibits only a small variation with \yjet\ in the region $|\yjet|<1.2$, and it decreases by approximately 10\% in the most forward \yjet\ interval.        

The contribution of reconstructed tracks which are not be matched to a generated primary particle in the MC samples of \pp\ collision events produced without
data overlay,
along with the residual contribution of tracks matched to secondary particles,
are together considered ``fake'' tracks. The fraction of fake tracks is less than 2\% over the full kinematic range of this measurement. A possible degradation of the tracking performance at high
occupancy is checked in the sample of \PbPb\ collision events simulated with the \hijing\ MC. No significant dependence of the rate of fake tracks on centrality is observed. The correction for the fake contribution is discussed in Section~\ref{sec:analysis}.

%-------------------------------------------------------------------------------
\section{Analysis procedure}
\label{sec:analysis}
The analysis procedure closely follows the one used in the measurement of jet fragmentation at $\sqrtsnn = 2.76$~\TeV~\cite{Aaboud:2017bzv}. Reconstructed  tracks are associated with a reconstructed jet if they 
fall within $\Delta R = 0.4$ of the jet axis and for each of
these particles the longitudinal momentum fraction, \z, is calculated.
The measured track yields,   $d \nchmeas/ d \z$ or $d \nchmeas/ d \pttrk$, are constructed as:
\begin{eqnarray*}
   \dfrac{d \nchmeas}{d \z} = \frac{\Delta \Nch(\z,\yjet,\ptjet)}{\Delta\z}
\label{eq:dzmeas}
\end{eqnarray*}
and
\begin{eqnarray*}
   \dfrac{d \nchmeas}{d \pttrk} = \frac{\Delta \Nch(\pttrk,\yjet,\ptjet)}{\Delta\pttrk},
\label{eq:dptmeas}
\end{eqnarray*}
where the quantities $\Delta \Nch(\z)$ and $\Delta \Nch(\pttrk)$ represent the number of associated tracks within the given \z\ or \pttrk\ range, respectively corrected for the track reconstruction efficiency. The efficiency correction is applied as a $1/\varepsilon(\pttrk, \ptjet, \yjet)$ weight on a track-by-track basis, assuming $\pttrk = \pTtrue$. While that assumption is not strictly valid, the efficiency varies sufficiently slowly with $\pTtrue$ that the error 
introduced by this assumption is less than 1\%.

Tracks which are not correlated with the jet need to be subtracted from the measured distributions; these tracks come from both fake tracks and the UE. In \PbPb\ collisions, contributions to the fragmentation functions from the charged particles originating from the UE in \pbpb\ collisions are subtracted. This contribution is evaluated as a function of charge particle \z\ or \pttrk, \yjet, \ptjet, and the collision centrality. Additionally, the measured track yields in \pp\ and \PbPb\ collisions are corrected for the presence of fake tracks. 

The UE contribution is determined for each measured jet using a grid of \RFour\ cones spanning the full coverage of the inner detector and following the method introduced in Ref.~\cite{Aad:2014wha}. The method is applied to events containing jets included in the analysis. The cones have a fixed distance between their centers chosen such that the inner detector acceptance is uniformly covered while avoiding overlaps. Any cone having a charged particle with $\pttrk>10$~\GeV\ or overlapping with a reconstructed jet with $\ptjet > 90$ \GeV\  is assumed to be associated with a hard process and is excluded from the UE estimation to avoid biasing it. The parameters defining the exclusion regions are evaluated in MC studies and are subjected to variations as part of the estimation of systematic uncertainties.
The resulting UE charged-particle yields,  $d \nchUE/ d \z$ or $d \nchUE/ d \pttrk$, are evaluated over $1 < \pttrk < 10$~\GeV\ according to:
	\begin{eqnarray*}
	\label{eq:background2}
		\dfrac{d \nchUE}{d \z} &=&
		\dfrac{1}{\Ncone}
	        \dfrac{1}{\varepsilon(\pttrk,\etatrk)}
	        \left.\dfrac{\Delta N_{\mathrm{ch}}^{\mathrm{cone}}(z,\ptjet,
		  \yjet)}{\Delta \z}\right\vert_{z=p_{\mathrm{T}}^{\mathrm{ch}} \cos{\Delta R} /p_{\mathrm{T}}^{\mathrm{jet}}},\\
	\label{eq:background1}
	\dfrac{d \nchUE}{d \pttrk} &=&
	\dfrac{1}{\Ncone}
        \dfrac{1}{\varepsilon(\pttrk,\etatrk)}
        \dfrac{\Delta N_{\mathrm{ch}}^{\mathrm{cone}}(\pttrk,\ptjet,
	  \yjet)}{\Delta \pttrk}.
	\end{eqnarray*}
Here \Ncone\ is the number of background cones used in the UE determination of a given jet, $\Delta N_{\mathrm{ch}}^{\mathrm{cone}}$ represents the number of charged particles summed over all background cones, and $\Delta R$ represents the distance between the center of a cone and the direction of a given charged particle. 
The term $\varepsilon(\pttrk,\etatrk)$ is the efficiency for reconstructing charged particles, estimated as a function of \pttrk\ and \etatrk\ without requiring track-to-jet matching.

The estimated contribution from the UE in each cone is corrected for the difference in the average yield of UE charged particles at a given \pttrk\ between the $\eta$ position of the cone and $\eta$ position of the jet. This correction is based on the centrality-, \pttrk-, and $\eta$-dependent distribution of charged-particle yields in minimum-bias
data events. An additional correction is applied to the charged-particle UE estimate to account for the difference in the azimuthal particle density, due to elliptic flow, between the $\phi$ angle of the cone and the $\phi$ angle of the jet. This  utilizes a centrality- and \pttrk-dependent parameterization of the measured elliptic flow coefficients~\cite{ATLAS:2012at}. 
 
The UE contribution is further corrected for the correlation between the actual UE charged particle yield underneath the jet and the jet energy resolution~\cite{Aad:2014wha}; in regions where the UE has an upward fluctuation, the jet energy resolution is worse. The smearing due to jet energy resolution leads to a net migration of jets from lower \ptjet\ to higher \ptjet\ values. The effect of the migration causes the actual UE contribution underneath the jet to be larger than that estimated from the procedure described above. This effect is corrected for by applying multiplicative correction factors, depending on \pTtrk\ or $z$, \yjet, \ptjet,  and collision centrality. 
The correction is estimated as a ratio of the UE charged particle yield evaluated by two different methods using the \PbPb\ MC samples. The first estimate uses the cone method discussed above. The second method calculates the UE contribution in the data overlay MC samples from tracks, within the area of a jet, that do not have an associated generated primary particle. The size of the correction is less than 2\% at low \z\ or \pttrk\ where the UE has the largest impact, and has only a small dependence on \ptjet.     

The contribution from fake tracks to the fragmentation functions is estimated from the MC samples without minimum-bias interactions overlaid. The fraction of these tracks is found to be below 2\% of the tracks that pass the selection in 
all track and jet kinematic regions in this analysis.

\begin{figure}[ht]
   \centerline{
      \begin{tabular}{c @{\hspace{0.1cm}} c @{\hspace{0.1cm}} c}
	 \includegraphics[width=0.333\textwidth]{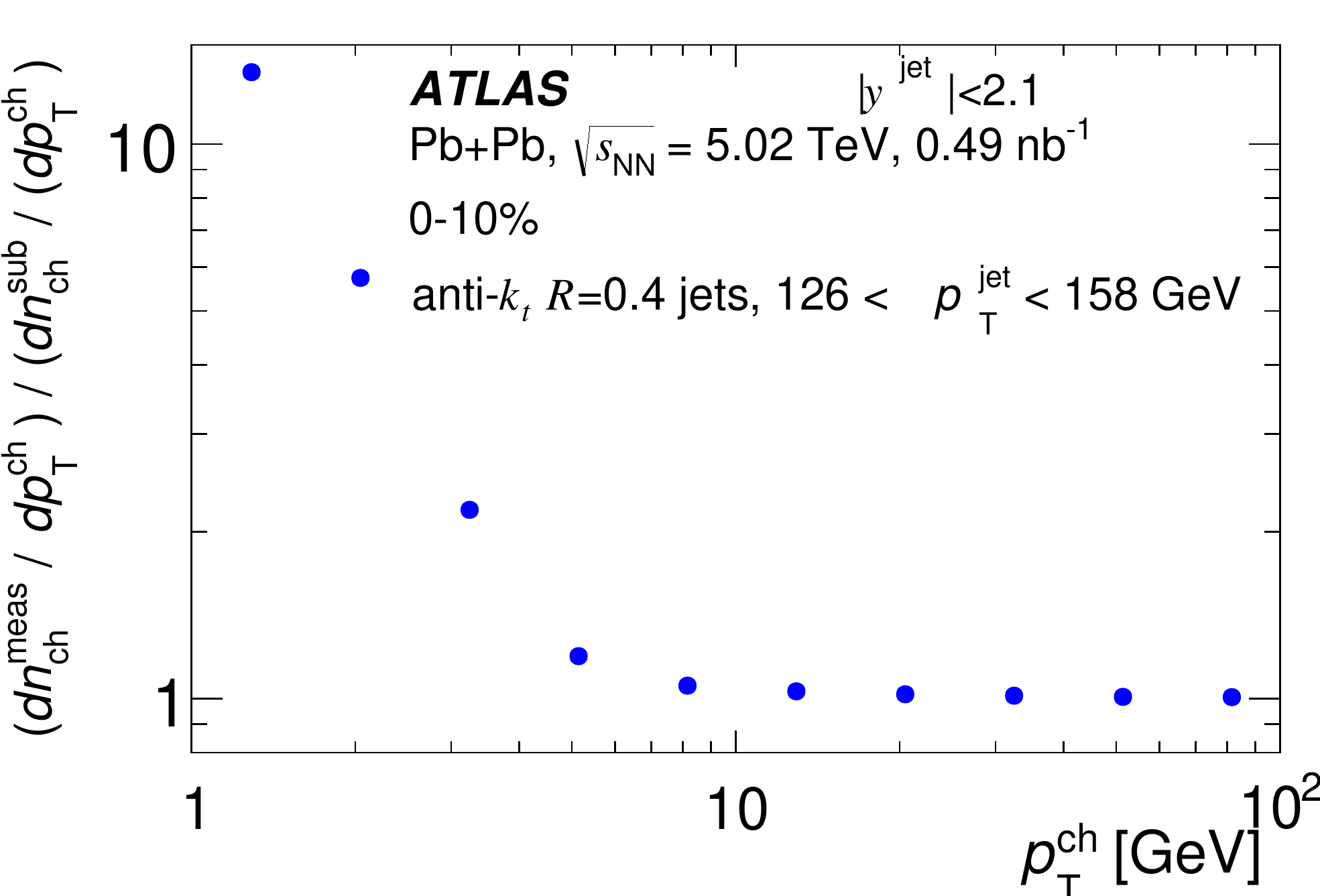} &
	 \includegraphics[width=0.333\textwidth]{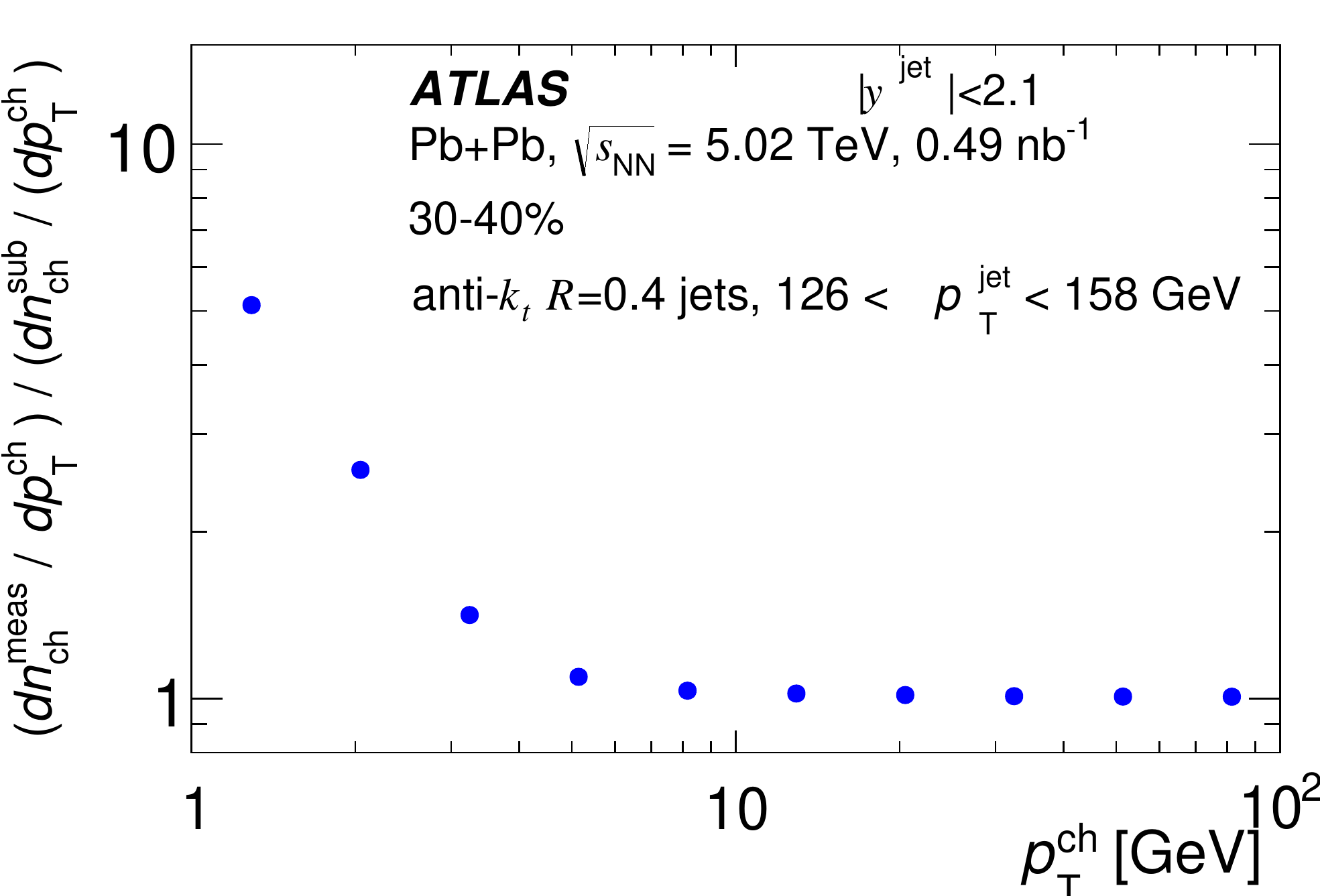} &
	 \includegraphics[width=0.333\textwidth]{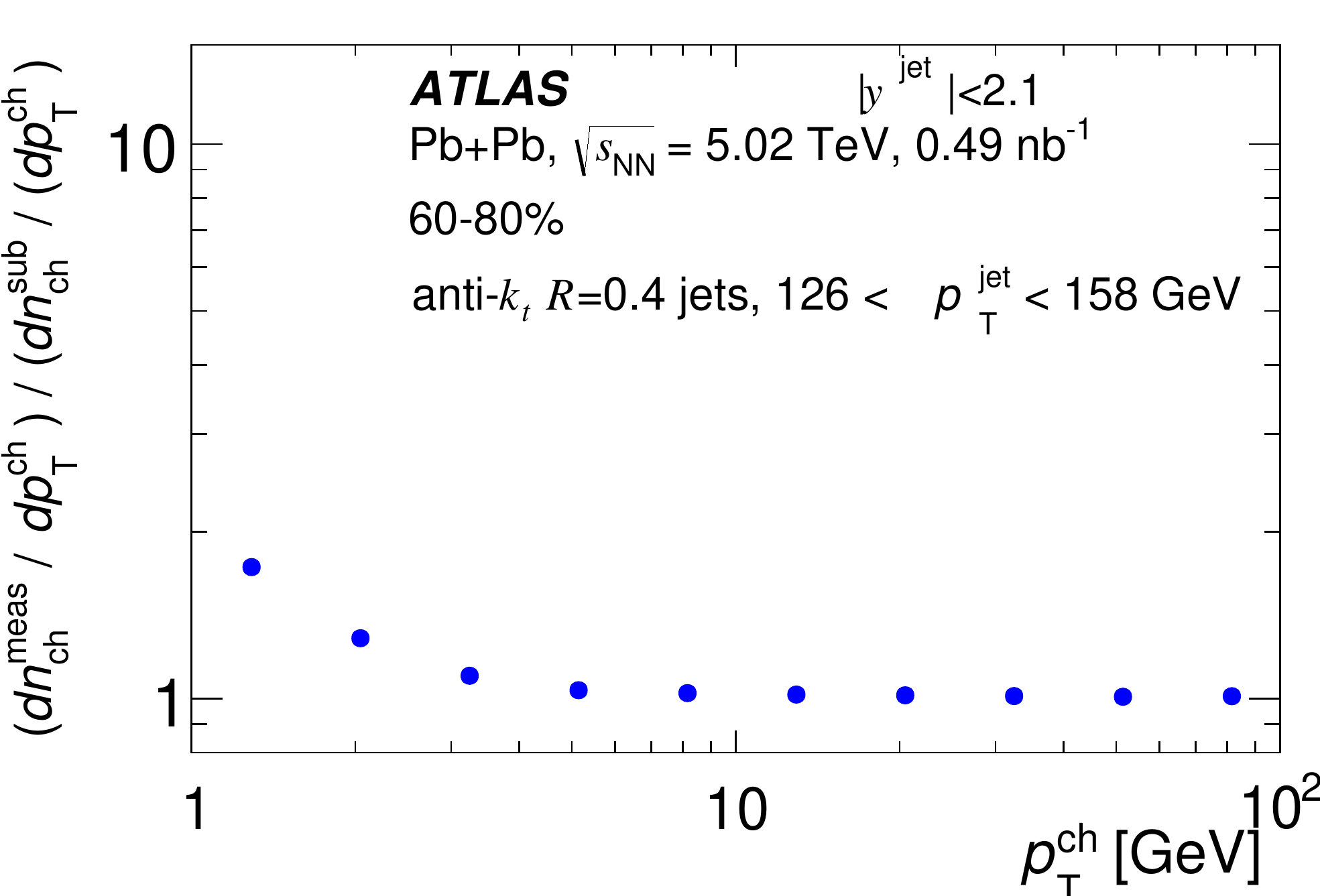} \\
	 \\
      \end{tabular} }
   \caption{Ratio of the measured charged-particle distributions before and after the subtraction of the UE and fake tracks as a function of \pttrk\ for \ptjet\ in the range 126--158~\GeV\
   for 0--10\% (left), 30-40\% (middle), and 60--80\% (right) centrality. The uncertainties are smaller than the marker size in all cases for which there is a significant UE.
   }
   \label{fig:UEsize}
\end{figure}

The UE distributions corrected for the additive contribution of fake tracks, $d \cnchUE / d \pttrk$ and $d \cnchUE / d z$, are then subtracted from 
the measured distributions, and the subtracted charged-particle yields and fragmentation functions are evaluated:

\begin{eqnarray*}
   \dfrac{d \nchsub}{d \z} &=& \frac{d \nchmeas}{d \z} - \frac{d \cnchUE}{d \z}, \\
   \Dzsub &=& \frac{1}{\Njetmeas}\frac{d \nchsub}{d \z},
\label{eq:dzmsub}
\end{eqnarray*}
and
\begin{eqnarray*}
   \dfrac{d \nchsub}{ d \pttrk} &=& \frac{d \nchmeas}{d \pttrk} - \frac{d \cnchUE}{d \pttrk}, \\
    \Dptsub &=& \frac{1}{\Njetmeas}\frac{d \nchsub}{d \pttrk},
\label{eq:dptsub}
\end{eqnarray*}
where $\Njetmeas$ is the total number of measured jets in a given \ptjet\ interval.
The signal-to-background ratio, $\nchsub/\nchUE$, strongly depends on the collision centrality and \pttrk. 
Figure~\ref{fig:UEsize} shows the distributions prior to the UE and fake-track subtraction,
$ \dfrac{d \nchmeas}{d \pttrk}$, divided by the distributions after the subtraction,
$ \dfrac{d \nchsub}{d \pttrk}$, as a function of \pttrk\ for three centrality selections.
In 0--10\% central collisions, the distributions prior to subtraction are over ten times larger
than the subtracted distributions
for the most extreme case of 1~\GeV\ charged particles.  This ratio is reduced to approximately 2
in peripheral collisions at the same charged-particle \pt.
The fake-track 
contribution to the fragmentation functions is subtracted from the measured fragmentation functions in both the \pp\ and \PbPb\ collisions; the UE subtraction is performed only for the \PbPb\ measurement as the UE contribution is negligible in the \pp\ collisions (less
than 2\% over the entire kinematic range measured).

\begin{figure}[ht]
\centerline{
\begin{tabular}{cc}
   \includegraphics[width=0.48\textwidth]{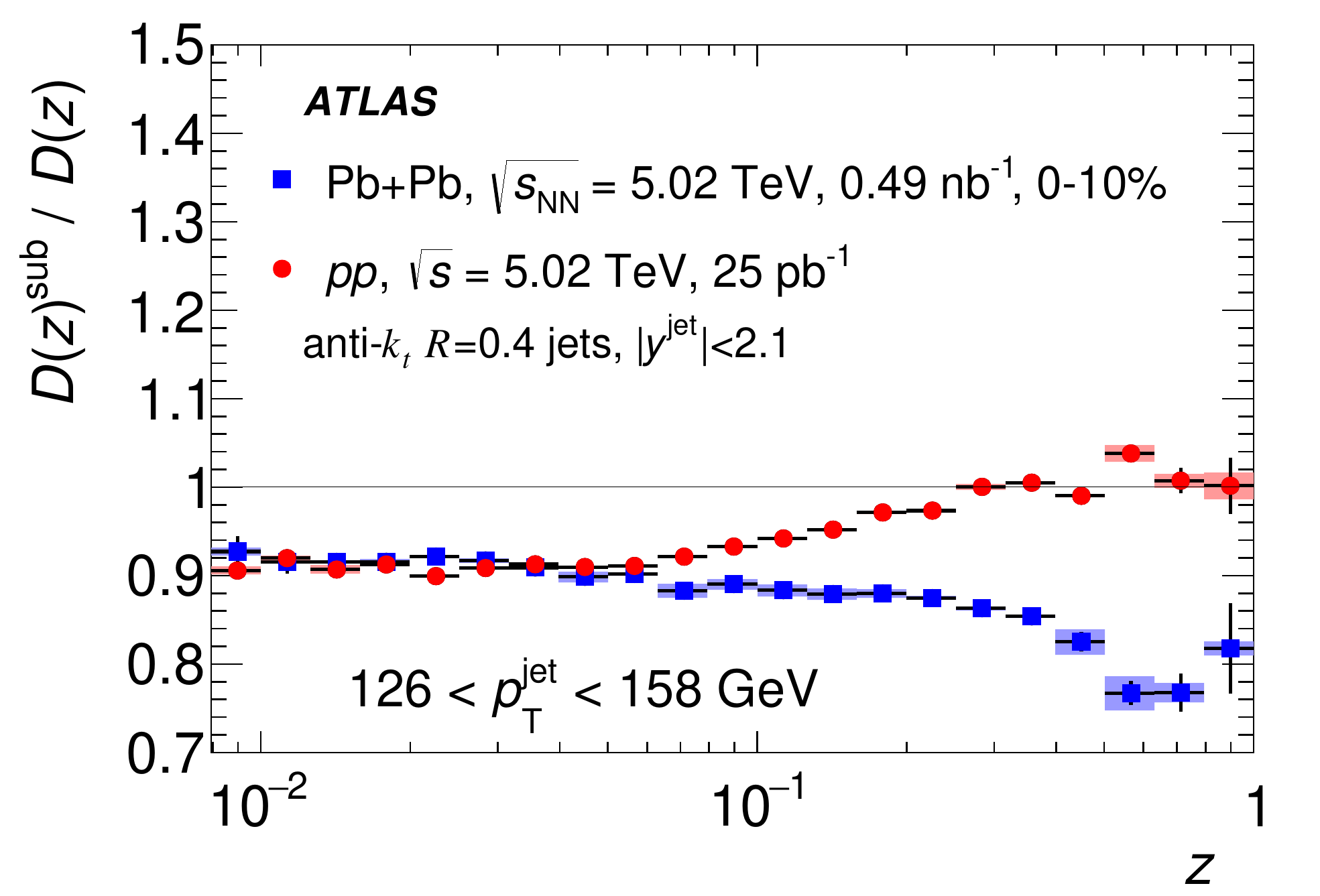} &
   \includegraphics[width=0.48\textwidth]{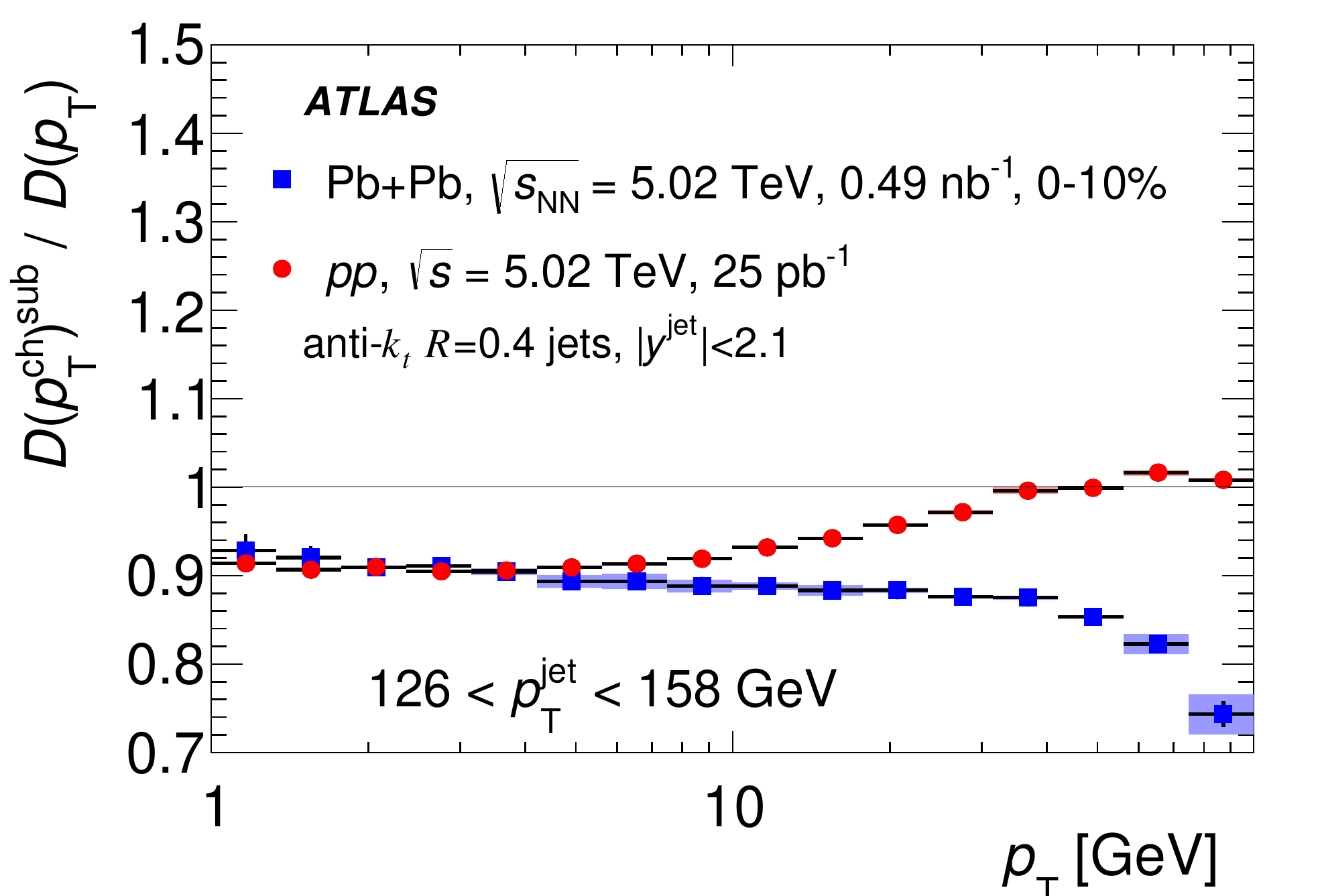} \\
   \includegraphics[width=0.48\textwidth]{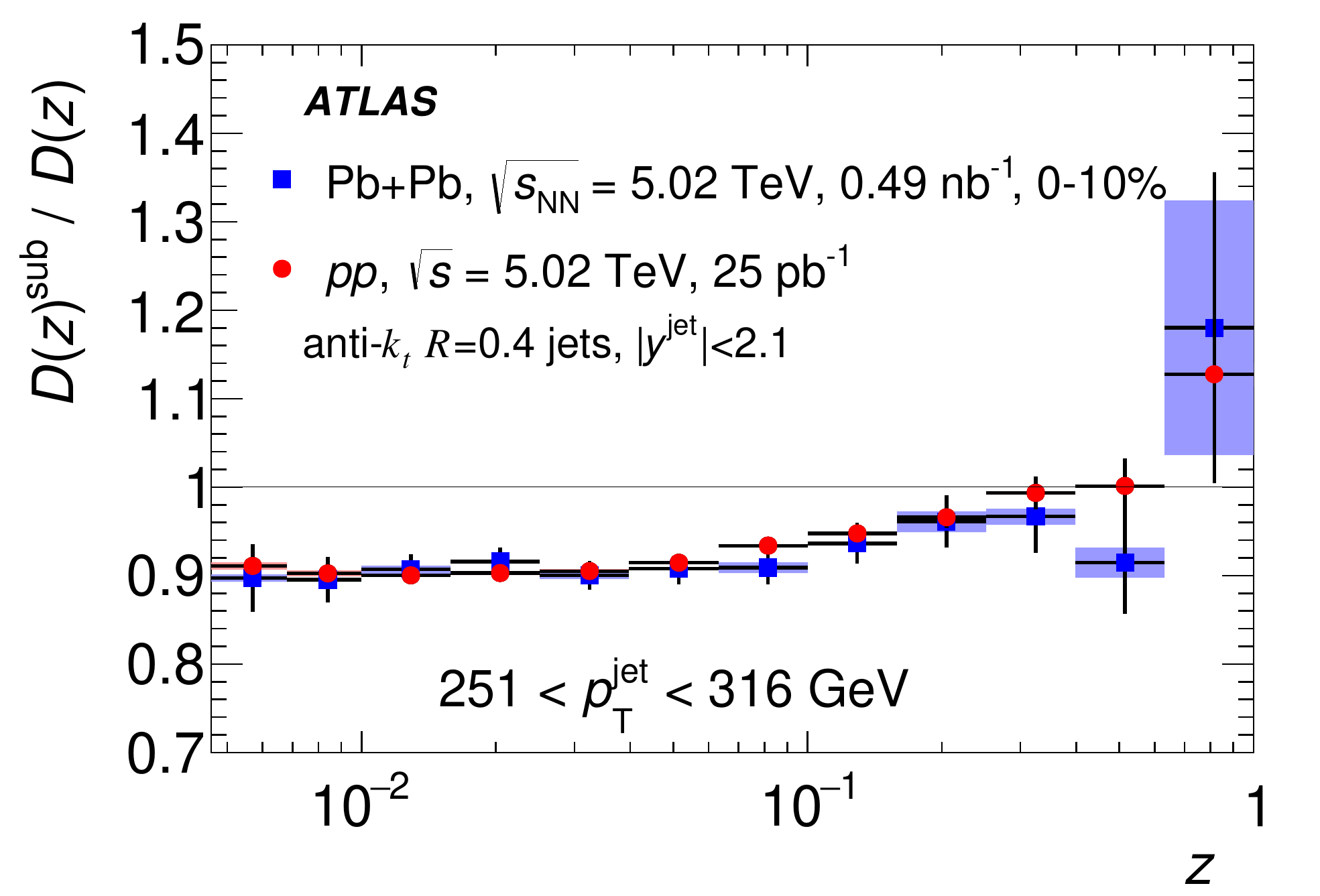} &
   \includegraphics[width=0.48\textwidth]{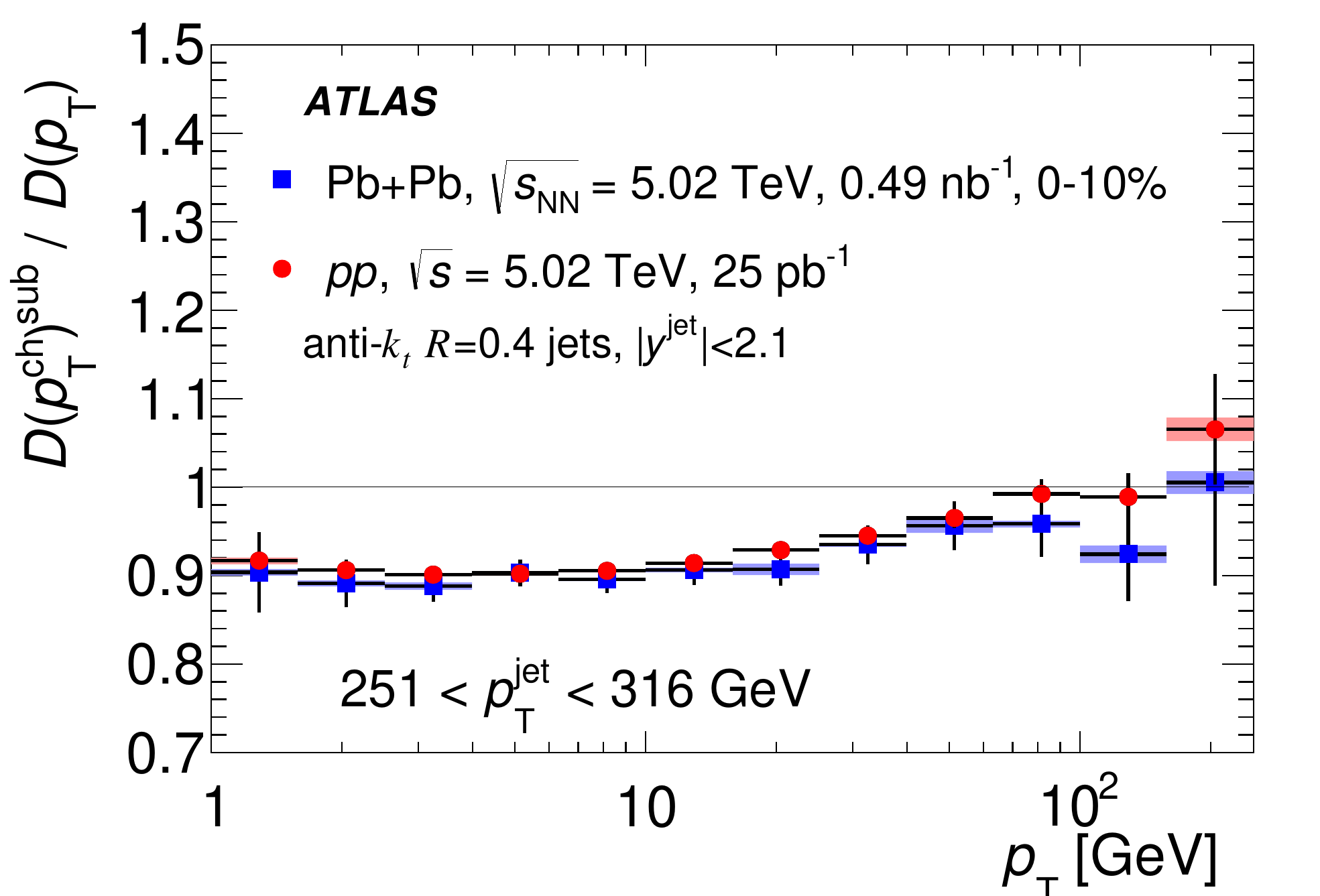} \\
  \\ 
\end{tabular} }
\caption{Ratios $\Dzsub/\Dz$ (left) and $\Dptsub/\Dpt$ (right) for \pp\ and 0--10\%
   central \pbpb\ collisions for 126~$<\ptjet<$~158~\GeV\ (top) and
   251~$<\ptjet<$~316~\GeV\ (bottom) for $|\yjet|<$~2.1.  The error bars show the statistical uncertainties and the
   boxes show the systematic uncertainties in the unfolding procedure.
   }
   \label{fig:unfolding}
\end{figure}

To remove the effects of bin migration due to the jet energy and track momentum resolution, the subtracted $d\nchsub/d\z$ and $d\nchsub/d\pttrk$ distributions are corrected by using a two-dimensional Bayesian unfolding procedure~\cite{DAgostini:1994zf}
in \z\ or \pT\ and \ptjet\ as implemented in the RooUnfold package~\cite{Adye:2011gm}.  
Two-dimensional unfolding is used because the calorimetric jet energy response depends on the fragmentation pattern of the jet~\cite{Aad:2011he}.
Using MC samples, four-dimensional response matrices are created using the generator-level and reconstructed \ptjet, and the generator-level and reconstructed charged-particle \z\ or \pt. Separate unfolding matrices are constructed for \pp\ data and each centrality interval in  \PbPb\ collisions. A separate one-dimensional Bayesian unfolding is used to correct the measured \ptjet\ spectra which are used to normalize the unfolded unnormalized fragmentation functions, $d \nchunf / d \pt$ and $d \nchunf / d z$.
To achieve better agreement with the data, the MC jet spectra and fragmentation functions are
reweighted to match the shapes in the reconstructed data. 
The Bayesian procedure requires a choice in the number of iterations. 
Additional iterations reduce the sensitivity to the choice of prior, but may
amplify statistical fluctuations in the distributions.
After four iterations for
both the one-dimensional and two-dimensional unfoldings 
the fragmentation functions are stable
for  both the \PbPb\ and \pp\ data. The final, particle-level corrected distributions are defined as:
\begin{eqnarray*}
 \label{eq:Dpt}
   D(z) &=& \frac{1}{N_\mathrm{jet}^\mathrm{unfolded}} \frac{d \nchunf}{d \z}, \\
   D(\pt) &=& \frac{1}{N_\mathrm{jet}^\mathrm{unfolded}} \frac{d \nchunf}{d \pt},
 \end{eqnarray*}
where $N_\mathrm{jet}^{\mathrm{unfolded}}$ is the unfolded number of jets in a given \ptjet\ interval.

The performance of the analysis procedure is tested by dividing the MC events in half and using one
half to generate response matrices with which the other half is unfolded and the ratio of unfolded to generator-level fragmentation functions\footnote{The generator-level fragmentation functions are constructed using generator-level jets and primary charged particles.} is evaluated. This procedure tests all the analysis corrections and the unfolding procedure. Good recovery of the generator-level (truth) MC distributions is observed for the unfolded events. The deviations from the exact recovery of the generator-level MC distributions, the non-closure, are included in the systematic uncertainties.
The ratios of \Dzsub\ and \Dptsub\ distributions to the unfolded \Dz\ and \Dpt\ distributions are shown in Figure~\ref{fig:unfolding} for \pp\ collisions and 0--10\% central \pbpb\ collisions. The magnitude of the unfolding effect varies as a function of \ptjet, \pTtrk, and centrality. The effect of the unfolding is similar in \pp\ and \pbpb\ collisions at low \z\ and \pT, but for higher-momentum particles within the jet, the effect of the unfolding in \pp\ and \pbpb\ collisions differs by up to 25\% 
between the two collision systems for 126~$<\ptjet <$~158~\GeV. 
This difference is due to UE fluctuations, which lead to poorer jet energy resolution in \pbpb\ collisions than in \pp\ collisions.  
With increasing \ptjet, the effect of UE fluctuations decreases;
 for 251~$<\ptjet <$~316~\GeV\
the effect of the unfolding is similar in \pbpb\ and \pp\ collisions at all value of \z\ and \pT.
The effect of the unfolding is larger at high \z\ and \pT\ due 
to the steepness of the fragmentation function near $\z=1$.
The shaded boxes in Figure~\ref{fig:unfolding} show the size of systematic uncertainties associated with the unfolding which originate from the sensitivity of the unfolding to the shape of input MC distributions, as described in the next section.

%-------------------------------------------------------------------------------
\section{Systematic uncertainties}
\label{sec:systematics}
  The following sources of systematic uncertainty are considered:
the jet energy scale (JES), the jet energy resolution (JER), 
the sensitivity of the  unfolding to the prior, 
the residual non-closure of the analysis procedure, UE contribution, and tracking-related uncertainties.
For each variation accounting for a source of systematic uncertainty, the fragmentation functions and ratios of $D(z)$ and $D(\pT)$ distributions in \PbPb\ and \pp\ collisions are re-evaluated. The difference between the varied and nominal distributions is used as an estimate of the resulting uncertainty.

The systematic uncertainty due to the JES in \PbPb\ collisions is composed of two parts: 
a centrality-independent baseline component and a centrality-dependent component. Only the centrality-independent baseline component is used in \pp\ collisions; 
it is determined from \textit{in situ} studies of the calorimeter 
response~\cite{Aaboud:2017jcu,HIjesnote,Run2jetpubnote}, and studies of the relative energy scale difference between the jet reconstruction procedure in heavy-ion collisions~\cite{HIjesnote} and the procedure in \pp\ collisions~\cite{Aaboud:2017jcu}. The centrality-dependent uncertainty reflects a modification of parton showers by the \PbPb\ environment. It is evaluated by comparing calorimeter \ptjet\ and the sum of \pt\ of tracks within the jet in data and MC simulation. The size of the centrality-dependent uncertainty in the JES reaches 0.5\% in the most central collisions. Each component that contributes to the JES uncertainty is varied separately by $\pm 1$ standard deviation for each interval in \ptjet, and the response matrix is recomputed accordingly. The data are unfolded with these matrices. The resulting uncertainty on the fragmentation functions increases with increasing $z$ and particle \pT\ at fixed \ptjet\ and decreases with increasing \ptjet.

The uncertainty in the fragmentation functions due to the JER is evaluated by repeating the unfolding procedure with modified response matrices, where an additional contribution is added to the resolution of the reconstructed \ptjet\ 
using a Gaussian smearing procedure. The smearing factor is evaluated using an \textit{in situ} technique in 13~\TeV\ \pp\ data involving studies of dijet energy balance~\cite{Aad:2012ag,JERConfNote}. An additional uncertainty is included to account for differences between the heavy-ion-style jet reconstruction and that used in analyses of 13~\TeV\ \pp\ data. The size of the resulting uncertainty on 
the fragmentation functions due to the JER typically reaches 10\% for the highest charged-particle \z\ and \pT\ bins and decreases with decreasing charged-particle \z\ and \pT\ at fixed \ptjet. The positive and negative  uncertainties from the JER are symmetrized.   

\begin{figure}[h]
\centerline{
\begin{tabular}{cc}
\includegraphics[width=0.49\textwidth]{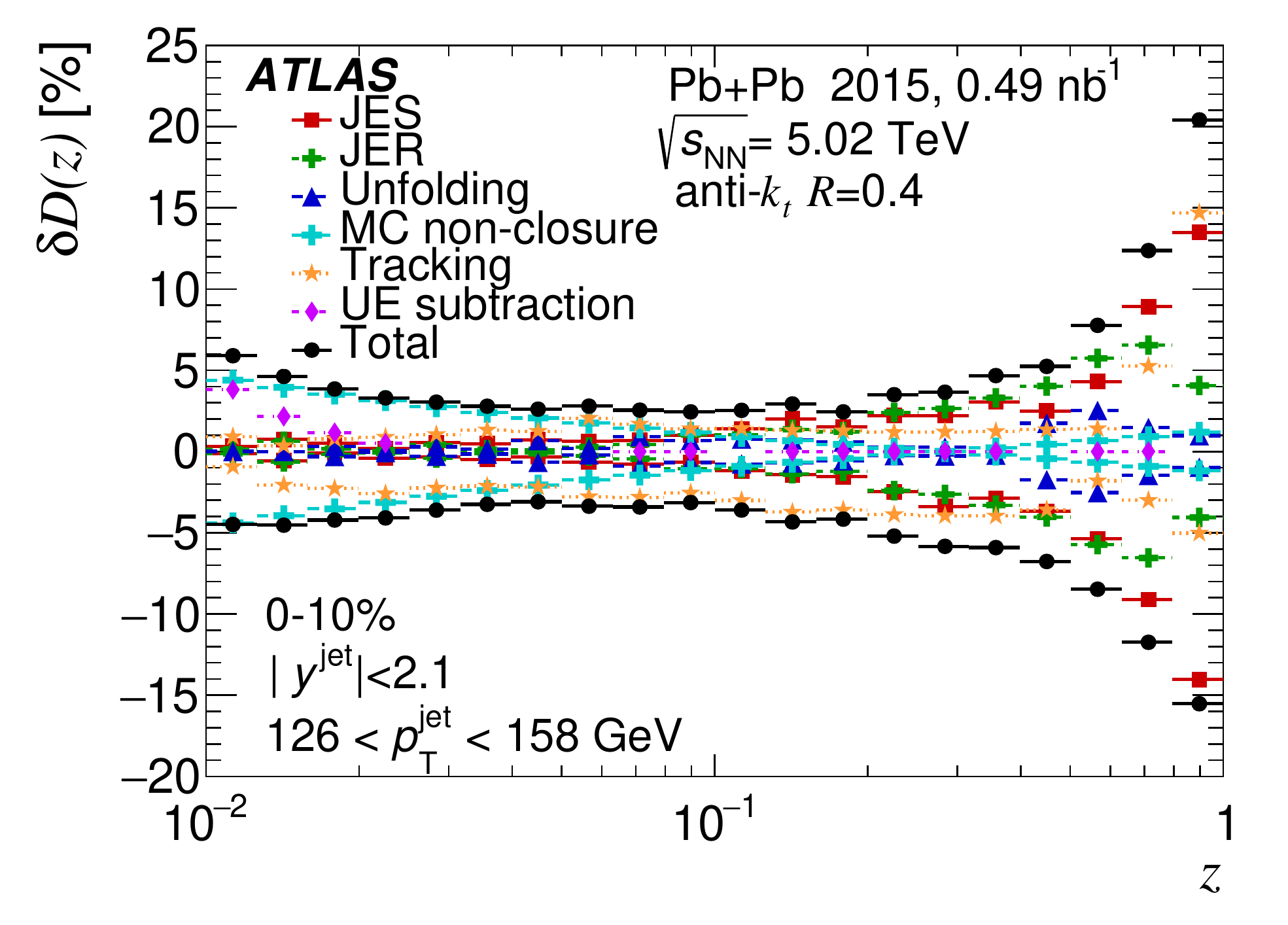} &
\includegraphics[width=0.49\textwidth]{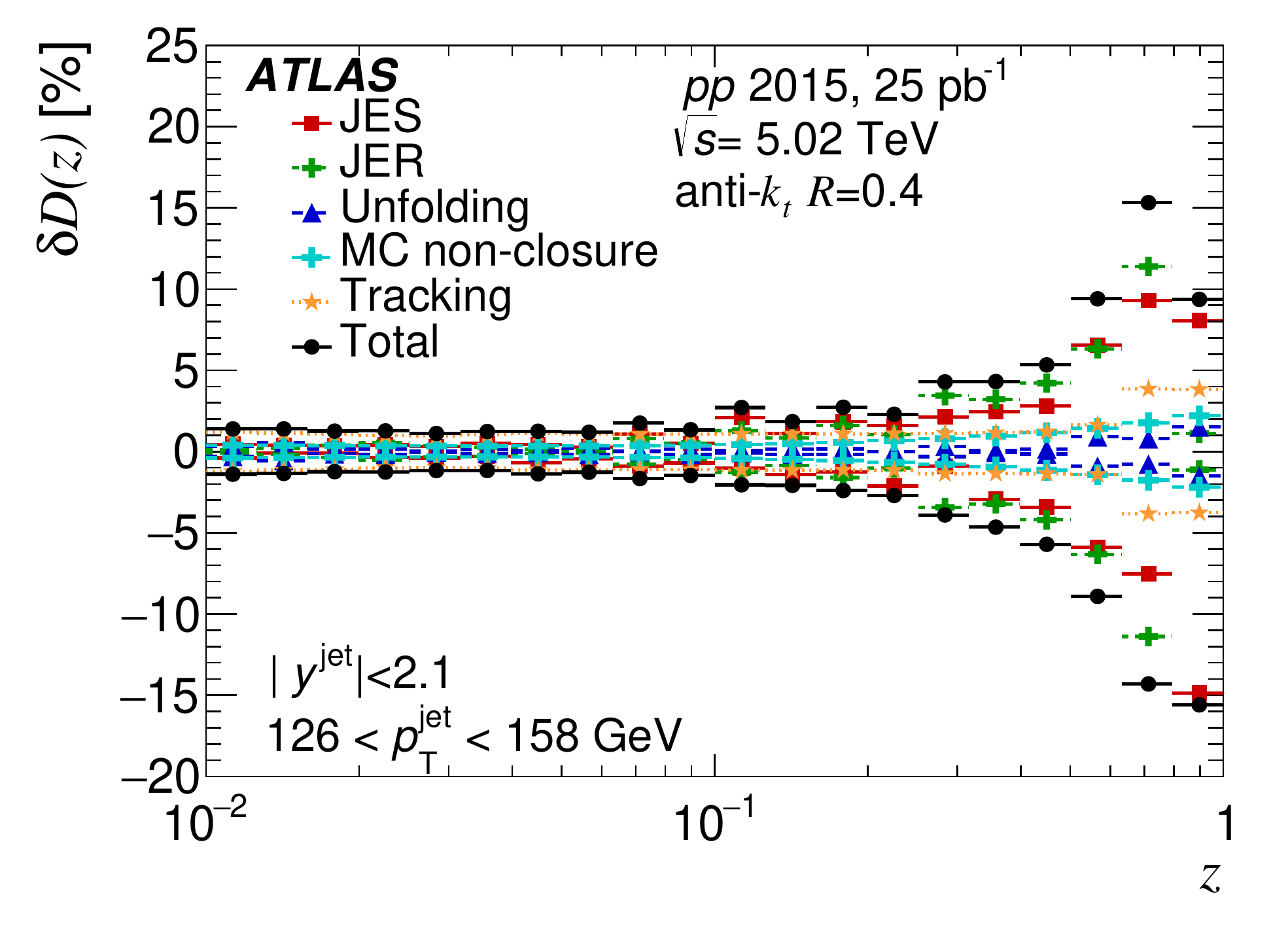} \\
\includegraphics[width=0.49\textwidth]{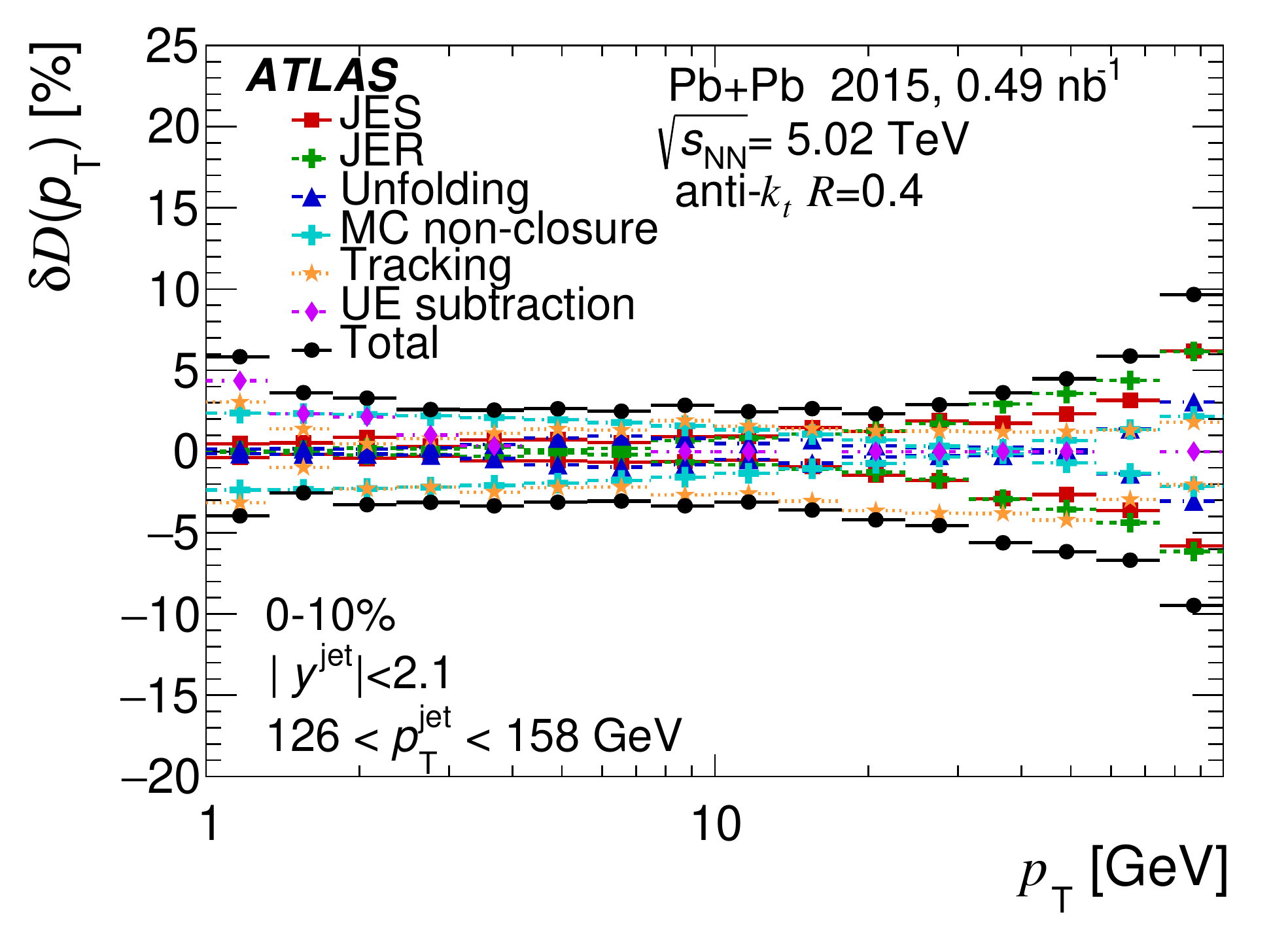} &
\includegraphics[width=0.49\textwidth]{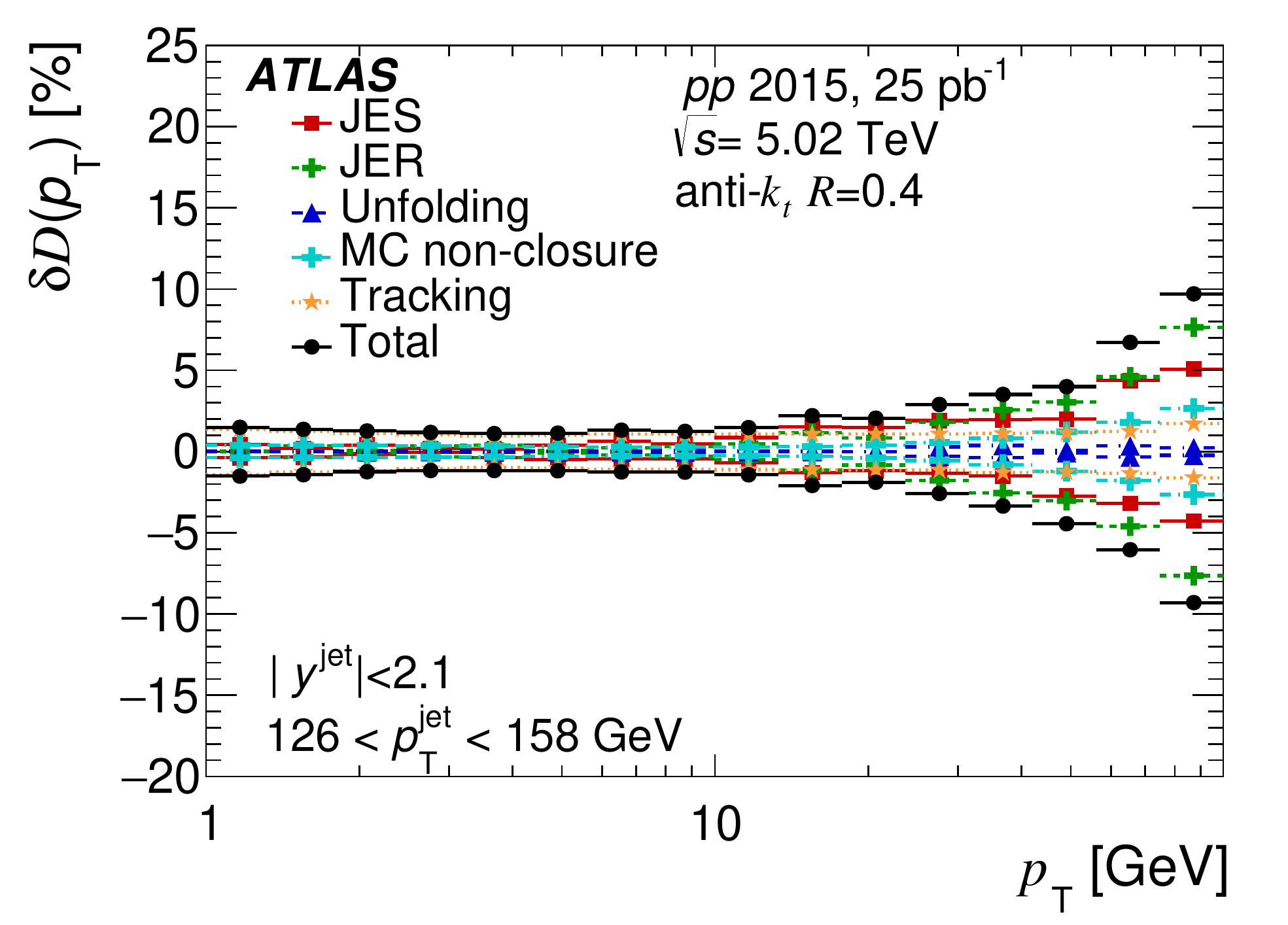} \\
\end{tabular}}
\caption{
Summary of the systematic uncertainties of the $D(z)$ (top) and \Dpt\ (bottom) distributions in 
0--10\% central \PbPb\ collisions (left) and \pp\ 
collisions (right) for jets in the 126--158~\GeV\ \ptjet\ interval. 
The systematic uncertainties due to JES, JER, unfolding, UE contribution, MC non-closure and tracking are shown along with the total systematic uncertainty from all sources.   
}
\label{Fig:Systematics_1}
\end{figure}

The unfolding uncertainty is estimated by generating the response matrices from the MC distributions without reweighting in \ptjet, $D(\z)$, and \Dpt. An additional uncertainty is assigned for the non-closure of the unfolded distributions in simulations, as described in
Section~\ref{sec:analysis}. The magnitude of the uncertainty due to the unfolding and the non-closure is typically below 2\% and 5\%, respectively.

The systematic uncertainty associated with the estimation of the UE contribution on the fragmentation functions has two components. First, the parameter that excludes random cones from the estimate is varied. Random cones are assumed to be associated with a hard process and excluded if the centroid of the cone is $\Delta R < 0.8$ from a reconstructed jet with $\pT > 90$~\GeV. The exclusion requirement is changed to $\Delta R < 1.2 $ to estimate the sensitivity of the UE contributions. The size of the resulting uncertainty on the fragmentation function is everywhere smaller than 3\% and it decreases in higher charged-particle \z\ or \pT. The second component of the UE uncertainty arises from a difference when the UE from the cone method is compared with an alternative UE estimation. The UE is alternatively evaluated using an efficiency-corrected differential yield of charged particles $d^4\nch/d\etatrk d\phitrk d\pTch d\Delta\Psi$, where $\Delta\Psi$ is the difference in azimuth of the charged particle 
from the second-order event
plane, evaluated in minimum-bias \PbPb\ events. To each event considered, a weight is assigned such that the event sample obtained from the minimum-bias trigger has the same centrality distribution as the sample collected by the jet trigger. The resulting uncertainty on the fragmentation functions is smaller than 10\% 
at low \z\ or \pT\ and it rapidly decreases in higher charged-particle \z\ or \pT\ bins.  

The uncertainties related to the track reconstruction and selection originate from several sources.
Uncertainties related to the fake rate, the material description in simulation, and the track transverse 
momentum are obtained from studies in data and simulation described in Ref.~\cite{ATL-PHYS-PUB-2015-051}.
The systematic uncertainty on the fake-track rate is 30\% in both collision 
systems~\cite{ATL-PHYS-PUB-2015-051}. The contamination of fake tracks is less than 2\%, and 
the resulting uncertainty on the fragmentation functions is at most 0.5\%.
The sensitivity of the tracking efficiency to the description of the 
inactive material in the MC samples is evaluated by varying the material description.
This resulting uncertainty in the track reconstruction efficiency is between
0.5\% and 2\% over the track \pT\ range used in the analysis. 
An additional uncertainty takes into account a possible residual misalignment of the tracking detectors
in \pp\ and \PbPb\ data-taking. The alignment in these data sets is checked \textit{in situ} using $Z\rightarrow \mu^{+}\mu^{-}$ events, and a track-\pT\-dependent uncertainty arises from the finite size of this sample. The resulting uncertainties on
the fragmentation functions are typically smaller than 1\%, except at large \z, where they are as large as 4\%. An additional  uncertainty on the tracking efficiency due to the high local 
track density in the core of jets is 0.4\%~\cite{ATL-PHYS-PUB-2016-007} for all \ptjet\ 
ranges in this analysis. The uncertainty due to the track selection is evaluated by repeating 
the analysis with an additional requirement on the significance of the distance of closest approach of the track to the primary vertex. This uncertainty affects 
the track reconstruction efficiencies, track momentum resolution, and rate of fake tracks. The resulting uncertainty typically varies from 1\% at low track \pT\ to 5\% at high track \pT.
Finally, the track-to-particle matching requirements
are varied. This variation affects the track reconstruction efficiency, track momentum resolution, and rate of fake tracks. The resulting systematic uncertainty in the fragmentation functions is less than 0.5\%. 

\begin{figure}
\centerline{
\begin{tabular}{cc}
\includegraphics[width=0.49\textwidth]{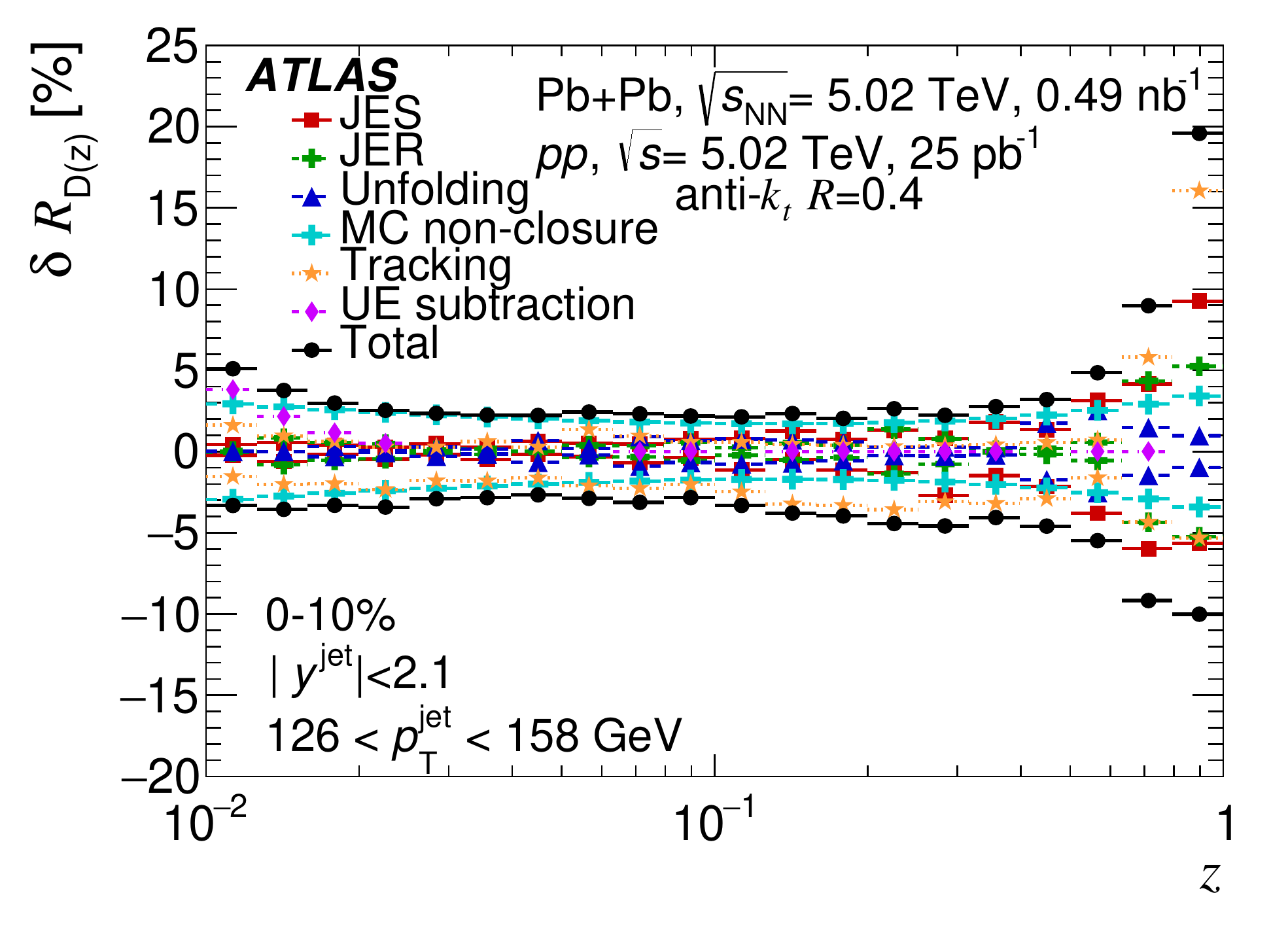} &
\includegraphics[width=0.49\textwidth]{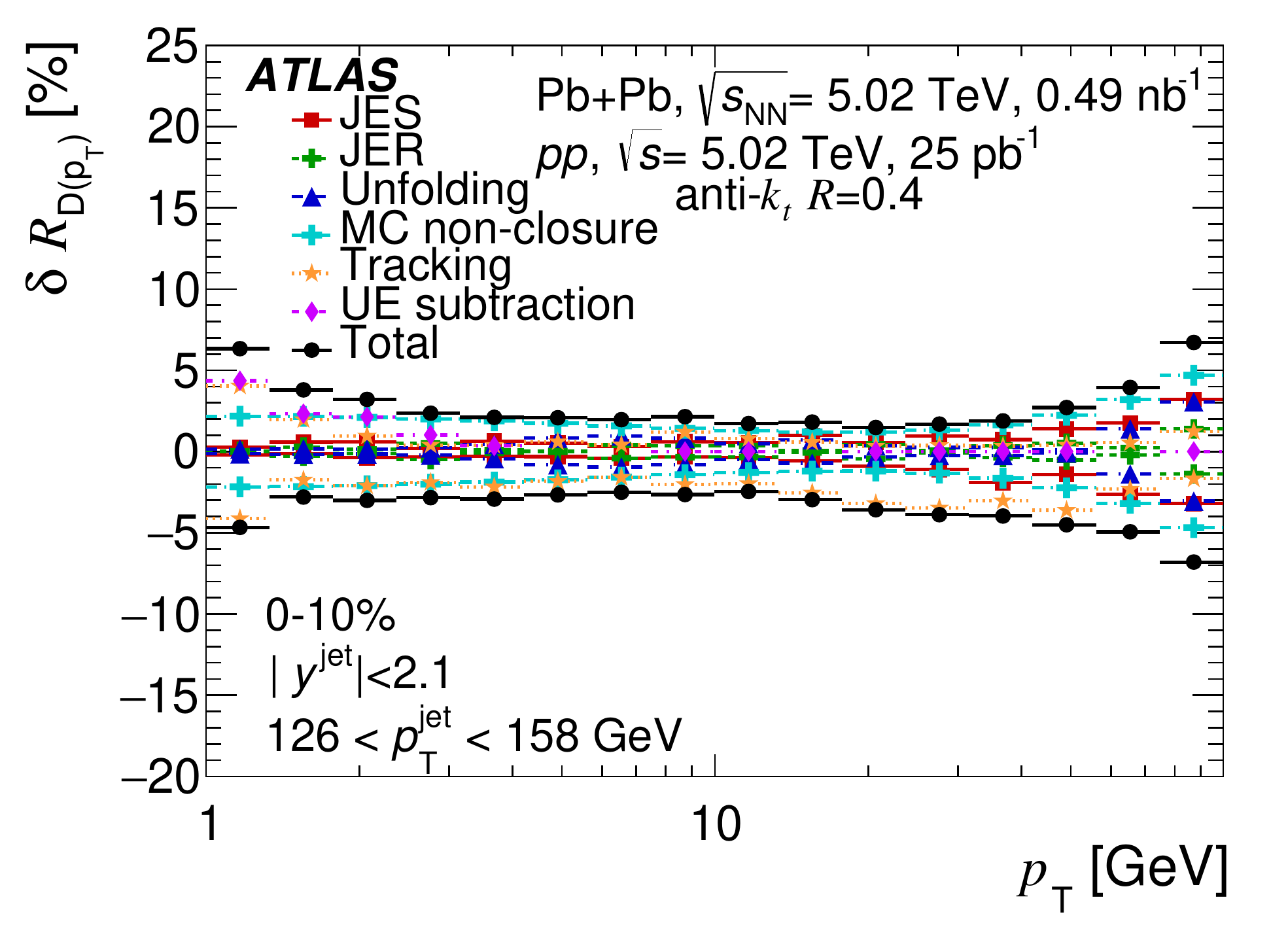} \\
\end{tabular}}
\caption{
Summary of the systematic uncertainties for 0--10\% central 
$R_{D(z)}$ (left) and $R_{D(\pT)}$ (right) ratios, for jets in the 126--158~\GeV\ \ptjet\ 
interval. The systematic uncertainties due to JES, JER, unfolding, UE contribution, MC non-closure and tracking are shown along with the total systematic uncertainty from all sources.
}
\label{Fig:Systematics_3}
\end{figure}

Example systematic uncertainties on the $D(z)$ and $D(\pT)$ distributions for jets in the 126--158~\GeV\ \ptjet\ 
range measured in the two collision systems are presented in Figure~\ref{Fig:Systematics_1}. 
All track-related systematic uncertainties are added in quadrature and presented as a total tracking uncertainty. The systematic uncertainties from each source are assumed to be uncorrelated, so they are combined in quadrature to obtain the total systematic uncertainty.

The correlations between the various systematic components are considered in evaluating the ratios of \PbPb\ 
to \pp\ fragmentation functions. The unfolding and the MC non-closure are each taken to be uncorrelated between the two collision systems. All other uncertainties are taken to be correlated. For the correlated uncertainties, the ratios are re-evaluated by applying the variation to both collision systems; the resulting variations of the ratios from their central values are used as the correlated systematic uncertainty. 
The uncorrelated uncertainties are added in quadrature. Each systematic uncertainty is assumed to be fully
correlated with itself between different rapidity bins. The systematic uncertainty from each source, except the non-closure of the unfolded distributions and the residual misalignment of the tracking detectors, is bin-to-bin correlated. The total systematic uncertainties of the $R_{D(z)}$ and $R_{D(\pT)}$ distributions are shown in Figure~\ref{Fig:Systematics_3} for one selected \ptjet\ range.

%-------------------------------------------------------------------------------
\section{Results}
\label{sec:results}
In this section, results are presented of the measurement of the \Dz\ and \Dpt\ 
distributions for jet \pt\ between 126 and 398~\GeV\ and six centrality intervals in \PbPb\ collisions; the same distributions are presented in \pp\ collisions for the same \ptjet\ ranges.  
In order to study the effects of hot dense matter on the jet fragmentation process, ratios of \pbpb\ fragmentation functions 
to \pp\ fragmentation functions are evaluated.

\begin{figure}[p!]
      \centerline{
	 \begin{tabular}{cc}
	    \includegraphics[width=0.4\textwidth]{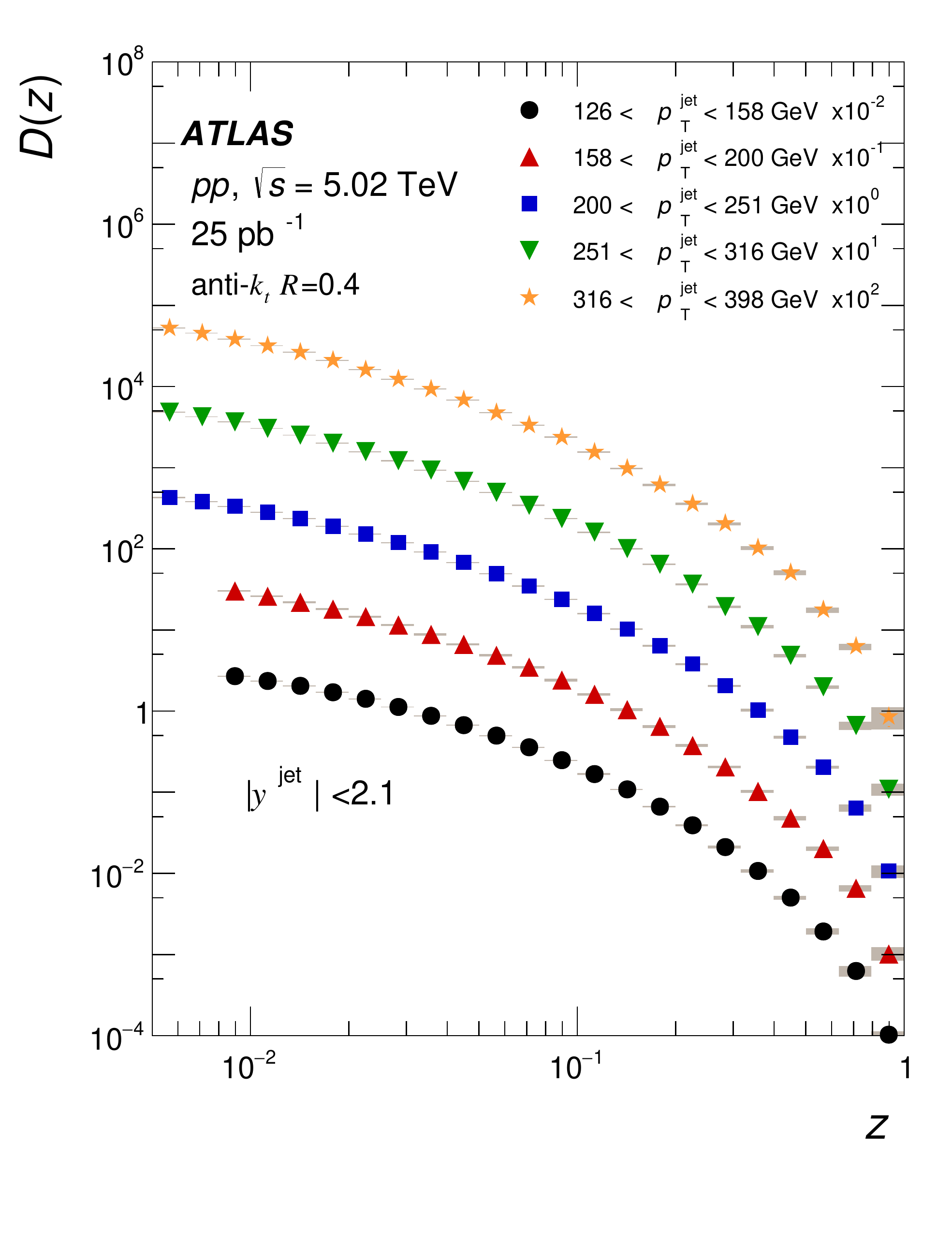} &
	    \includegraphics[width=0.4\textwidth]{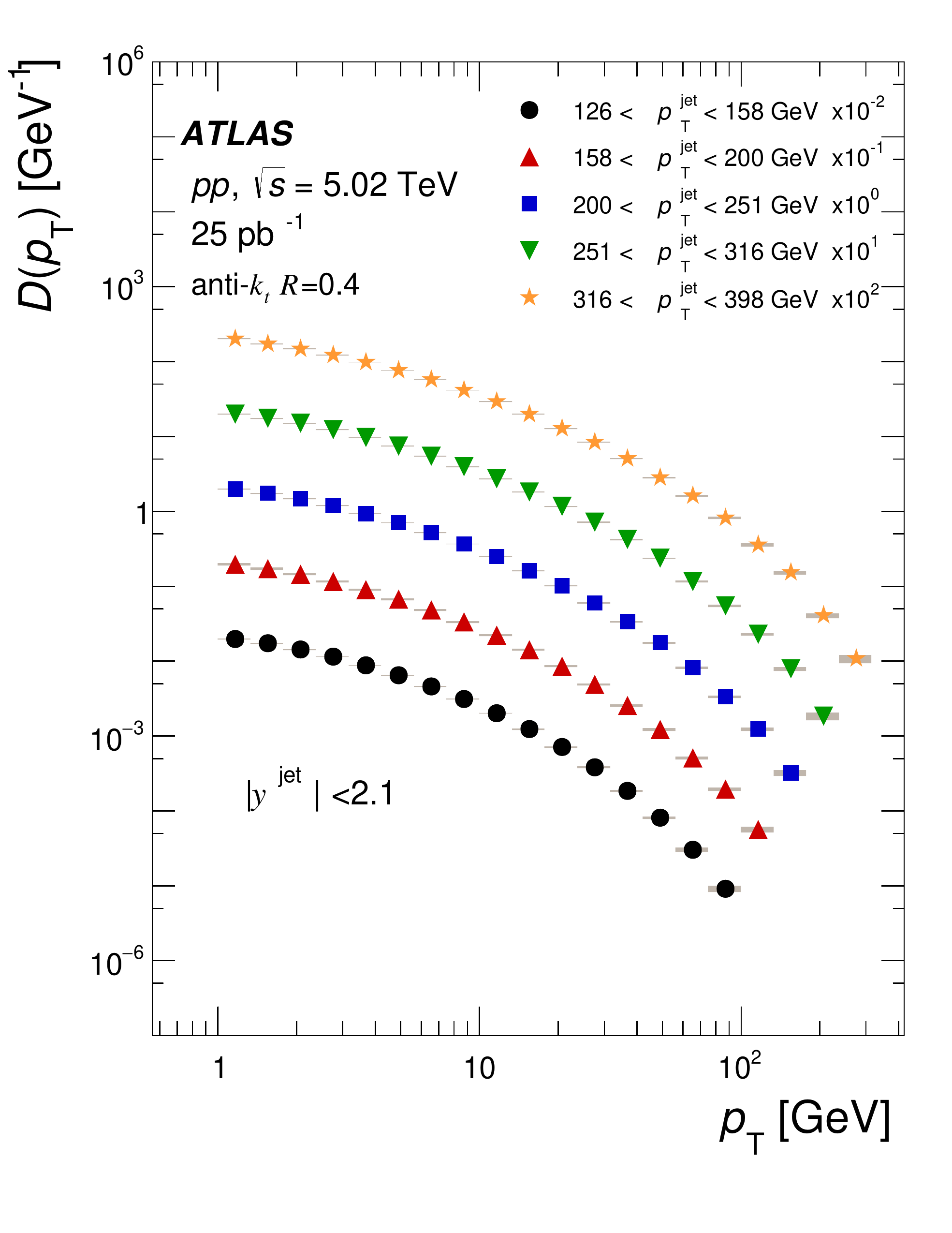} \\
      \end{tabular}
      }
      \caption{Fragmentation functions,  \Dz\ (left) and \Dpt\ (right), in \pp\ collisions measured in five 
       \ptjet\ ranges from 126 to 398~\GeV. The vertical bars on the data points indicate statistical uncertainties, while the shaded bands indicate systematic uncertainties. In most cases, the statistical uncertainties are smaller than the marker size.}
      \label{fig:ppff}
   \end{figure}

\begin{figure}[ht]
      \centerline{
	 \begin{tabular}{cc}
	    \includegraphics[width=0.4\textwidth]{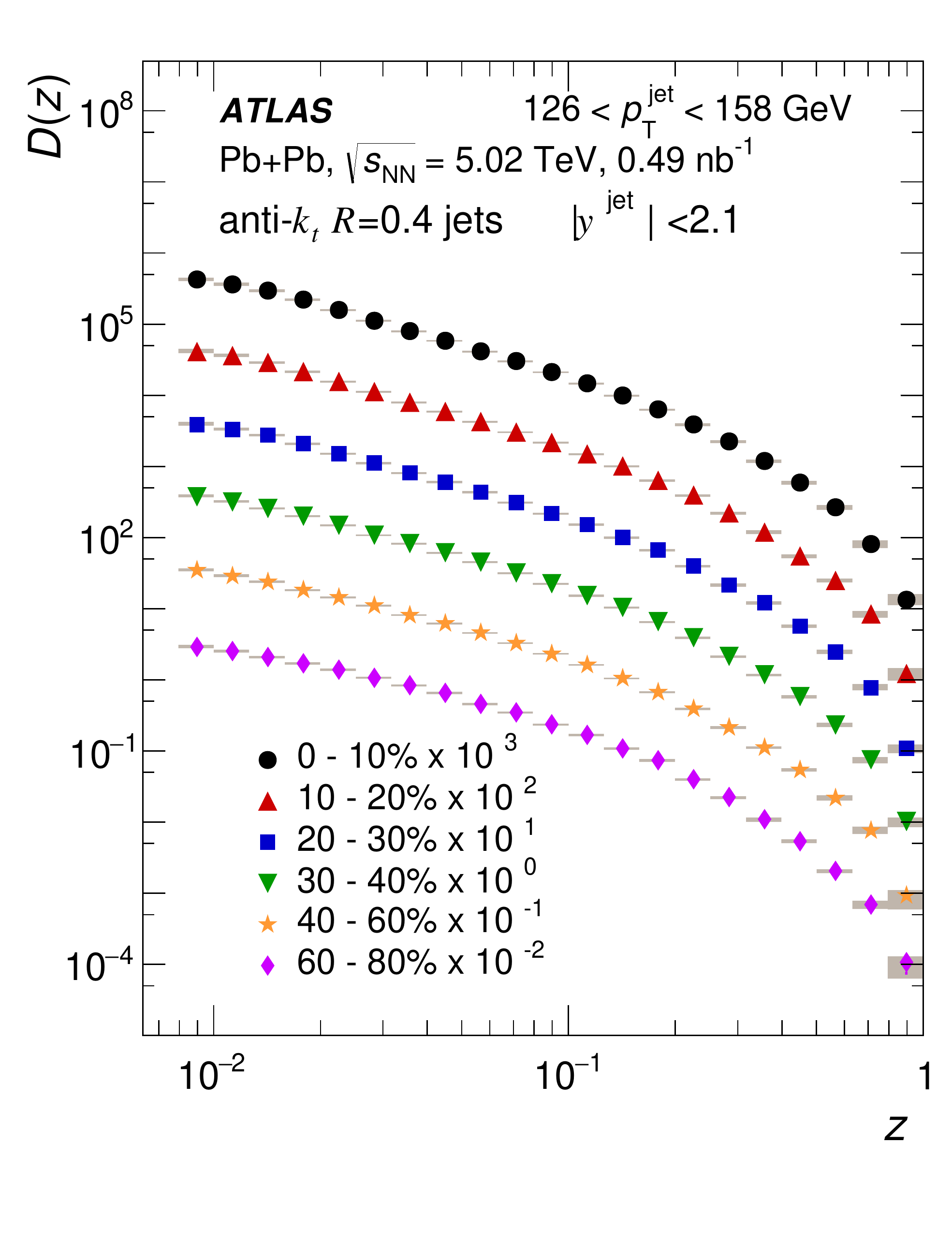} &
	 \includegraphics[width=0.4\textwidth]{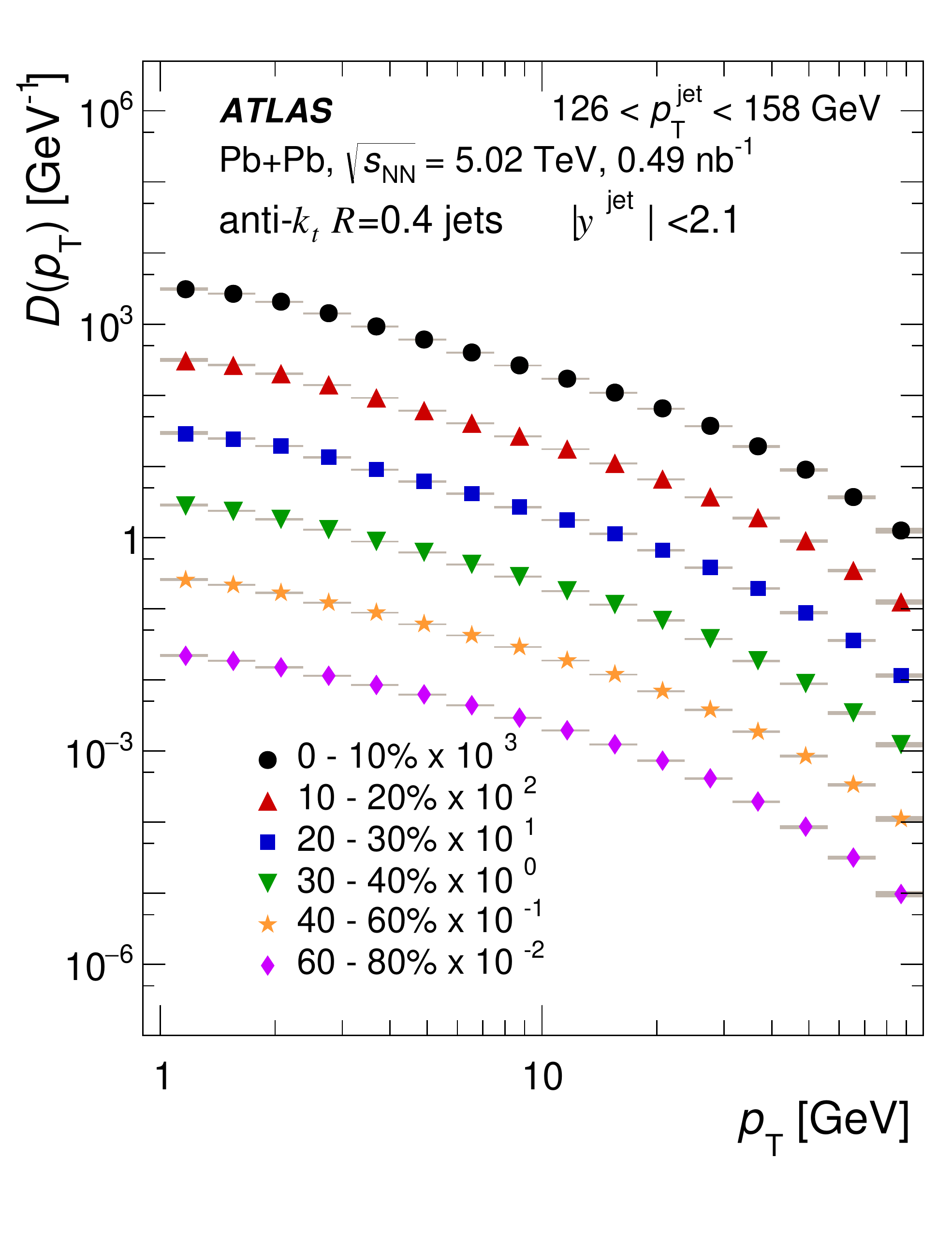} \\
      \end{tabular}
      }
      \caption{Fragmentation functions,  \Dz\ (left) and \Dpt\ (right), in \pbpb\ collisions measured in six different centrality classes for \ptjet\ of 126 to 158~\GeV. The vertical bars on the data points indicate statistical uncertainties, while the shaded bands indicate systematic uncertainties. In most cases, the statistical uncertainties are smaller than the marker size.}
     \label{fig:centdep7}
   \end{figure}

\begin{figure}[ht]
      \centerline{
	 \begin{tabular}{cc}
	    \includegraphics[width=0.4\textwidth]{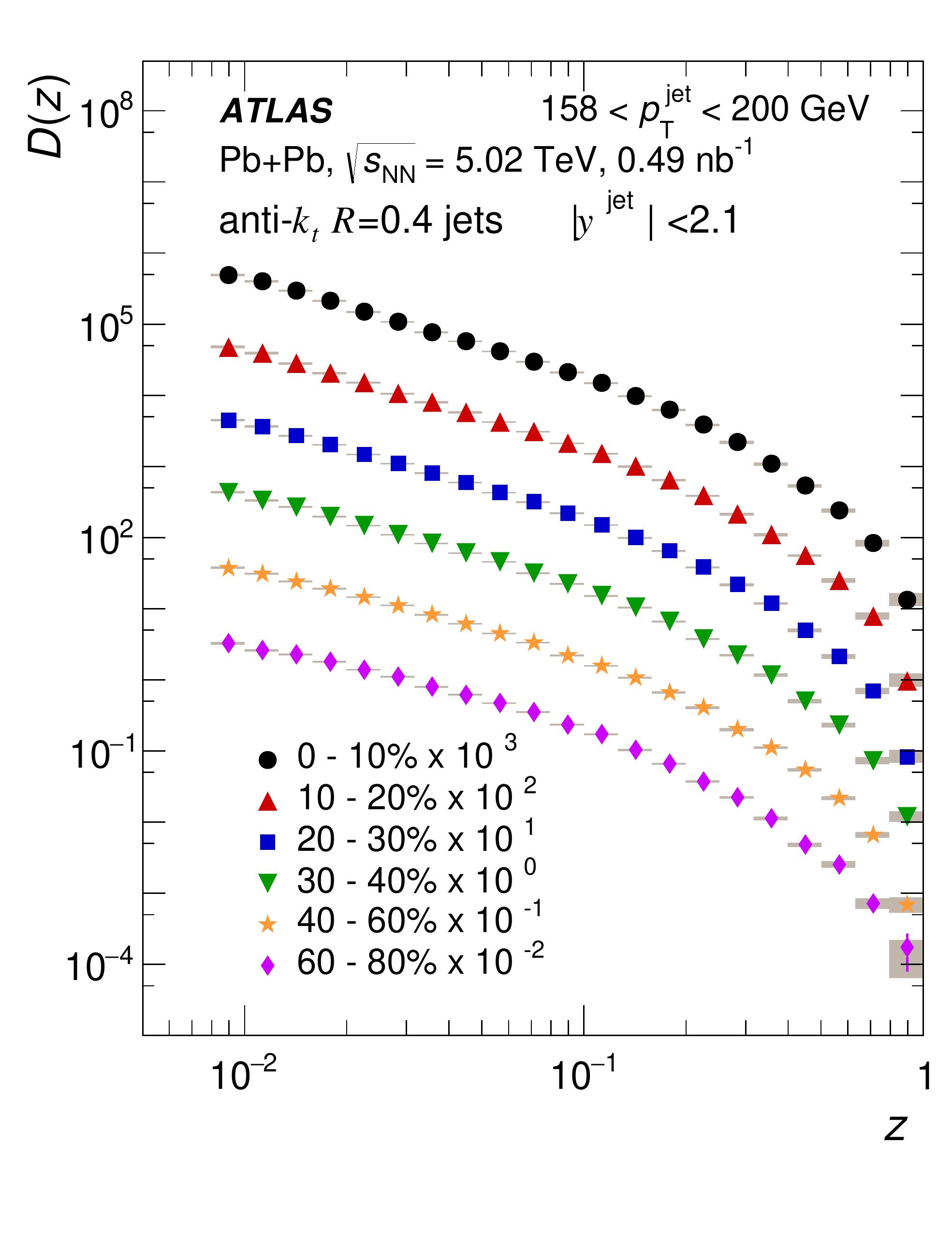} &
	 \includegraphics[width=0.4\textwidth]{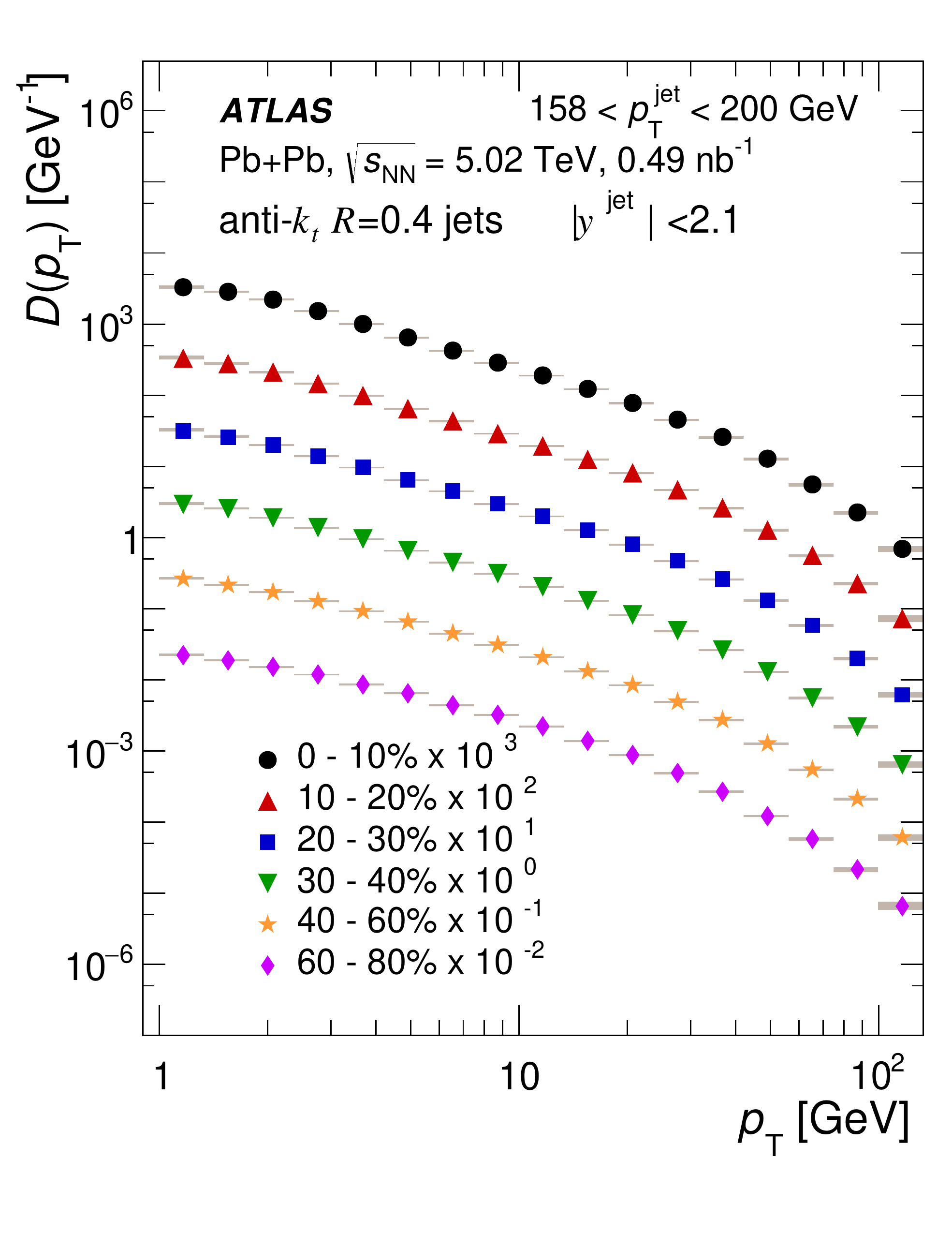} \\
      \end{tabular}
      }
      \caption{Fragmentation functions,  \Dz\ (left) and \Dpt\ (right), in \pbpb\ collisions measured in six different centrality classes for \ptjet\ of 158 to 200~\GeV. The vertical bars on the data points indicate statistical uncertainties, while the shaded bands indicate systematic uncertainties. In most cases, the statistical uncertainties are smaller than the marker size.}
      \label{fig:centdep8}
   \end{figure}

\begin{figure}[ht]
      \centerline{
	 \begin{tabular}{cc}
	    \includegraphics[width=0.4\textwidth]{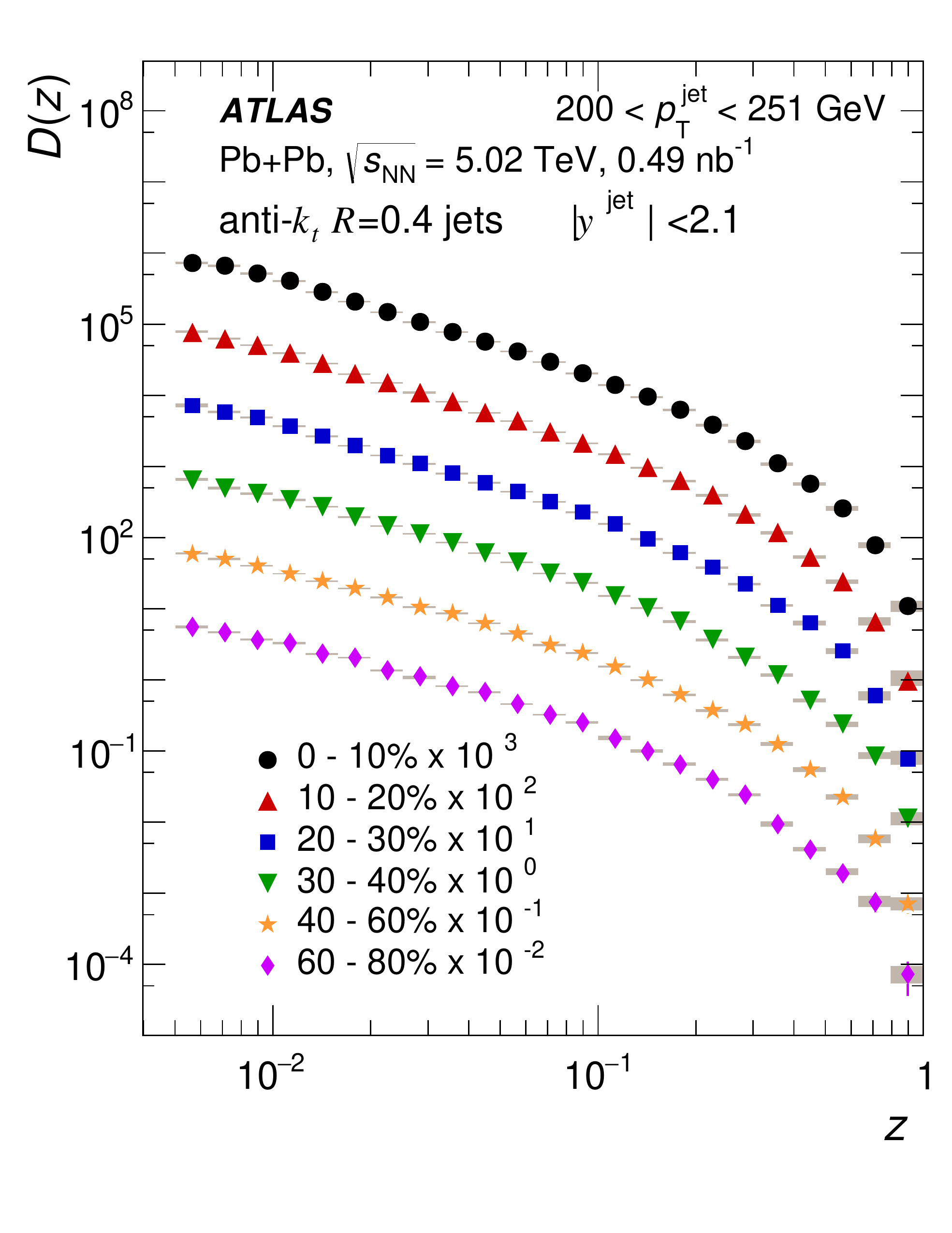} &
	 \includegraphics[width=0.4\textwidth]{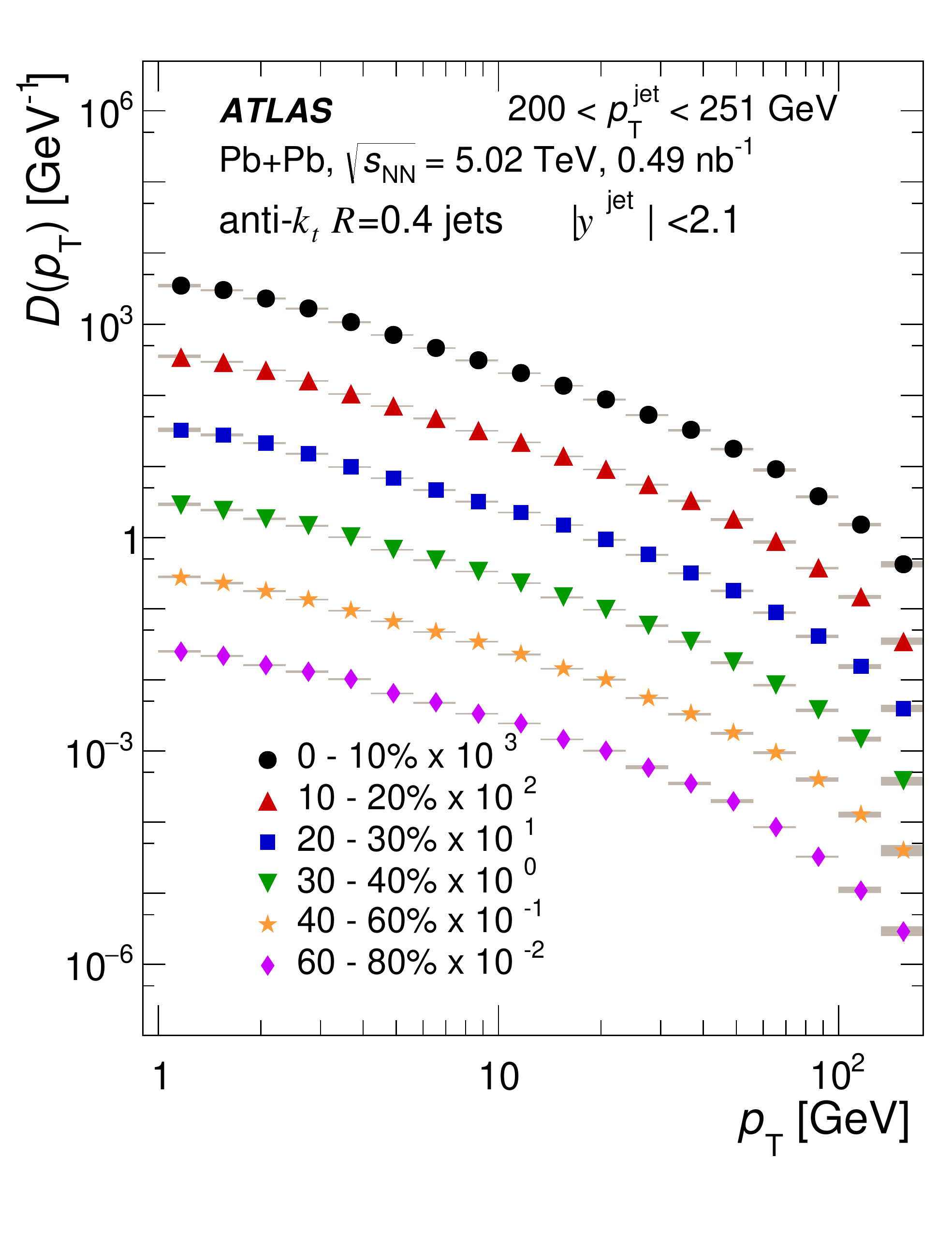} \\
      \end{tabular}
      }
      \caption{Fragmentation functions,  \Dz\ (left) and \Dpt\ (right), in \pbpb\ collisions measured in six different centrality classes for \ptjet\ of 200 to 251~\GeV. The vertical bars on the data points indicate statistical uncertainties, while the shaded bands indicate systematic uncertainties. In most cases, the statistical uncertainties are smaller than the marker size.}
      \label{fig:centdep9}
   \end{figure}

\begin{figure}[ht]
      \centerline{
	 \begin{tabular}{cc}
	    \includegraphics[width=0.4\textwidth]{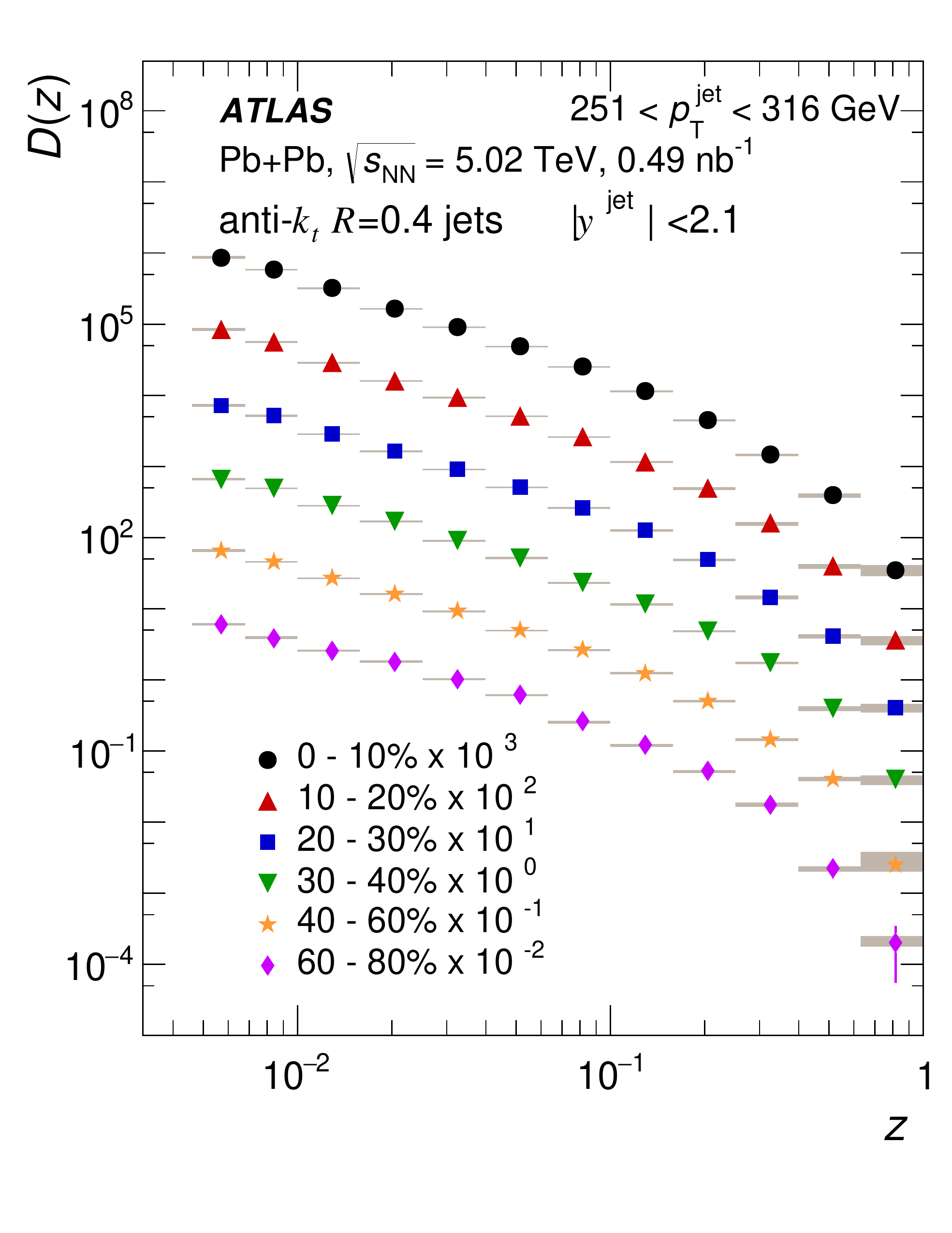} &
	 \includegraphics[width=0.4\textwidth]{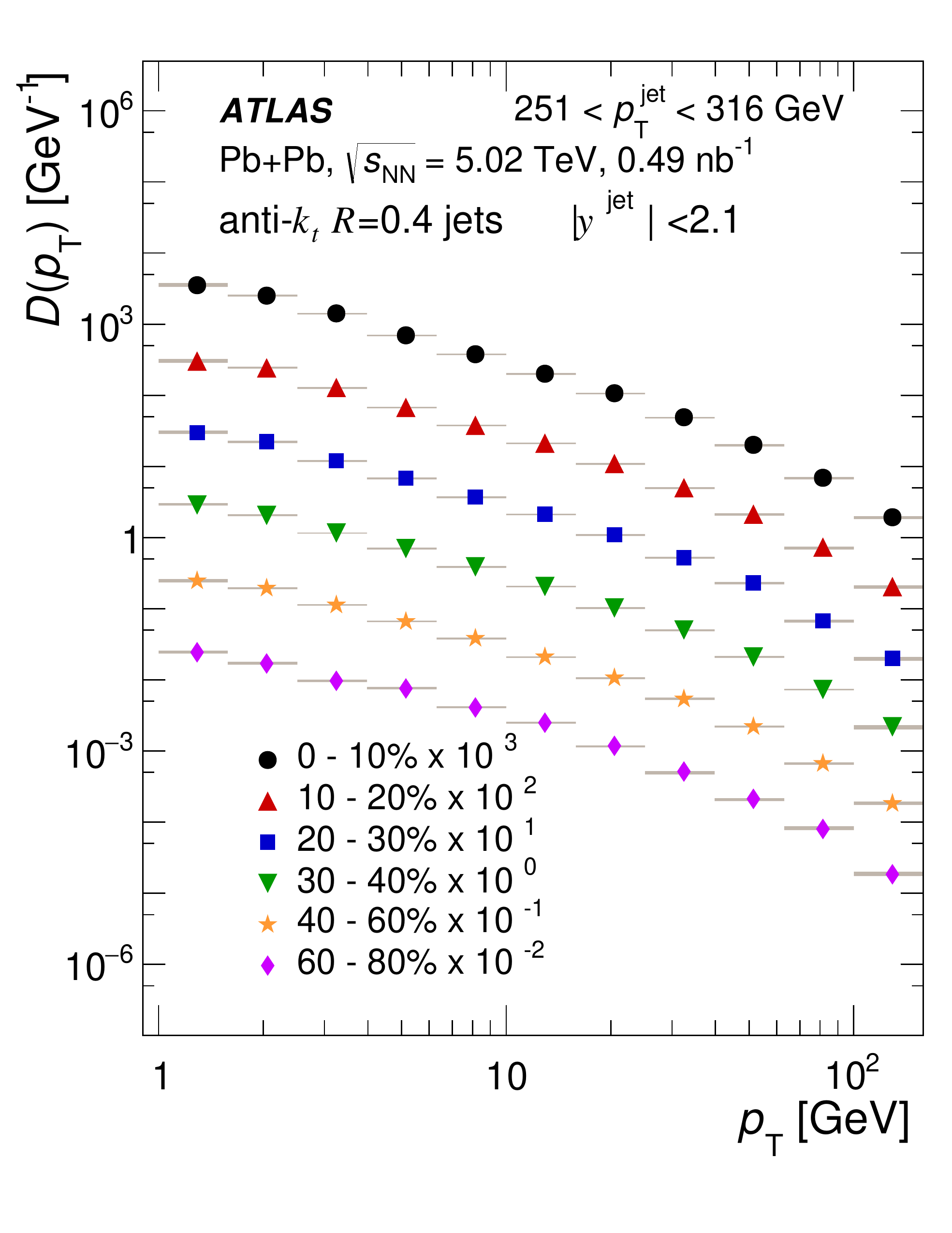} \\
      \end{tabular}
      }
      \caption{Fragmentation functions,  \Dz\ (left) and \Dpt\ (right), in \pbpb\ collisions measured in six different centrality classes for \ptjet\ of 251 to 316~\GeV. The vertical bars on the data points indicate statistical uncertainties, while the shaded bands indicate systematic uncertainties. In most cases, the statistical uncertainties are smaller than the marker size.}
      \label{fig:centdep10}
   \end{figure}

\begin{figure}[ht]
      \centerline{
	 \begin{tabular}{cc}
	    \includegraphics[width=0.4\textwidth]{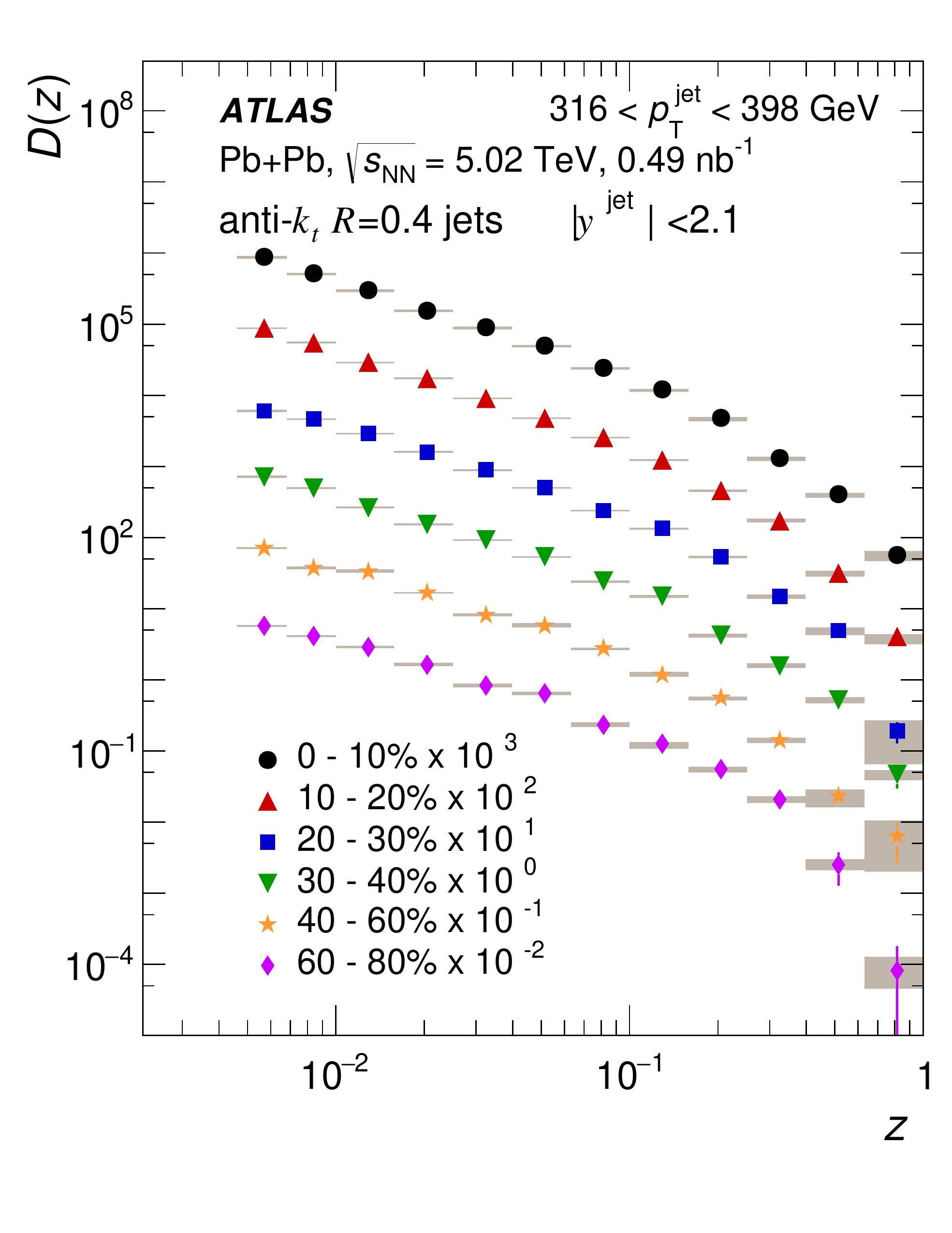} &
	 \includegraphics[width=0.4\textwidth]{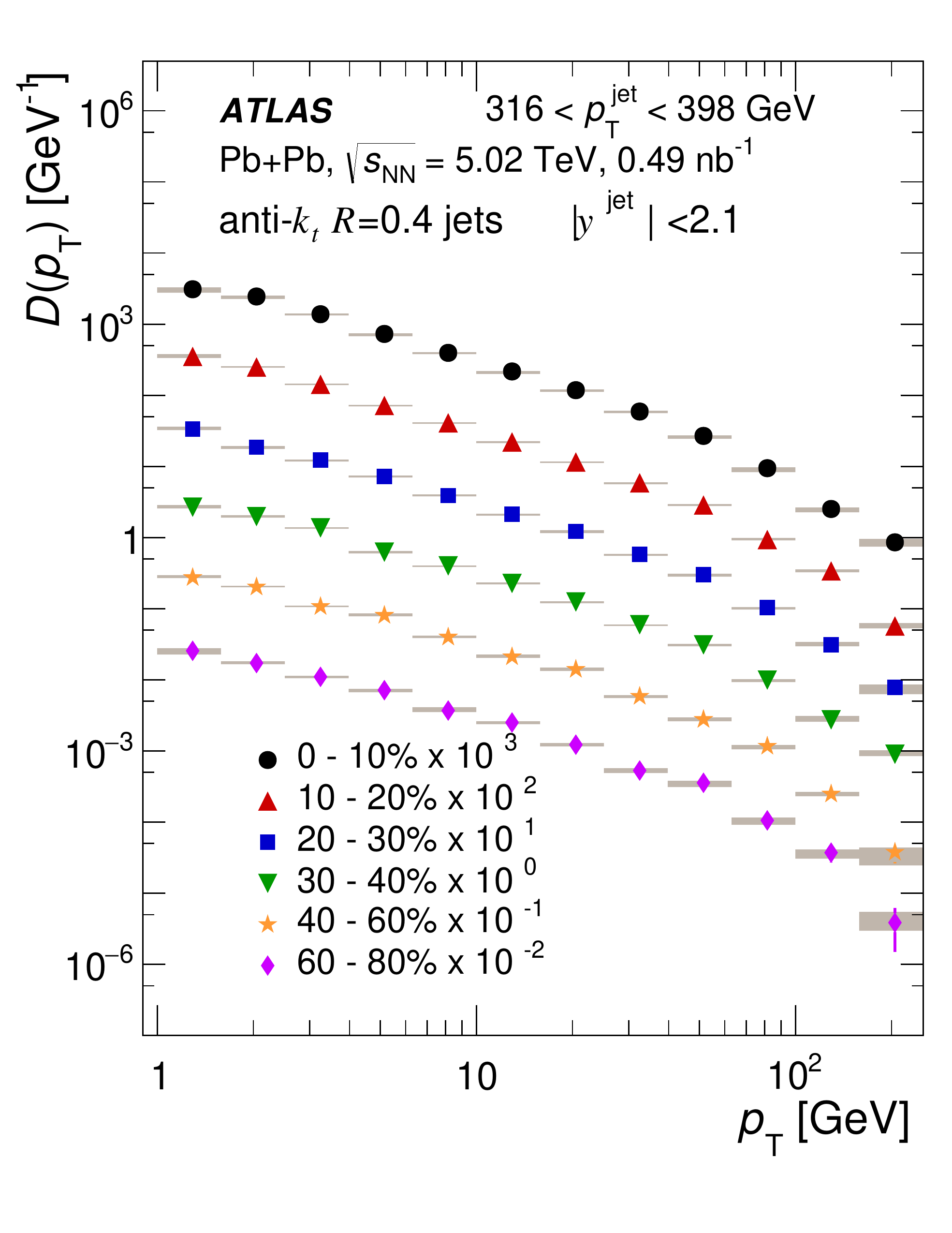} \\
      \end{tabular}
      }
      \caption{Fragmentation functions,  \Dz\ (left) and \Dpt\ (right), in \pbpb\ collisions measured in six different centrality classes for \ptjet\ of 316 to 398~\GeV. The vertical bars on the data points indicate statistical uncertainties, while the shaded bands indicate systematic uncertainties. In most cases, the statistical uncertainties are smaller than the marker size.}
      \label{fig:centdep11}
   \end{figure}

The \Dz\ and \Dpt\ distributions in \pp\ collisions are shown in Figure~\ref{fig:ppff}.
The corresponding distributions
in \pbpb\ collisions
are shown in Figures~\ref{fig:centdep7} through~\ref{fig:centdep11}. 

   \FloatBarrier

\begin{figure}[ht]
      \centerline{
	 \includegraphics[width=\textwidth]{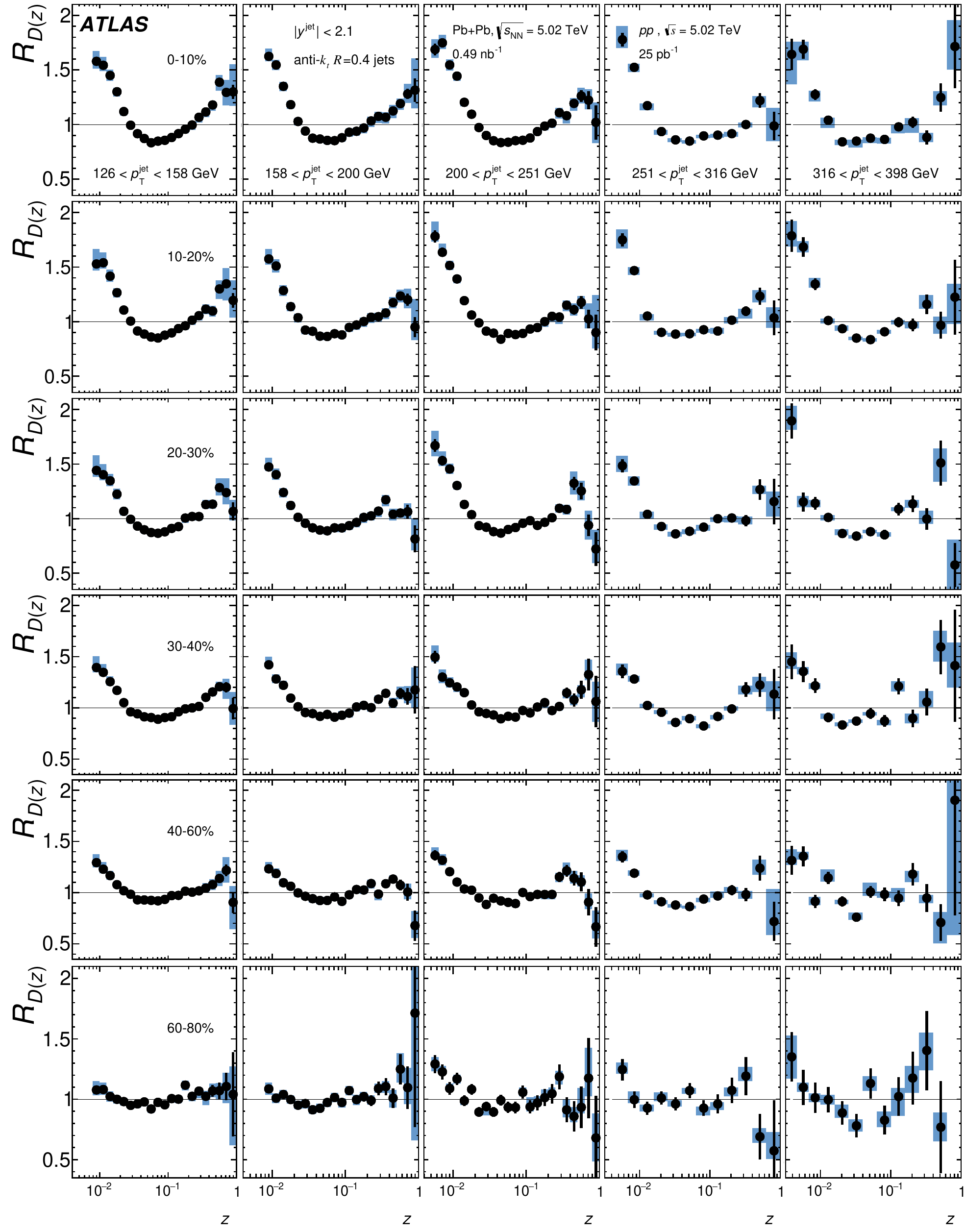}
      }
      \caption{Ratios of \Dz\  distributions in six centrality intervals of \pbpb\
      collisions to \pp\ collisions evaluated for five \ptjet\ ranges for jets with $|\yjet| < 2.1$. The vertical bars on the data points indicate statistical uncertainties, while the shaded bands indicate systematic uncertainties.
      Centrality decreases from top to bottom panels and \ptjet\ increases from left to right panels.}
      \label{fig:ratiosdz}
   \end{figure}

\begin{figure}[ht]
      \centerline{
	 \includegraphics[width=\textwidth]{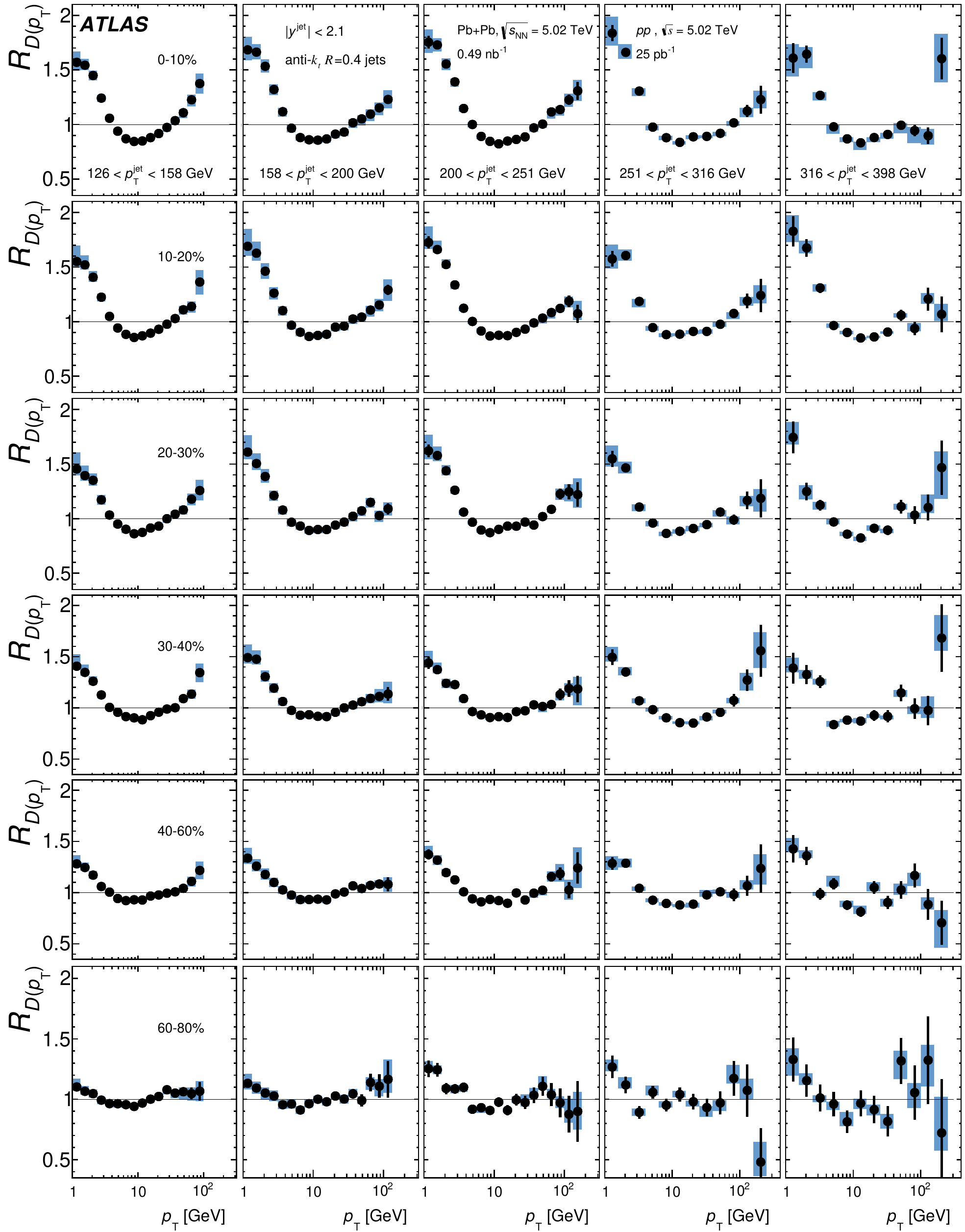}
      }
      \caption{Ratios of \Dpt\  distributions in six centrality intervals of \pbpb\
      collisions to \pp\ collisions evaluated for five \ptjet\ ranges for jets with $|\yjet| < 2.1$. The vertical bars on the data points indicate statistical uncertainties, while the shaded bands indicate systematic uncertainties.
      Centrality decreases from top to bottom panels and \ptjet\ increases from left to right panels.}
      \label{fig:ratiosdpt}
   \end{figure}

In order to quantify the difference in the fragmentation functions between \PbPb\ and \pp\ collisions,
the ratios of \Dz\ and \Dpt\ distributions measured in \pbpb\ 
collisions to those measured in \pp\ collisions, $R_{D(\z)}$ and $R_{D(\pT)}$, are shown
in Figures~\ref{fig:ratiosdz} and~\ref{fig:ratiosdpt}, respectively. 
In each figure, the shaded boxes indicate systematic uncertainties and the vertical bars show the statistical 
uncertainties. 
	
The shapes of the  $R_{D(z)}$ and $R_{D(\pt)}$ distributions are similar for all centralities: inside the jets; the yields of particles with low \pT\ or \z\  are enhanced; there is a reduction for particles with intermediate
\pT\ or \z; and the yields of particles with high \pT\ or \z\ are enhanced. 
This is qualitatively consistent with previous
measurements of jet fragmentation
at $\sqrtsnn=2.76$~\TeV~\cite{Aad:2014wha,Chatrchyan:2014ava,Aaboud:2017bzv};
a quantitative comparison is provided in Section~\ref{sec:discussion}.
The magnitudes of the deviations of the ratios from unity decrease with 
decreasing collision centrality. In the most central collisions, the size of the enhancement is as large as 70\% at low \pT\ or \z\  and 30\% at high \pT or \z.
The depletion of charged-particle yields at intermediate \pT\ and \z\ is as large as 20\%. In some centrality and \ptjet\ ranges there is a decrease of the fragmentation functions at the highest $z$ values.  In this region the statistical and systematic uncertainties are the largest; more precise measurements are needed to determine if a significant decrease exists. 

\begin{figure}[ht]
      \centerline{
	 \includegraphics[width=\textwidth]{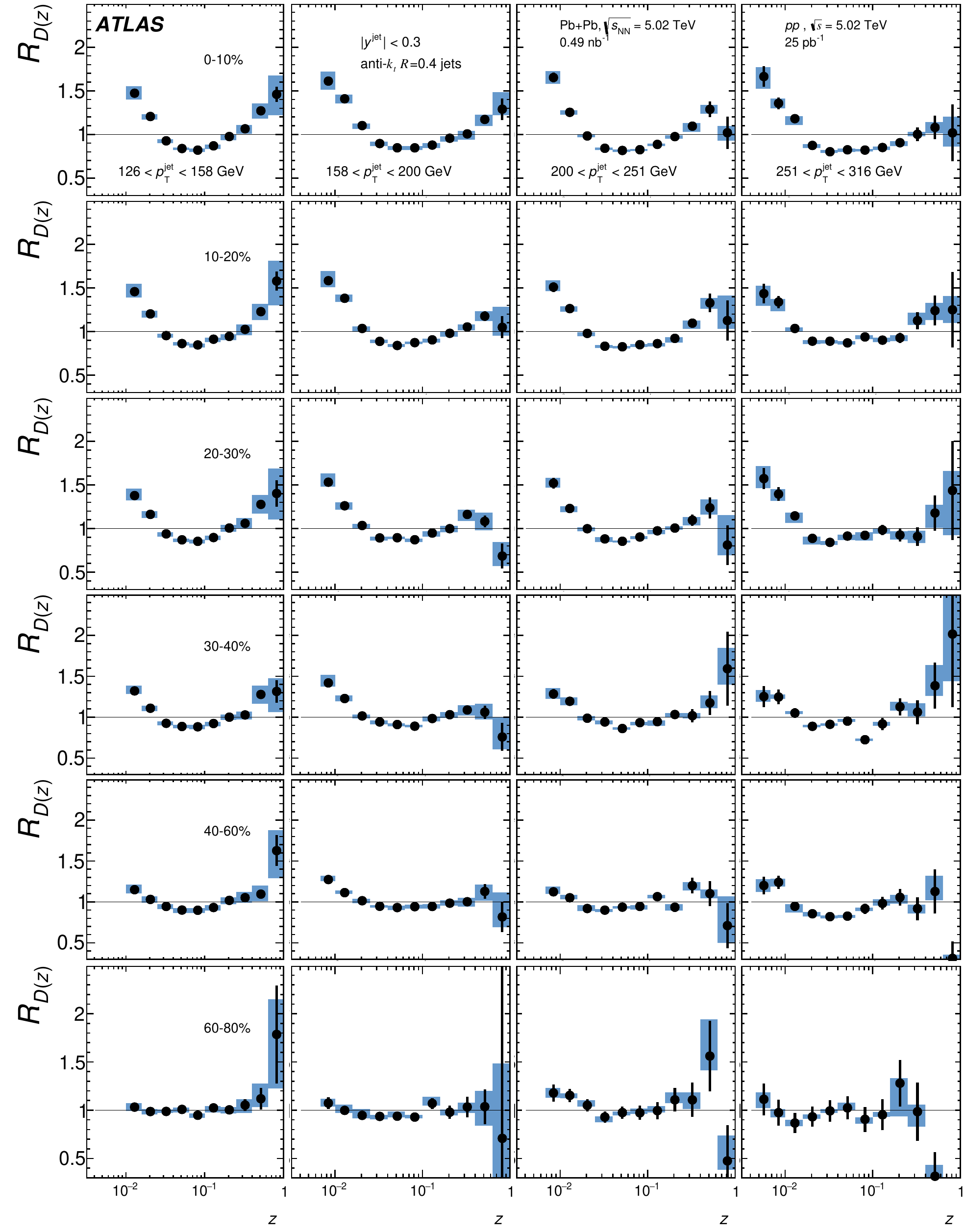}
      }
      \caption{Ratios of \Dz\  distributions in six centrality intervals of \pbpb\
      collisions to \pp\ collisions evaluated in four \ptjet\ ranges for jets with $|\yjet| <$~0.3. The vertical bars on the data points indicate statistical uncertainties, while the shaded bands indicate systematic uncertainties.
      Centrality decreases from top to bottom panels and \ptjet\ increases from left to right panels.}
      \label{fig:ratiosdzeta0}
   \end{figure}

\begin{figure}[ht]
      \centerline{
	 \includegraphics[width=\textwidth]{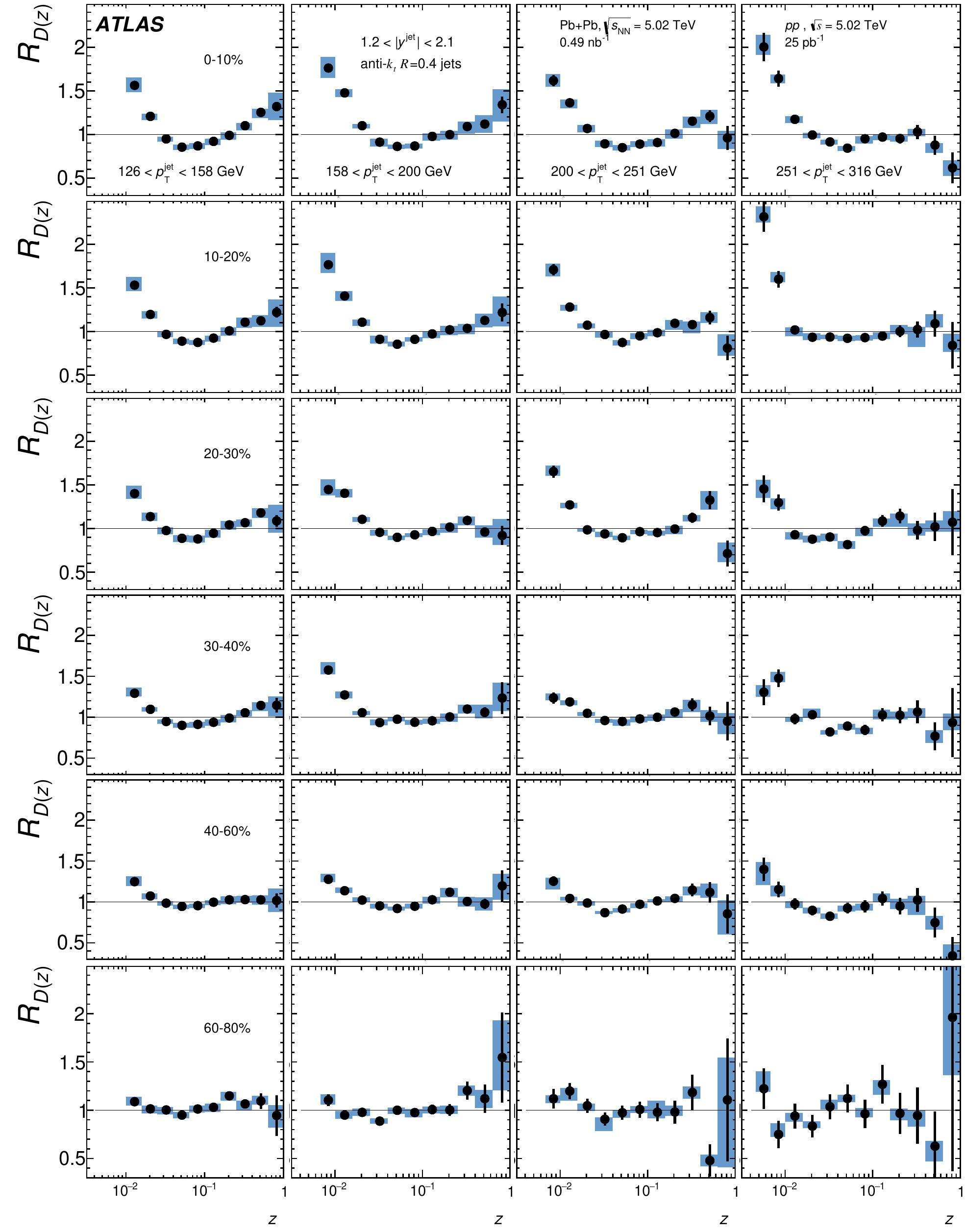}
      }
      \caption{Ratios of \Dz\  distributions in six centrality intervals of \pbpb\
      collisions to \pp\ collisions evaluated in four \ptjet\ ranges for jets with 1.2~$<|\yjet| <$~2.1. The vertical bars on the data points indicate statistical uncertainties, while the shaded bands indicate systematic uncertainties.
      Centrality decreases from top to bottom panels and \ptjet\ increases from left to right panels.}
      \label{fig:ratiosdzeta3}
   \end{figure}

\begin{figure}[ht]
      \centerline{
	 \includegraphics[width=\textwidth]{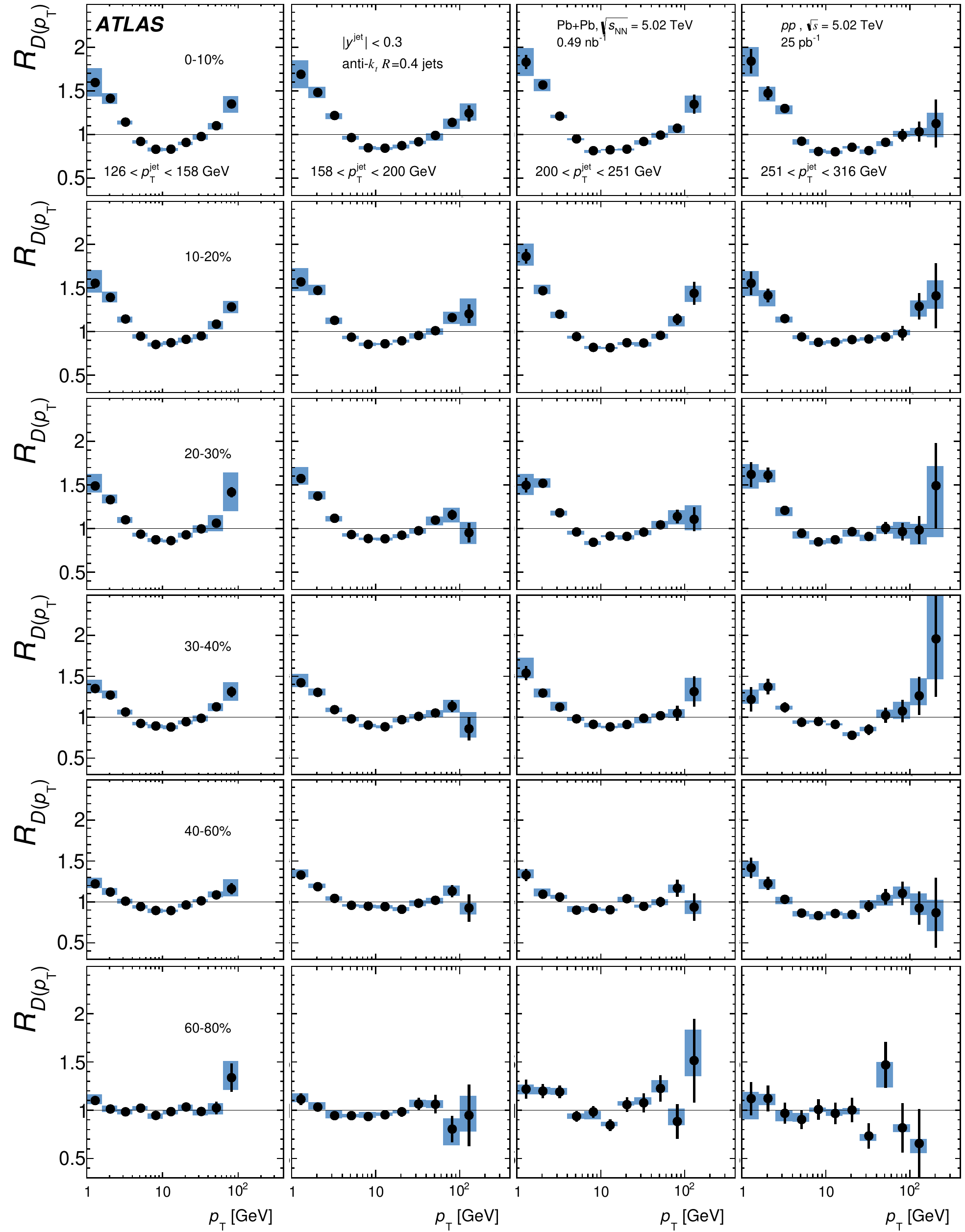}
      }
      \caption{Ratios of \Dpt\  distributions in six centrality intervals of \pbpb\
   collisions to \pp\ collisions evaluated in four \ptjet\ ranges for jets with $|\yjet|<$~0.3. The vertical bars on the data points indicate statistical uncertainties, while the shaded bands indicate systematic uncertainties.
      Centrality decreases from top to bottom panels and \ptjet\ increases from left to right panels.}
      \label{fig:ratiosdpteta0}
   \end{figure}

\begin{figure}[ht]
      \centerline{
	 \includegraphics[width=\textwidth]{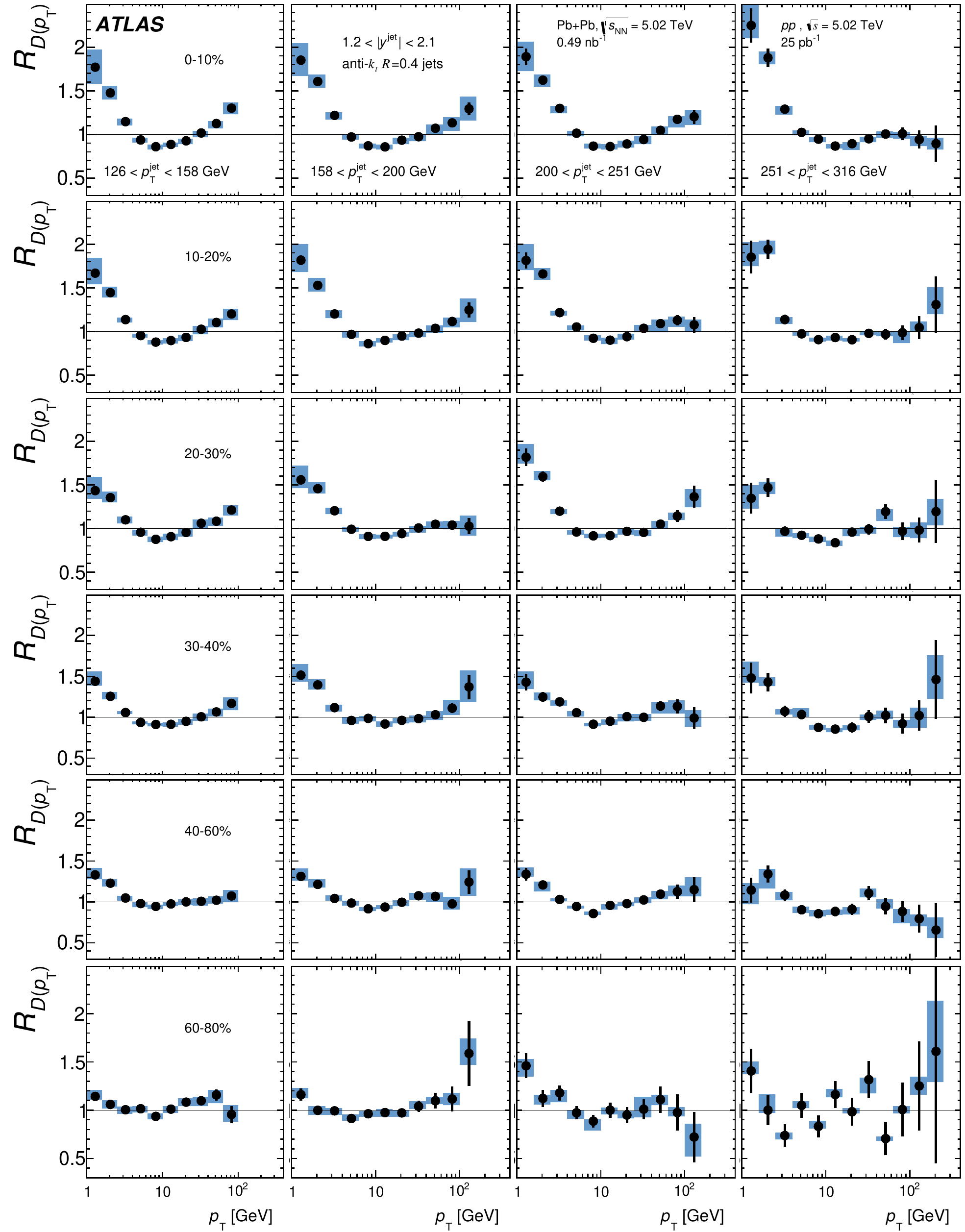}
      }
      \caption{Ratios of \Dpt\  distributions in six centrality intervals of \pbpb\
   collisions to \pp\ collisions evaluated in four \ptjet\ ranges for jets with 1.2~$<|\yjet|<$~2.1. The vertical bars on the data points indicate statistical uncertainties, while the shaded bands indicate systematic uncertainties.
      Centrality decreases from top to bottom panels and \ptjet\ increases from left to right panels.}
      \label{fig:ratiosdpteta3}
   \end{figure}

Figures~\ref{fig:ratiosdzeta0} and ~\ref{fig:ratiosdzeta3} show the $R_{D(z)}$ distributions for jets 
in the most central and most forward rapidity intervals, 0.0--0.3 and 1.2--2.1, respectively, for the six centrality intervals used in this
analysis and for four \ptjet\ intervals: 126--158~\GeV, 158--200~\GeV, 200--251~\GeV, and 251--316~\GeV.  
Figures~\ref{fig:ratiosdpteta0} and~\ref{fig:ratiosdpteta3} show $R_{D(\pT)}$ distributions for the same jet 
rapidity, centrality,  and \ptjet\ ranges. In all rapidity ranges, the $R_{D(z)}$ and $R_{D(\pT)}$ distributions have the 
same qualitative shape and centrality dependence as the rapidity-inclusive results presented above.

\FloatBarrier
%-------------------------------------------------------------------------------
\section{Discussion}
\label{sec:discussion}
In this section, the results from the previous section are further discussed and compared to theoretical models.

\begin{figure}[ht]
         \centerline{
   	 \begin{tabular}{c @{\hspace{0.1cm}} c}
	  \includegraphics[width=0.51\textwidth]{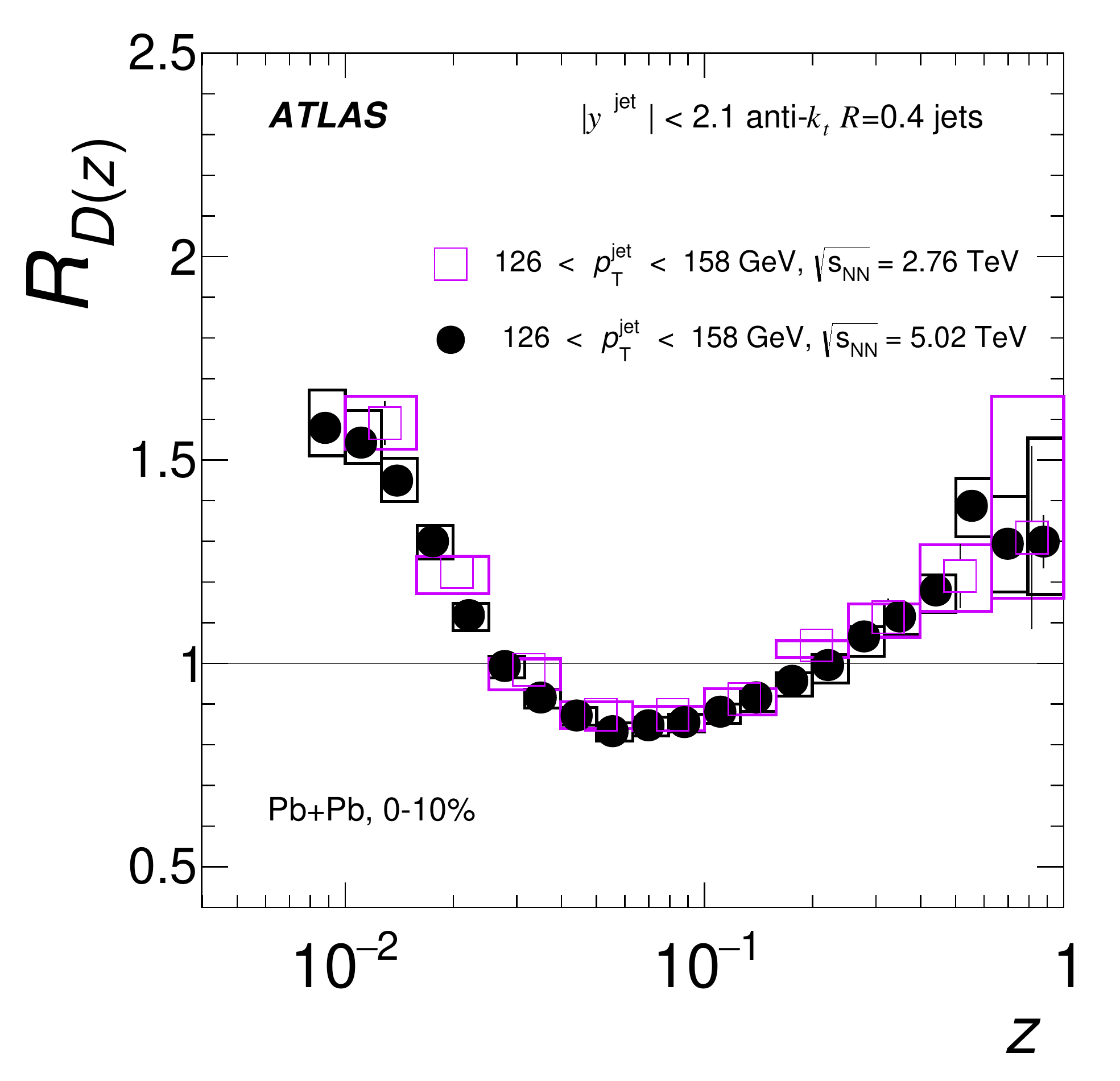} & 
	  \includegraphics[width=0.51\textwidth]{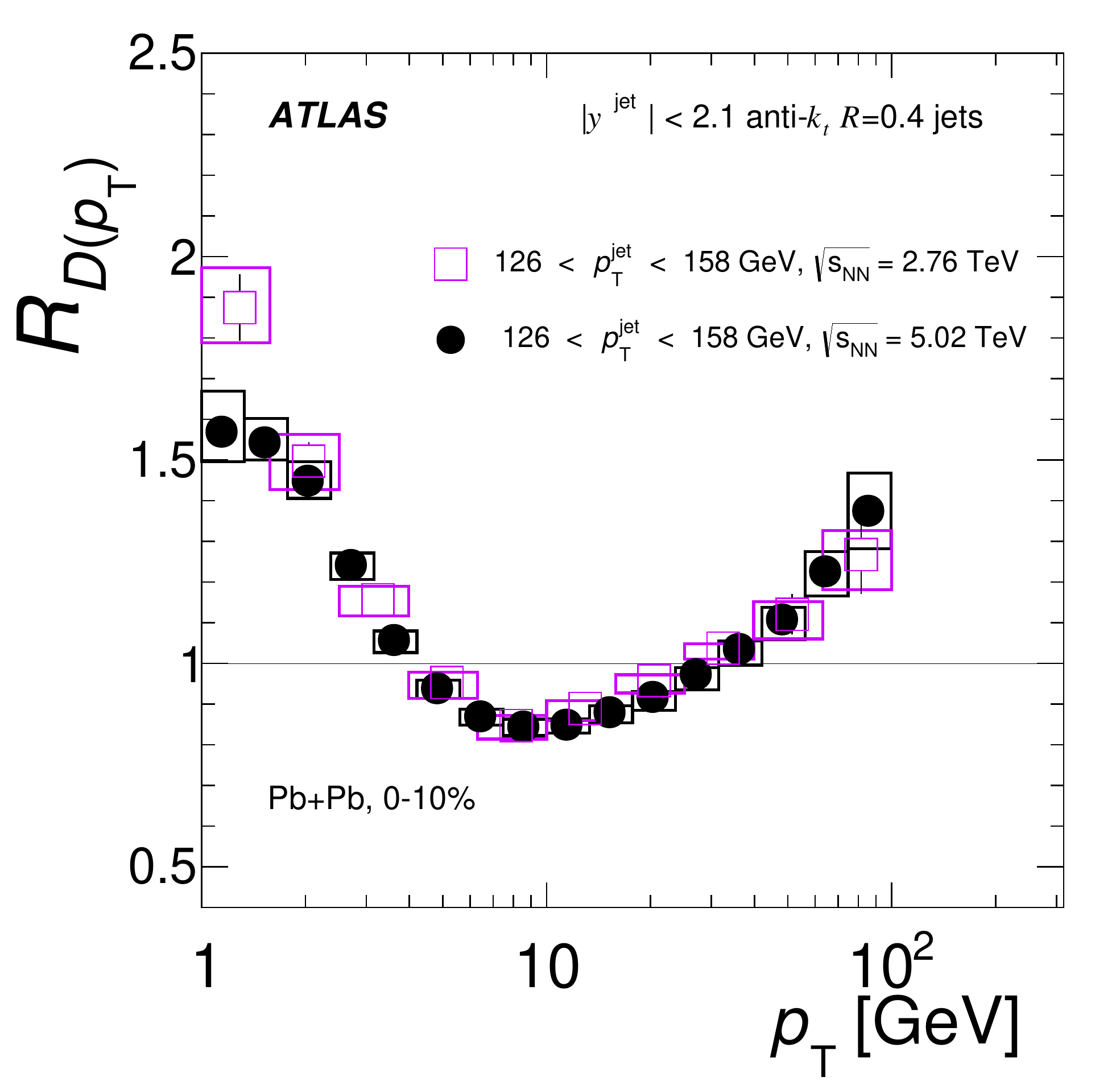} \\ 
      \end{tabular}
            }
      \caption{ $R_{D(z)}$ (left) and $R_{D(\pT)}$ (right) for 126--158~\GeV\ jets for
       collision energies of 5.02~\TeV\ (this analysis) and 2.76 \TeV~\cite{Aaboud:2017bzv}. The vertical bars on the data points indicate statistical uncertainties while the boxes indicate systematic uncertainties.}
	 \label{fig:energydep}
      \end{figure}

In order to make a direct comparison with measurements at 2.76~\TeV, Figure~\ref{fig:energydep} overlays the $R_{D(z)}$ and 
 $R_{D(\pT)}$ distributions measured in 2.76~\TeV\ collisions~\cite{Aaboud:2017bzv} on those obtained in this analysis at 5.02~\TeV. 
 The two measurements at the two collision energies quantitatively agree over the entire \z\ and
 charged-particle \pT\ range of the measurement; no significant collision energy dependence is observed
 (the lowest point in the \Dpt\ ratios differs by less than two standard deviations when the statistical and systematic 
 uncertainties are combined).

\begin{figure}[ht]
         \centerline{
   	 \begin{tabular}{c @{\hspace{0.1cm}} c}
	 \includegraphics[width=0.51\textwidth]{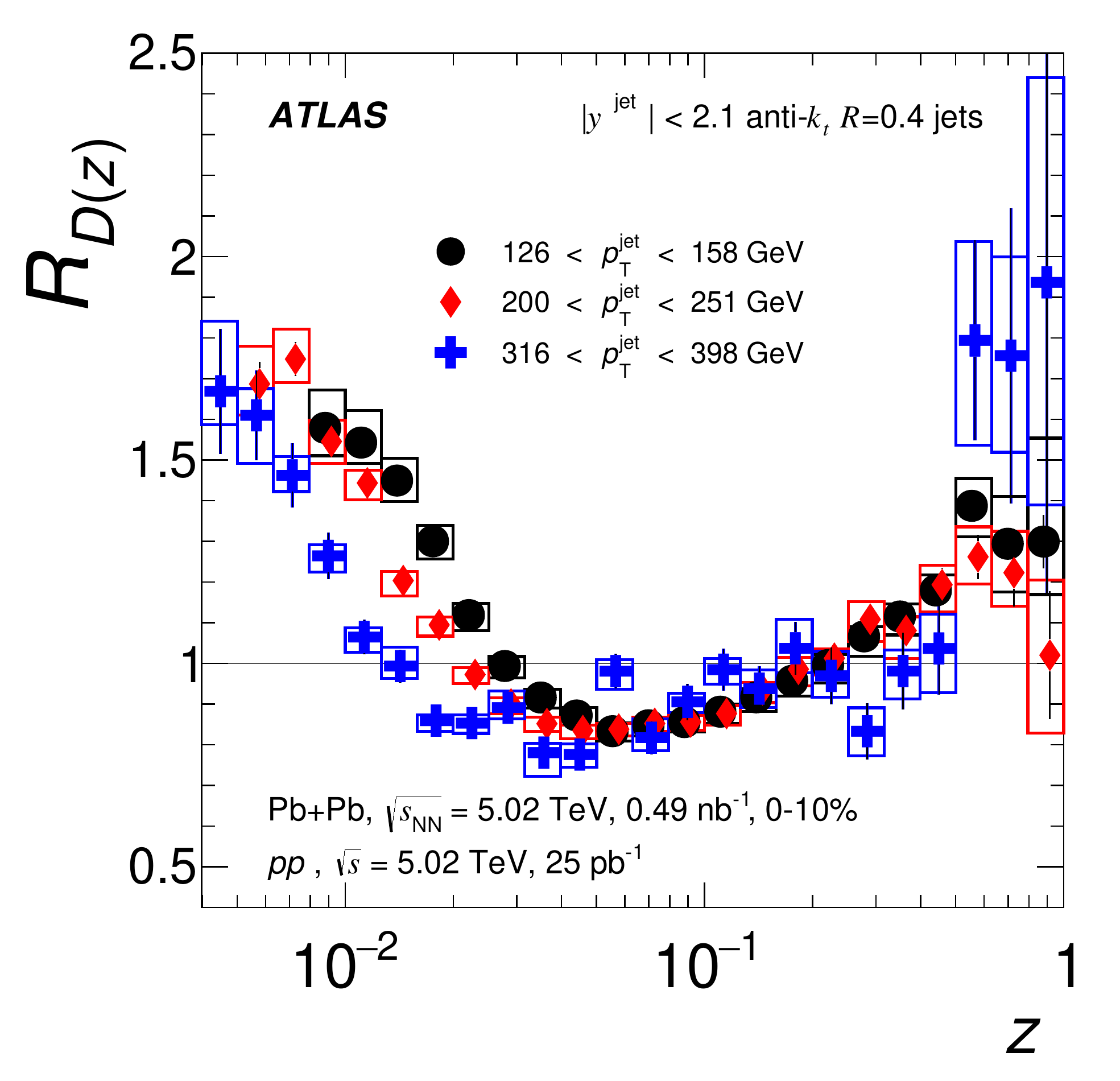} &
	 \includegraphics[width=0.51\textwidth]{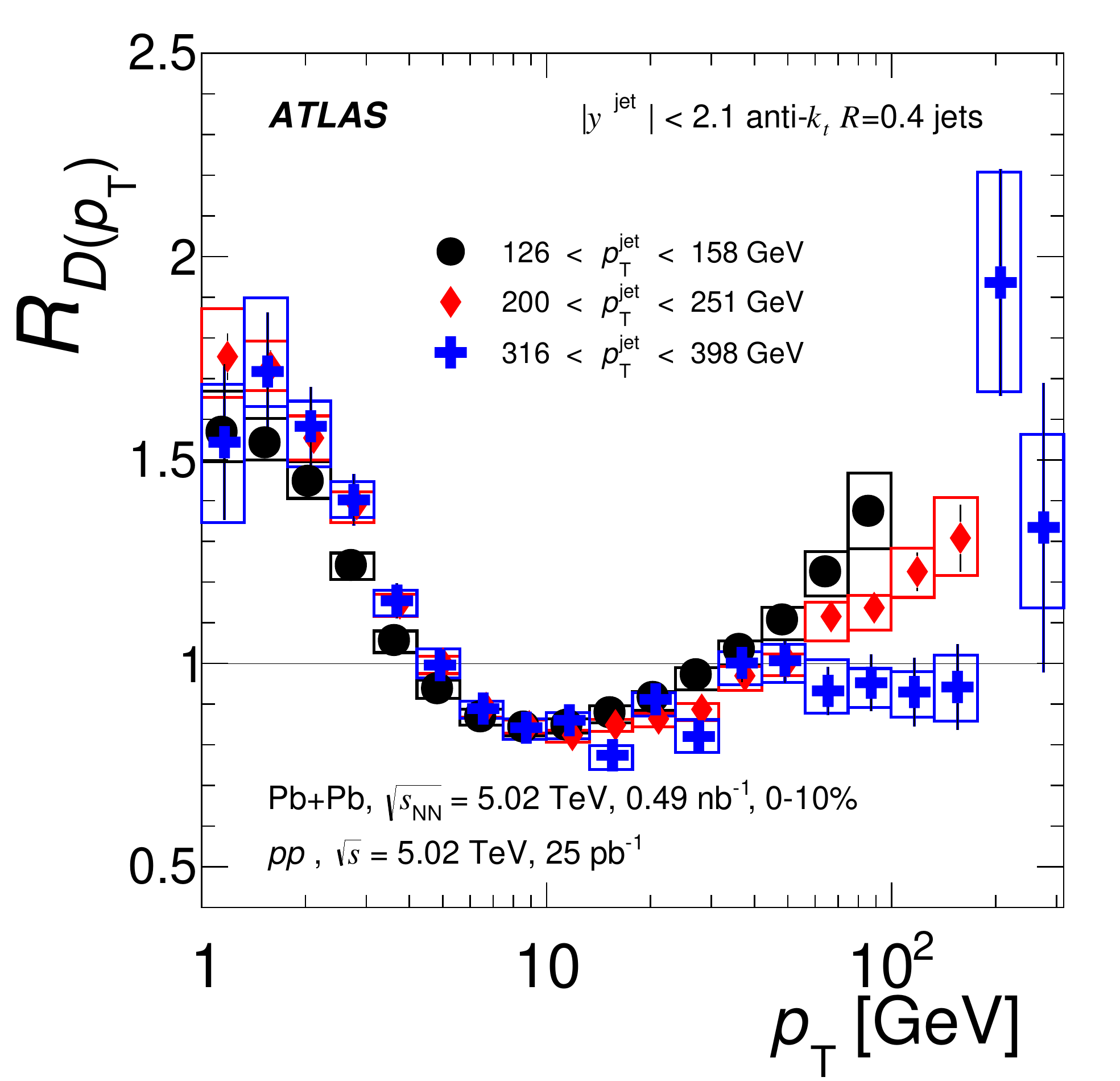} \\ 
      \end{tabular}
            }
       \caption{$R_{D(z)}$ (left) and $R_{D(\pT)}$ (right) 
       ratios for three \ptjet\ ranges: 126--158~\GeV\ (circles), 200--251~\GeV\ 
	 (diamonds) and 316--398~\GeV\ (crosses).  The statistical uncertainties
       are shown as bars and the systematic uncertainties as outlined boxes. }
	 \label{fig:jetptdep}
      \end{figure}

In order to determine how the fragmentation functions depend on \ptjet,
the fragmentation functions from three \ptjet\ intervals are compared in Figure~\ref{fig:jetptdep}.
The \Dpt\ and \Dz\ distributions are closely related to 
each other, differing, primarily, in the normalization by 
\ptjet\ in the definition of \z\ (see Eq.~(\ref{eq:zdef})). 
Therefore, a comparison of the 
modifications of the fragmentation functions as a function of \ptjet\ can show whether the size of modifications 
scales with charged-particle \z\ or with \pT.  The former would be expected for fragmentation effects, and the latter
 might indicate some scale in the QGP.
The large \ptjet\ range available in this measurement allows these two scenarios to be distinguished.
Figure~\ref{fig:jetptdep} shows that the excess of soft particles observed in central \pbpb\ collisions
exhibits a much smaller \ptjet\ dependence for the \Dpt\ ratios than for the \Dz\ ratios;
the transition from enhancement to suppression for soft fragments occurs at \pT\ around 4~\GeV\ for all \ptjet\ values investigated in this analysis.
The same comparison can be made for the hard particles.
In this case, 
Figure~\ref{fig:jetptdep} shows that the enhancement of hard fragments with $\z\gtrsim0.3$
is nearly independent of \ptjet.

\begin{figure}[ht]
	 \centerline{
	    \begin{tabular}{cc}
	       \includegraphics[width=0.7\textwidth]{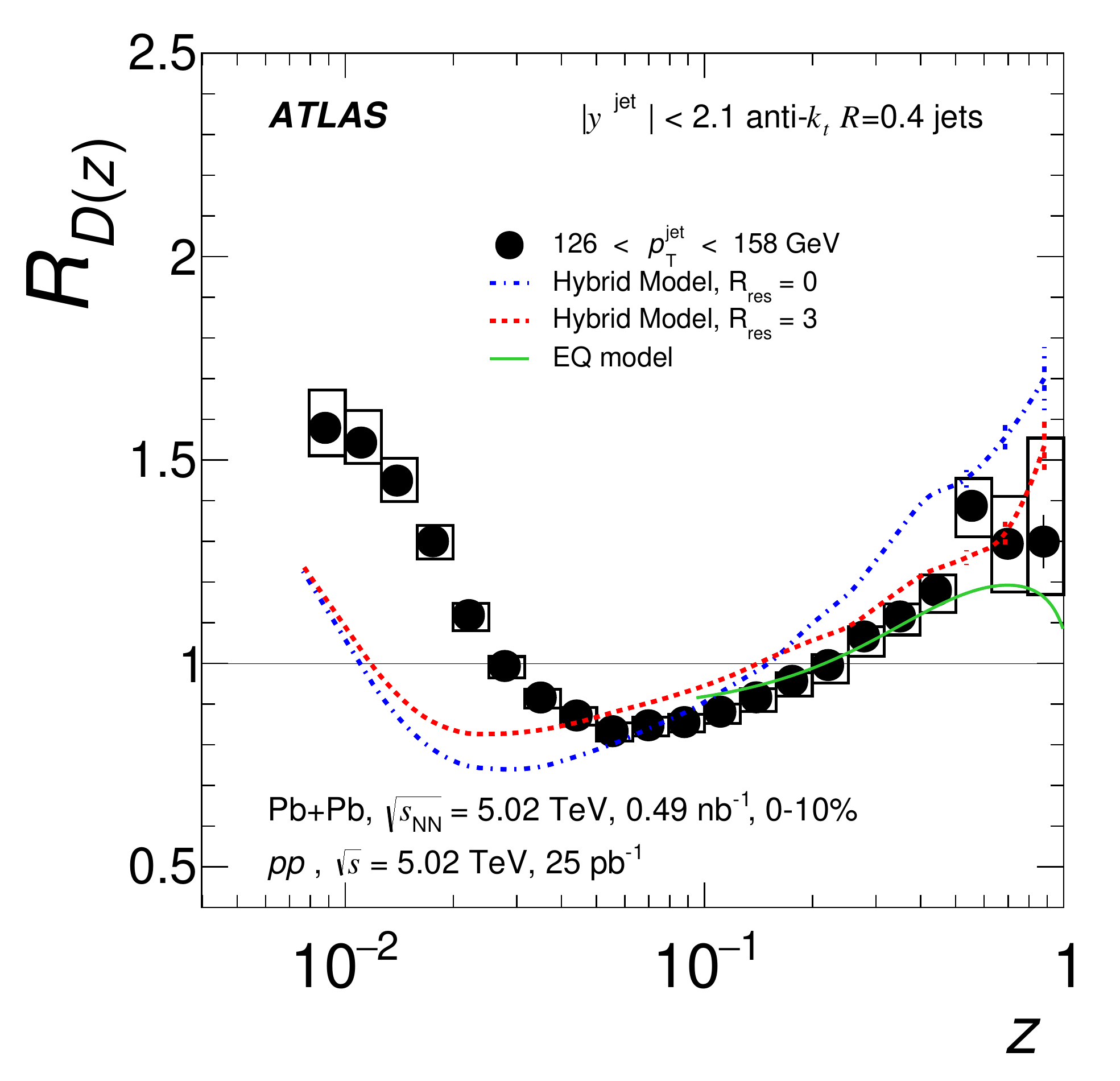}
	    \end{tabular}
	    }
	 \caption{ $R_{D(z)}$ for jets with 126~$<\ptjet<$~158~GeV
	 compared with calculations from Ref.~\cite{Hulcher:2017cpt} (Hybrid model) for $\Rres =$~0
	 (dot-dashed curve), $\Rres =$~3 (dashed curve), and to calculations from Ref.~\cite{Spousta:2015fca} (EQ model).
	 }
	 \label{fig:theorycomp1}
      \end{figure}

The fragmentation functions have been calculated within a hybrid model of jet quenching, which uses
perturbative techniques for the high-$Q^2$ processes in jet evolution
and strong coupling for the low momentum scales associated with the QGP~\cite{Casalderrey-Solana:2014bpa,Hulcher:2017cpt}. 
Within this model, there is a length scale, \Lres, which can be interpreted as the minimum distance required to resolve a parton as separate from the others in the showering process when it occurs in the QGP medium. The scale \Lres\ can be expressed in terms of the temperature of QGP, $T$, as $\Lres = \Rres/\pi T$
where \Rres\ is a parameter of the model. The fragmentation functions measured here are compared with calculations
from this model in Figure~\ref{fig:theorycomp1} for two values of
\Rres.  The calculations with $\Rres=$~3 are qualitatively consistent
with the measurement at high \z\ and \pt. 
At low \z\ and \pt, the results of the calculations are below
the data, in agreement with prior observations in comparisons to related observables~\cite{Casalderrey-Solana:2016jvj}.
Also shown in Figure~\ref{fig:theorycomp1} is a calculation from Ref.~\cite{Spousta:2015fca}
which is a phenomenological model, the \textit{Effective Quenching (EQ) Model},
incorporating energy-loss effects through two downward shifts
in the \ptjet\ spectrum: one for quark-initiated jets and a larger one for gluon-initiated jets.  In this case,
the jets fragment as in vacuum, but $R_{D(z)}$ differs from unity due to an increase in the fraction of
quark jets in \pbpb\ collisions relative to \pp\ collisions at a fixed \ptjet.  Since quark jets
are more likely to produce high-\z\ particles than gluon jets~\cite{Barate:1998cp,Abbiendi:1999pi} this causes $R_{D(z)}  > $~1 at high~\z\ in the model predictions.  The EQ model does not have a description of the soft processes from soft gluon radiation or the response
of the hot QCD matter to the jet passing through it, so the comparison with data is only appropriate
at $\z >$~0.1. 

\begin{figure}[ht]
	 \centerline{
	    \begin{tabular}{cc}
	       \includegraphics[width=0.7\textwidth]{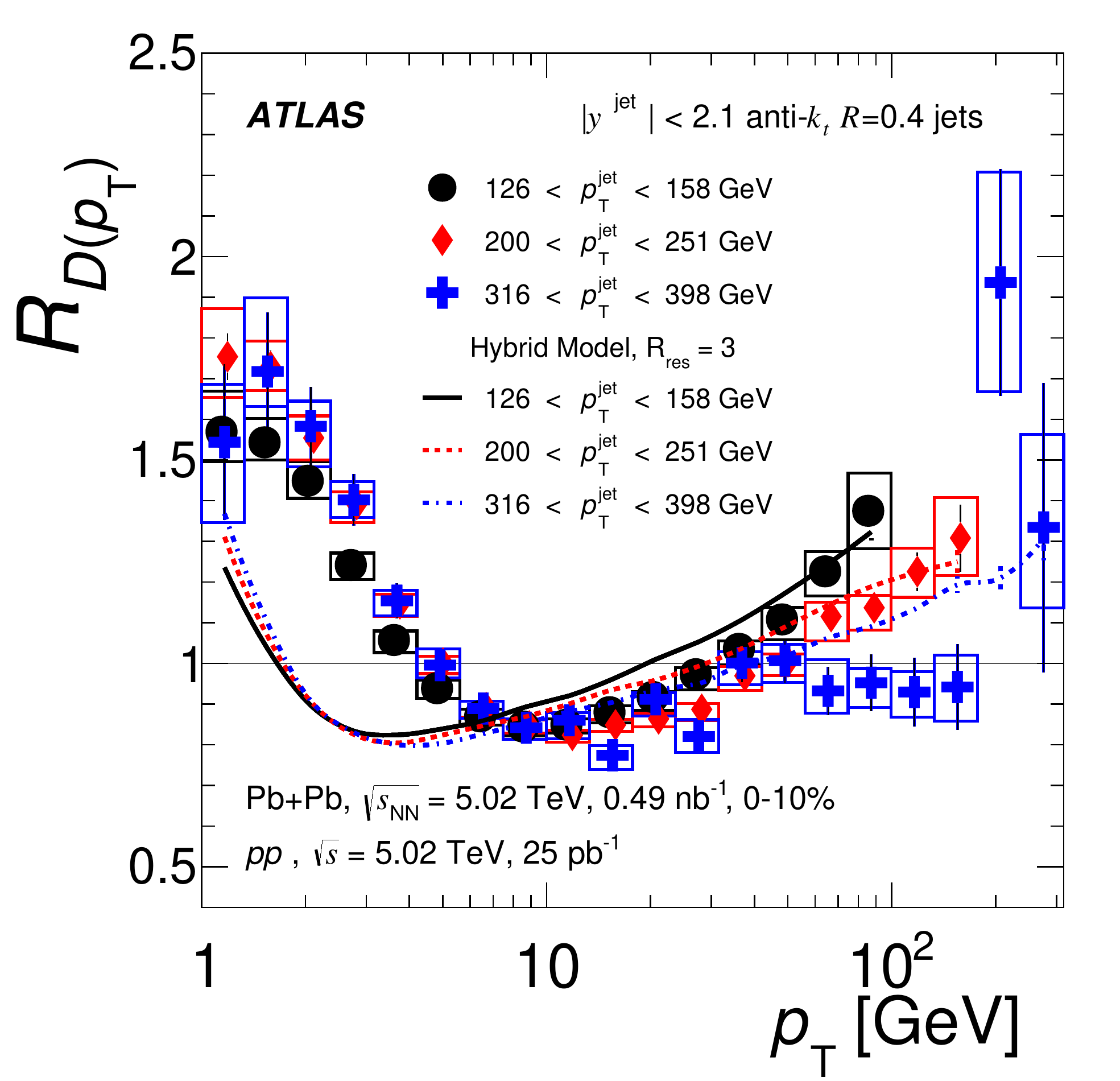}
	    \end{tabular}
	    }
	 \caption{
	 $R_{D(\pT)}$ 
       ratios for three \ptjet\ ranges: 126--158~\GeV\ (circles), 200--251~\GeV\ 
	 (diamonds) and 316--398~\GeV\ (crosses) compared with calculations from the hybrid model~\cite{Hulcher:2017cpt} with $\Rres =$~3.}
	 \label{fig:theorycomp3}
      \end{figure}

\begin{figure}[ht]
	 \centerline{
	    \begin{tabular}{cc}
	       \includegraphics[width=0.7\textwidth]{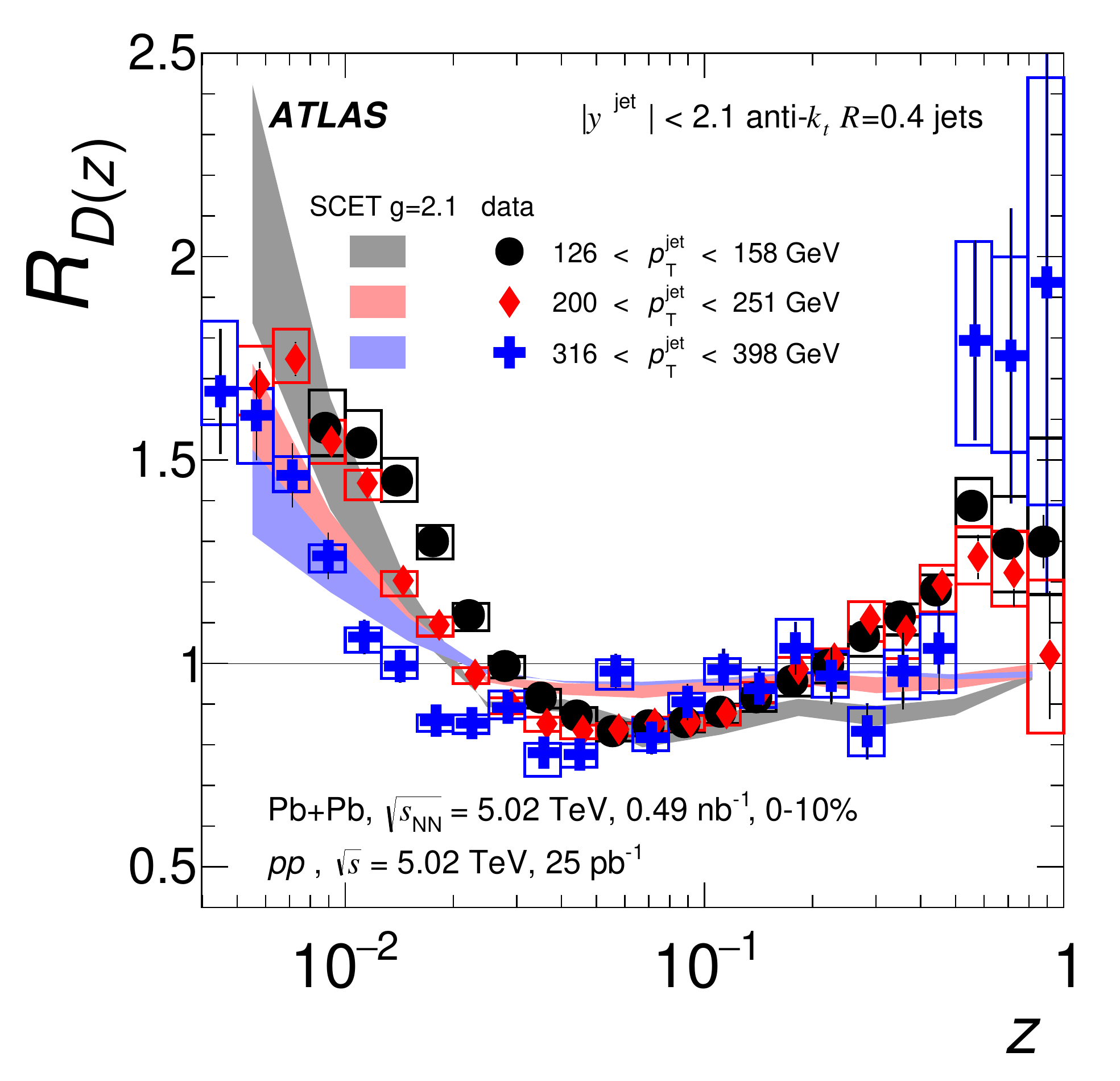}
	    \end{tabular}
	    }
	 \caption{
	 $R_{D(z)}$ for three \ptjet\ ranges: 126--158~\GeV\ (circles), 200--251~\GeV\
	 (diamonds) and 316--398~\GeV\ (crosses) compared with calculations from the SCET 
	 model~\cite{Chien:2015vja,Kang:2017frl}.
	 }
	 \label{fig:theorycomp2}
      \end{figure}

Figure~\ref{fig:theorycomp3} shows a comparison between measured $R_{D(\pT)}$ and the hybrid model calculation with $\Rres=$~3 for three \ptjet\ intervals.  The magnitude of the enhancement of high-\pT\ particles in the calculation 
agrees with the observations for \ptjet\ in the ranges 126--158~\GeV and 200--251~\GeV.
The $R_{D(z)}$ values are also compared in Figure~\ref{fig:theorycomp2} with a third model which uses 
calculations based on Soft Collinear Effective Theory (SCET)~\cite{Chien:2015vja,Kang:2017frl}.
This model well describes $R_{D(z)}$ in the low and intermediate $z$ regions, but does not reproduce the enhancement in the high-\z\ region observed in the data.

\begin{figure}[ht]
	 \centerline{
	    \includegraphics[width=\textwidth]{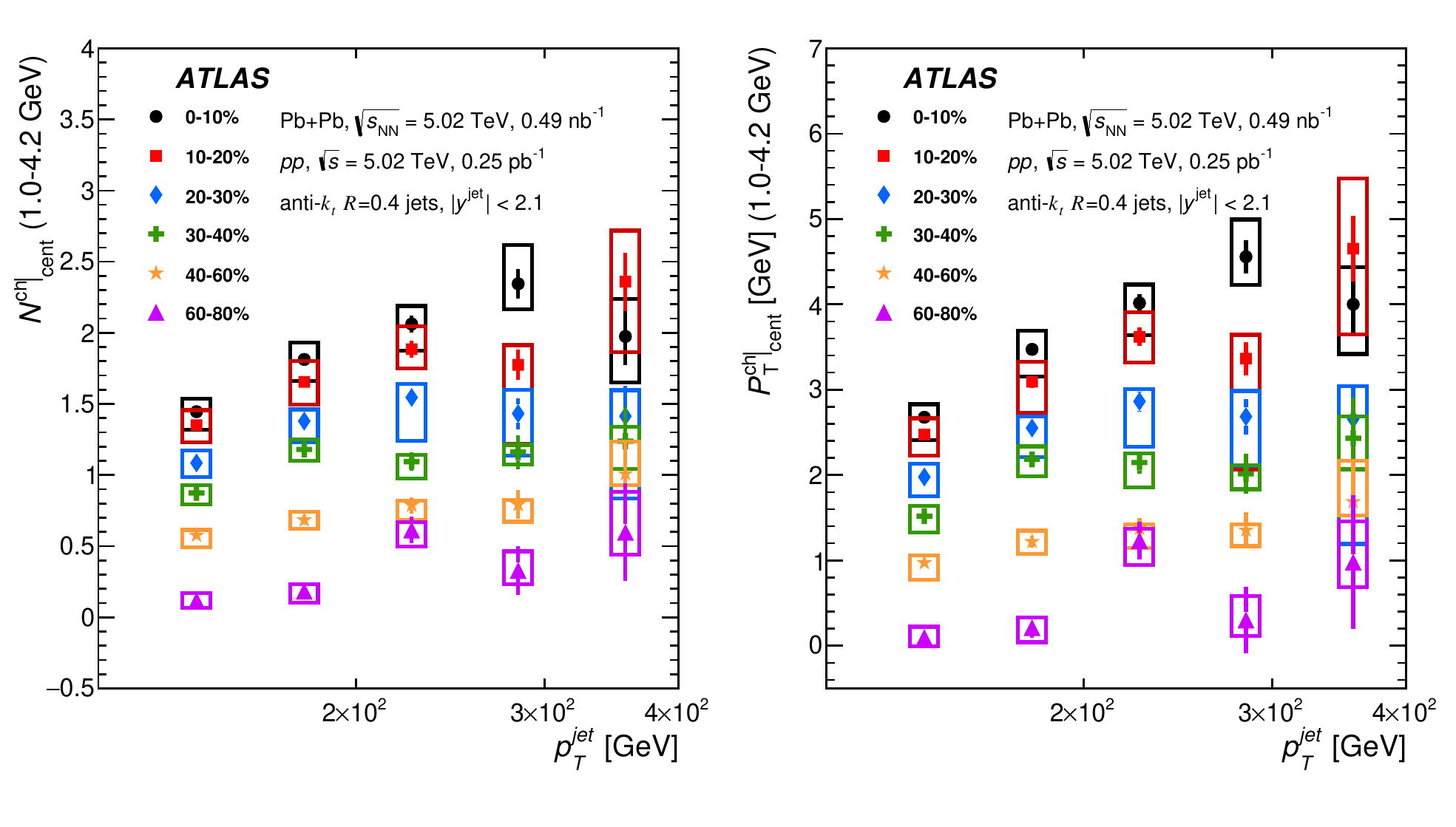}
	 }
	 \caption{Difference between \pbpb\ collisions and \pp\ collisions in the total yield of charged particles, $N^{\mathrm{ch}}\rvert_{\mathrm{cent}}$, (left) and difference in the total transverse momentum carried by charged particles, $P_{\mathrm{T}}^{\mathrm{ch}}\rvert_{\mathrm{cent}}$, (right) for particles with \pt\ from 1~$<\pT<$~4.2~\GeV\ evaluated as a function of \ptjet\ for six centrality intervals. The vertical bars on the data points indicate statistical uncertainties while the boxes indicate systematic uncertainties.}
	 \label{fig:integrals}
      \end{figure}

In order to quantify the magnitude of the low-\pT\ enhancement in the \Dpt\ distributions in \pbpb\ collisions compared to \pp collisions, the difference between the two distributions is evaluated for the \ptjet\ and centrality intervals used in this analysis:

   \begin{equation*}
      N^{\mathrm{ch}}\rvert_{\mathrm{cent}} \equiv \int_{\pTmin}^{\pTmax} \left( \Dpt\rvert_{\mathrm{cent}} - \Dpt\rvert_{\pp}\right) d \pT ,
      \label{eq:deltaN}
   \end{equation*}
where ``cent'' represents one of the six centrality intervals, and the values of \pTmin\ and \pTmax\ are boundaries of the low \pT\ enhancement region, chosen to be 1.0 and 4.2 GeV, respectively.
In addition, the \pT-weighted difference between the same quantities is also computed:
   \begin{equation*}
      P_{\mathrm{T}}^{\mathrm{ch}}\rvert_{\mathrm{cent}} \equiv \int_{\pTmin}^{\pTmax} \left( \Dpt\rvert_{\mathrm{cent}} - \Dpt\rvert_{\pp}\right) \pT d \pT.
      \label{eq:deltapt}
   \end{equation*}
The $P_{\mathrm{T}}^{\mathrm{ch}}\rvert_{\mathrm{cent}}$ represents the total transverse momentum carried by particles in the low \pT\ enhancement region. The dependence of $N^{\mathrm{ch}}\rvert_{\mathrm{cent}}$ and $P_{\mathrm{T}}^{\mathrm{ch}}\rvert_{\mathrm{cent}}$ on \ptjet\ and centrality is presented in Figure~\ref{fig:integrals}.  Overall, both quantities are found to increase as a function of \ptjet\ and collision centrality.  In the most central collisions, $N^{\mathrm{ch}}$ increases from approximately 1.5 to 2.0 particles over the \ptjet\ range of this measurement.  The amount of transverse momentum carried by these particles increases from approximately 2.5~\GeV\ to 4~\GeV\ over the same \ptjet\ range.  In peripheral collisions, the number of particles contributing to the enhancement is much smaller, approximately 0.2 particles carrying less than 0.5~\GeV\ of transverse momentum in the lowest \ptjet\ range. 
These results are in qualitative agreement with measurements of the same quantities in $\sqrtsnn~=$~2.76~\TeV\ \pbpb\ collisions~\cite{Aaboud:2017bzv}; however, the \ptjet\ ranges
are not the same as used in this analysis and the \ptjet\ dependence is not reported in that measurement.

\begin{figure}[ht]
	 \centerline{
	    \begin{tabular}{cc}
	       \includegraphics[width=1.0\textwidth]{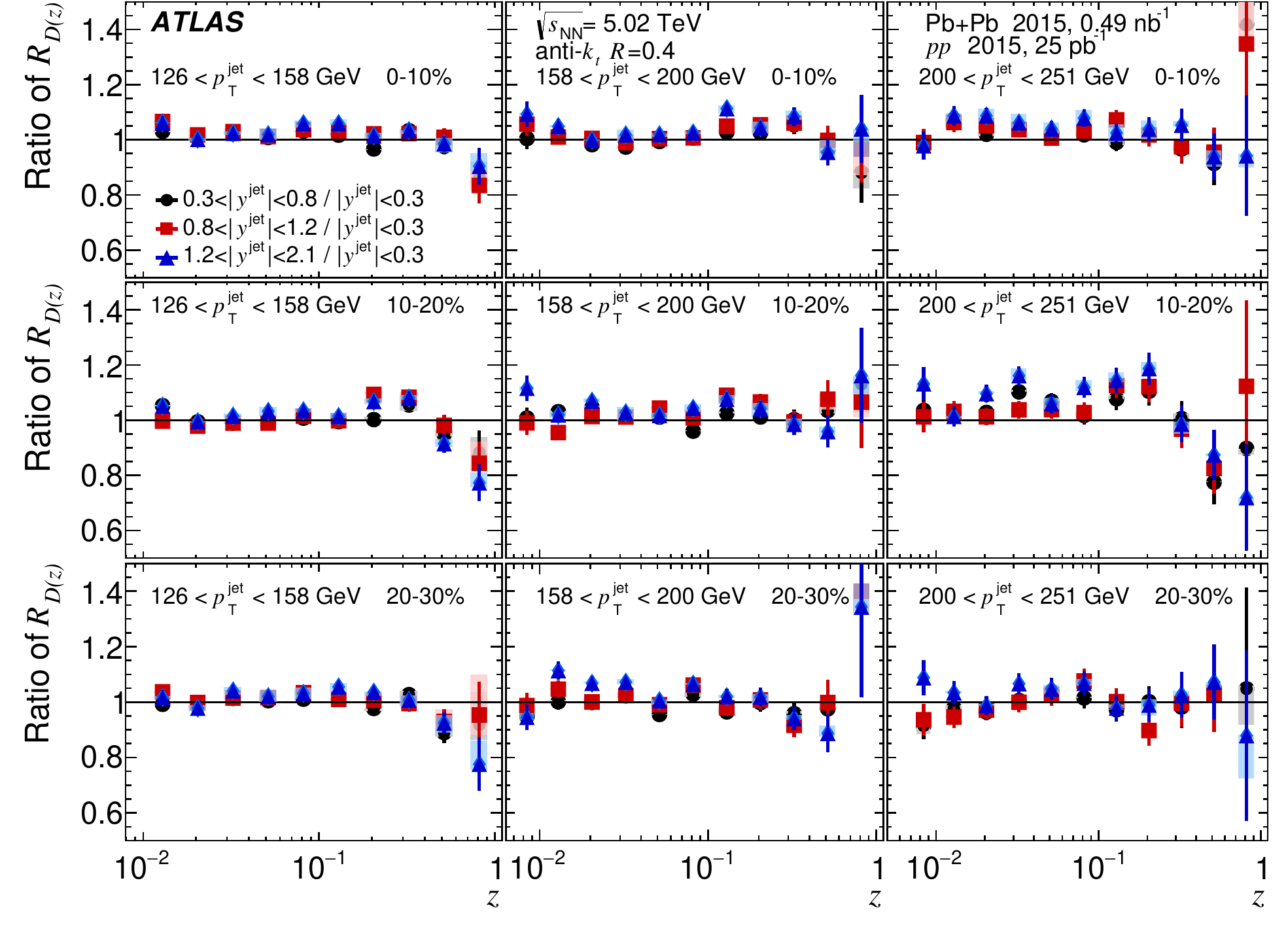}
	    \end{tabular}
	    }
   \caption{Ratio of the rapidity-selected $R_{D(z)}$ distributions to the $R_{D(z)}$ distributions measured in $|\yjet|<0.3$ for three \ptjet\ ranges and three centrality intervals. The vertical bars on the data points indicate statistical uncertainties while the shaded bands indicate systematic uncertainties.}
   \label{fig:rapiditydoubleratiosaltref}
\end{figure}

\begin{figure}[ht]
	 \centerline{
	    \begin{tabular}{cc}
	       \includegraphics[width=1.0\textwidth]{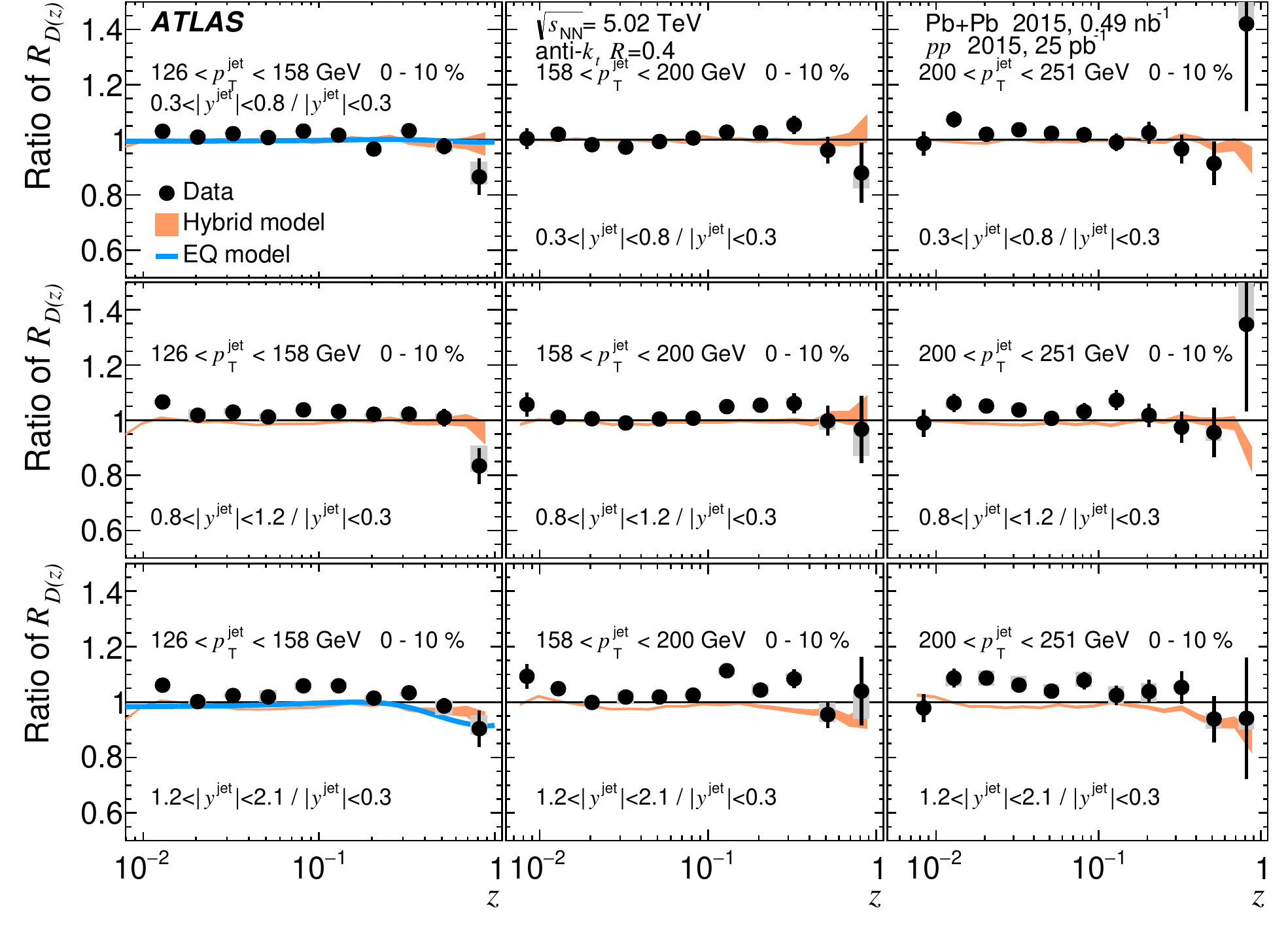}
	    \end{tabular}
	    }
   \caption{Comparison of the measured ratio of the rapidity-selected $R_{D(z)}$ distributions to the $R_{D(z)}$ distributions measured in $|\yjet|<0.3$ and the same quantity evaluated in the hybrid model~\cite{Hulcher:2017cpt} for $\Rres =$~3 and in the EQ model~\cite{Spousta:2015fca}. The comparison with the hybrid model is done for three \ptjet\ ranges in 0--10\% central collisions. The comparison with the EQ model is shown for 126--158~\GeV\ \ptjet\ interval.  
   The vertical bars on the data points indicate statistical uncertainties while the shaded bars indicate systematic uncertainties. The band represents the statistical uncertainty of the 
   calculations.}
   \label{fig:rapiditydoubleratiosaltreftheory}
\end{figure}

   In order to quantify the rapidity dependence, the ratio of $R_{D(z)}$ in the rapidity intervals 0.3--0.8, 0.8--1.2, and 1.2--2.1 to the $R_{D(z)}$ in $|\yjet|<$~0.3 is shown in Figure~\ref{fig:rapiditydoubleratiosaltref} for \ptjet\ intervals of 126--158~\GeV, 158--200~\GeV, and 200--251~\GeV\ and for 0--10\%, 10--20\%, and 20--30\% central collisions. 
   A similar quantity was reported in Ref.~\cite{Aaboud:2017bzv}
   for 100--398~\GeV\ jets at 2.76~\TeV.  In that measurement, a small rapidity dependence for $R_{D(z)}$
   is observed at high \z\ for jets with $|\yjet|<$~0.8;
   however, no strong conclusion could be drawn due to the size of the uncertainties.
   The \ptjet\ intervals used in the measurement presented here 
   are selected to be similar to those used in the measurement of fragmentation functions at 2.76~\TeV. 
   Furthermore, jets populating the 200--251~\GeV\ \ptjet\ interval in collisions at 5.02~\TeV\ have similar fractions of quark- and gluon-initiated jets as jets having \pt between 126 and 158~\GeV\ in 2.76~\TeV\ collisions.     
   The ratios of $R_{D(z)}$ evaluated in various rapidity intervals to the  most central rapidity $R_{D(z)}$ in different \ptjet\ intervals suggest with a low significance a small enhancement of yields of fragments with low and intermediate \z\ and reduction of yields of high-\z\ fragments for more forward jets in the most central \PbPb\ collisions. However, the observation for high-\z\ fragments is of limited significance due to the limited size of the available data sample. Figure~\ref{fig:rapiditydoubleratiosaltreftheory} shows the same ratios for the 0--10\% centrality
    interval compared with calculations from the Hybrid model~\cite{Hulcher:2017cpt} and
    the Effective Quenching model~\cite{Spousta:2015fca}. Both calculations are consistent with the data for jets with $|\yjet|<$~1.2 with larger deviations in rapidity interval 1.2~$<|\yjet|<$~2.1.

\FloatBarrier

%-------------------------------------------------------------------------------
\section{Summary}
\label{sec:summary}
This paper presents an analysis of 0.49~nb$^{-1}$ of \PbPb\ and 25~pb$^{-1}$ of \pp\ collisions at $\sqrtsnn = 5.02$~\TeV\ using data collected with the ATLAS detector at the LHC in 2015.
The analysis measures the fragmentation functions of jets into charged particles
and the distributions of charged-particle transverse momenta within \RFour\ \antikt\ jets with $|\yjet| <$2.1 and with \ptjet\ from 126 to 398~\GeV. The studies are
 performed as a function of the event centrality, jet rapidity 
and jet transverse momentum for charged particles with transverse momentum greater than 1~\GeV\ .  

Centrality-dependent modifications to these fragmentation functions in \PbPb\ collisions are observed when compared with those measured in \pp\ collisions. The magnitude of these modifications increases with increasing collision centrality. 
The ratios of fragmentation functions evaluated in \PbPb\ collisions to those in \pp\ collisions exhibit 
enhancements both for transverse momentum less than 4~\GeV\
 and for $z \gtrsim 0.3$. 
Between these two enhancements there is a suppression of the fragmentation functions in \pbpb\ collisions compared
to \pp\ collisions.
 The enhancement of yields of low and high transverse momentum fragments is as large as 70\% and 30\%, 
respectively, in  central collisions.  The depletion of fragment yields with intermediate \pT\ and \z\ is as large as 20\%. 
The difference in charged-particle multiplicity and total transverse momentum in \pbpb\ compared to
\pp\ collisions for $1.0 < \pT < 4.2$~\GeV\ range
 increases with increasing centrality
and jet transverse momentum.  No significant dependence of the high-\z\ enhancement on the transverse momentum
of the jet is observed. The SCET model describes the low \pT\ excess and the EQ and Hybrid models 
describe the high-\z\ excess, but none of the models describes the modification of the full fragmentation functions. A small increase in the modification of yields of fragments with low and intermediate \z\ is observed in forward jets compared to those at central rapidity.  These measurements provide new information about the jet transverse momentum and rapidity dependence of the modifications to jet fragmentation in \pbpb\ collisions and, together with other jet measurements in heavy-ion collisions, will constrain 
models of jet quenching in the QGP created in heavy-ion collisions.

\clearpage
%-------------------------------------------------------------------------------

%-------------------------------------------------------------------------------
\section*{Acknowledgements}
%-------------------------------------------------------------------------------

We thank CERN for the very successful operation of the LHC, as well as the
support staff from our institutions without whom ATLAS could not be
operated efficiently.

We acknowledge the support of ANPCyT, Argentina; YerPhI, Armenia; ARC, Australia; BMWFW and FWF, Austria; ANAS, Azerbaijan; SSTC, Belarus; CNPq and FAPESP, Brazil; NSERC, NRC and CFI, Canada; CERN; CONICYT, Chile; CAS, MOST and NSFC, China; COLCIENCIAS, Colombia; MSMT CR, MPO CR and VSC CR, Czech Republic; DNRF and DNSRC, Denmark; IN2P3-CNRS, CEA-DRF/IRFU, France; SRNSFG, Georgia; BMBF, HGF, and MPG, Germany; GSRT, Greece; RGC, Hong Kong SAR, China; ISF, I-CORE and Benoziyo Center, Israel; INFN, Italy; MEXT and JSPS, Japan; CNRST, Morocco; NWO, Netherlands; RCN, Norway; MNiSW and NCN, Poland; FCT, Portugal; MNE/IFA, Romania; MES of Russia and NRC KI, Russian Federation; JINR; MESTD, Serbia; MSSR, Slovakia; ARRS and MIZ\v{S}, Slovenia; DST/NRF, South Africa; MINECO, Spain; SRC and Wallenberg Foundation, Sweden; SERI, SNSF and Cantons of Bern and Geneva, Switzerland; MOST, Taiwan; TAEK, Turkey; STFC, United Kingdom; DOE and NSF, United States of America. In addition, individual groups and members have received support from BCKDF, the Canada Council, CANARIE, CRC, Compute Canada, FQRNT, and the Ontario Innovation Trust, Canada; EPLANET, ERC, ERDF, FP7, Horizon 2020 and Marie Sk{\l}odowska-Curie Actions, European Union; Investissements d'Avenir Labex and Idex, ANR, R{\'e}gion Auvergne and Fondation Partager le Savoir, France; DFG and AvH Foundation, Germany; Herakleitos, Thales and Aristeia programmes co-financed by EU-ESF and the Greek NSRF; BSF, GIF and Minerva, Israel; BRF, Norway; CERCA Programme Generalitat de Catalunya, Generalitat Valenciana, Spain; the Royal Society and Leverhulme Trust, United Kingdom.

The crucial computing support from all WLCG partners is acknowledged gratefully, in particular from CERN, the ATLAS Tier-1 facilities at TRIUMF (Canada), NDGF (Denmark, Norway, Sweden), CC-IN2P3 (France), KIT/GridKA (Germany), INFN-CNAF (Italy), NL-T1 (Netherlands), PIC (Spain), ASGC (Taiwan), RAL (UK) and BNL (USA), the Tier-2 facilities worldwide and large non-WLCG resource providers. Major contributors of computing resources are listed in Ref.~\cite{ATL-GEN-PUB-2016-002}.

\printbibliography

\clearpage
% ATLAS Collaboration author list
% Reference date of HION-2017-04 is 2018-03-06
% Author list last updated on date 25-JUL-18
% Data extracted on 25-Jul-2018 for paper reference HION-2017-04
% at 9:43am
 
\begin{flushleft}
{\Large The ATLAS Collaboration}

\bigskip

M.~Aaboud$^\textrm{\scriptsize 34d}$,    
G.~Aad$^\textrm{\scriptsize 99}$,    
B.~Abbott$^\textrm{\scriptsize 124}$,    
O.~Abdinov$^\textrm{\scriptsize 13,*}$,    
B.~Abeloos$^\textrm{\scriptsize 128}$,    
D.K.~Abhayasinghe$^\textrm{\scriptsize 91}$,    
S.H.~Abidi$^\textrm{\scriptsize 164}$,    
O.S.~AbouZeid$^\textrm{\scriptsize 39}$,    
N.L.~Abraham$^\textrm{\scriptsize 153}$,    
H.~Abramowicz$^\textrm{\scriptsize 158}$
% ...
\bigskip
\\

% ...
$^{13}$Institute of Physics, Azerbaijan Academy of Sciences, Baku; Azerbaijan.\\
% ...
$^{34}$$^{(a)}$Facult\'e des Sciences Ain Chock, R\'eseau Universitaire de Physique des Hautes Energies - Universit\'e Hassan II, Casablanca;$^{(b)}$Centre National de l'Energie des Sciences Techniques Nucleaires (CNESTEN), Rabat;$^{(c)}$Facult\'e des Sciences Semlalia, Universit\'e Cadi Ayyad, LPHEA-Marrakech;$^{(d)}$Facult\'e des Sciences, Universit\'e Mohamed Premier and LPTPM, Oujda;$^{(e)}$Facult\'e des sciences, Universit\'e Mohammed V, Rabat; Morocco.\\
% ...
$^{39}$Niels Bohr Institute, University of Copenhagen, Copenhagen; Denmark.\\
% ...
$^{91}$Department of Physics, Royal Holloway University of London, Egham; United Kingdom.\\
% ...
$^{99}$CPPM, Aix-Marseille Universit\'e, CNRS/IN2P3, Marseille; France.\\
% ...
$^{124}$Homer L. Dodge Department of Physics and Astronomy, University of Oklahoma, Norman OK; United States of America.\\
% ...
$^{128}$LAL, Universit\'e Paris-Sud, CNRS/IN2P3, Universit\'e Paris-Saclay, Orsay; France.\\
% ...
$^{153}$Department of Physics and Astronomy, University of Sussex, Brighton; United Kingdom.\\
% ...
$^{158}$Raymond and Beverly Sackler School of Physics and Astronomy, Tel Aviv University, Tel Aviv; Israel.\\
% ...
$^{164}$Department of Physics, University of Toronto, Toronto ON; Canada.\\
% ...
$^{a}$ Also at Borough of Manhattan Community College, City University of New York, NY; United States of America.\\
$^{b}$ Also at Centre for High Performance Computing, CSIR Campus, Rosebank, Cape Town; South Africa.\\
$^{c}$ Also at CERN, Geneva; Switzerland.\\
$^{d}$ Also at CPPM, Aix-Marseille Universit\'e, CNRS/IN2P3, Marseille; France.\\
$^{e}$ Also at D\'epartement de Physique Nucl\'eaire et Corpusculaire, Universit\'e de Gen\`eve, Gen\`eve; Switzerland.\\
% ...
$^{*}$ Deceased

\end{flushleft}

% Created with Glance <Atlas.Glance@cern.ch>

\end{document}